\documentclass[12pt]{article}
\usepackage{a4wide}
\usepackage{latexsym}
\usepackage{amsmath}
\usepackage{amsfonts}
\usepackage{amscd}
\usepackage{amssymb}
\usepackage{cite}
\usepackage{slashed} 
\usepackage{axodraw} 

\usepackage{pslatex}
\usepackage{graphicx}
\usepackage[latin1,utf8]{inputenc}
\usepackage[T1]{fontenc}

\allowdisplaybreaks

\newcommand{\bq}{\begin{eqnarray}}
\newcommand{\eq}{\end{eqnarray}}
\newcommand{\eps}{\varepsilon}

\newcommand{\iterintmodular}[2]{F\left(#1;#2\right)}

\newcommand{\qorder}{50}

\begin{document}

\thispagestyle{empty}

\begin{flushright}
  MITP/18-117 
% \\ version of \today
\end{flushright}

\vspace{1.5cm}

\begin{center}
  {\Large\bf The electron self-energy in QED at two loops revisited \\
  }
  \vspace{1cm}
  {\large Ina H\"onemann ${}^{a}$, Kirsten Tempest ${}^{a,b}$ and Stefan Weinzierl ${}^{a}$ \\
  \vspace{1cm}
      {\small ${}^{a}$ \em PRISMA Cluster of Excellence, Institut f{\"u}r Physik, }\\
      {\small \em Johannes Gutenberg-Universit{\"a}t Mainz,}\\
      {\small \em D - 55099 Mainz, Germany}\\
  \vspace{2mm}
      {\small ${}^{b}$ \em Department of Physics, University of Toronto,} \\
      {\small \em 60 St George St.,Toronto, Ontario, M5S 1A7, Canada} \\
  } 
\end{center}

\vspace{2cm}

% abstract ---------------------------------------
\begin{abstract}\noindent
  {
We reconsider the two-loop electron self-energy in quantum electrodynamics.
We present a modern calculation, where all relevant two-loop integrals are expressed
in terms of iterated integrals of modular forms.
As boundary points of the iterated integrals we consider the four cases $p^2=0$, $p^2=m^2$, $p^2=9m^2$ and $p^2=\infty$.
The iterated integrals have $q$-expansions, which can be used for the numerical evaluation.
We show that a truncation of the $q$-series to order ${\mathcal O}(q^{30})$ gives numerically for the finite part 
of the self-energy a relative precision better than $10^{-20}$ for all real values $p^2/m^2$.
   }
\end{abstract}

\vspace*{\fill}

% -----------------------------------------------------------------------------
\newpage

\section{Introduction}
\label{sec:intro}

The two-loop contribution to the electron self-energy in quantum electrodynamics (QED) has been
computed for the first time by A. Sabry \cite{Sabry:1962} in 1962.
Already at that time it was noticed that the calculation involves certain elliptic integrals.
For lack of better techniques at that time the integrands have been approximated by series expansions, and 
the analysis has been restricted to the region above the threshold.
We are now in a better position: Feynman integrals associated to elliptic curves are now a current topic of research
interest and our techniques to compute these integrals have 
evolved \cite{Laporta:2004rb,MullerStach:2011ru,Adams:2013nia,Bloch:2013tra,Remiddi:2013joa,Adams:2014vja,Bloch:2014qca,Adams:2015gva,Adams:2015ydq,Sogaard:2014jla,Bloch:2016izu,Remiddi:2016gno,Adams:2016xah,Bonciani:2016qxi,vonManteuffel:2017hms,Adams:2017tga,Adams:2017ejb,Bogner:2017vim,Ablinger:2017bjx,Remiddi:2017har,Bourjaily:2017bsb,Hidding:2017jkk,Broedel:2017kkb,Broedel:2017siw,Broedel:2018iwv,Adams:2018yfj,Adams:2018bsn,Adams:2018kez,Broedel:2018qkq,Bourjaily:2018yfy,Bourjaily:2018aeq,Besier:2018jen,Mastrolia:2018uzb,Ablinger:2018zwz}.
It is therefore natural to revisit the two-loop electron self-energy in QED and to present the result in a modern language.
The two-loop electron self-energy is of course 
the central piece for the determination of the $\alpha^2$-term of the 
electron mass renormalisation constant $Z_m$ and the electron field renormalisation constant $Z_2$.
Let us note that in the $\overline{\mathrm{MS}}$-scheme only the pole terms are relevant whereas in the on-shell scheme
all loop integrals are evaluated on-shell at $p^2=m^2$.
In both cases, we do not encounter elliptic integrals for the determination of the renormalisation constants.
In this paper we are interested in the finite part of the two-loop electron self-energy for generic values $p^2/m^2$
where elliptic integrals do occur.
We view the electron self-energy as a (off-shell and gauge-dependent) building block entering two-loop scattering amplitudes
in QED.
We are interested in an analytic expression for the two-loop self-energy 
and in efficient methods for the numerical evaluation of the occurring transcendental functions.
As a new result our final answer allows a numerical evaluation with arbitrary precision for all real
values of $p^2/m^2$.
Other methods for the numerical evaluation of some of the relevant Feynman integrals 
have been discussed in \cite{Broadhurst:1993mw,Fleischer:1994ef,Caffo:2008aw,Pozzorini:2005ff,Passarino:2016zcd}.

The electron self-energy is a gauge-dependent object. We perform the calculation in a covariant gauge with gauge parameter $\xi$.

The renormalised electron self-energy is a renormalisation-scheme dependent quantity.
For this reason we present the bare electron self-energy independently of the contributions from the counterterms.
In QED the on-shell scheme is the conventional choice. We give the contributions from the counterterms
in the on-shell scheme.
Renormalisation removes ultraviolet divergences, infrared divergences remain or are introduced through 
infrared poles in the renormalisation constants.
The latter case already occurs at one-loop in the on-shell scheme in QED.

The bare electron self-energy is not a pure function.
We may associate a weight to the iterated integrals and the transcendental constants appearing in our calculation.
A function is pure if each term in the $\eps$-expansion has uniform weight, 
where $\eps$ denotes the dimensional regularisation parameter.
Although we may choose our master integrals as pure functions, the bare electron self-energy is not pure (and is
not expected to be pure).

We follow the standard approach for loop calculations: The two-loop electron self-energy is expressed as a linear
combination of master integrals.
The coefficients of the master integrals are given in appendix~\ref{sect:coefficients_masters}
(for Feynman gauge $\xi=1$) and in a supplementary electronic file attached to this article (for a general covariant gauge $\xi \neq 1$).
The master integrals are computed from their 
differential equations \cite{Kotikov:1990kg,Kotikov:1991pm,Remiddi:1997ny,Gehrmann:1999as,Argeri:2007up,MullerStach:2012mp,Henn:2013pwa,Henn:2014qga,Tancredi:2015pta,Ablinger:2015tua,Adams:2017tga,Bosma:2017hrk}.
For the case at hand, they can be expressed as iterated integrals of modular forms \cite{Adams:2017ejb}.
In order to give the reader some orientation on the level of complexity of the calculation, let us classify loop integrals along two criteria:
(i) the number of the involved dimensionless variables and
(ii) whether the loop integrals may be expressed in terms of multiple polylogarithms or not.
The simplest case is given by loop integrals depending on a single dimensionless variable and expressible in terms of (multiple) polylogarithms.
Loop integrals associated to two-loop $2 \rightarrow 2$-scattering amplitudes in massless theories are 
a typical example \cite{Bern:2000ie,Bern:2000dn,Anastasiou:2000kg,Anastasiou:2000ue,Anastasiou:2000mv,Anastasiou:2001sv,Glover:2001af,Bern:2002tk}, 
and the class of functions reduces to the sub-class of harmonic polylogarithms.
The next difficult case is given by loop integrals depending on several dimensionless variables and expressible in terms of multiple polylogarithms.
Loop integrals associated to the two-loop scattering amplitudes for the process $e^+e^- \rightarrow q g \bar{q}$ in massless QCD are an example \cite{Garland:2001tf,Garland:2002ak,Moch:2002hm}.
In both cases, the standard approach is to transform the system of differential equations to an $\eps$-form \cite{Henn:2013pwa} through an algebraic change 
of the kinematic variables
and an algebraic change of the basis of the master integrals.
If we leave the class of multiple polylogarithms, we first have the case of Feynman integrals beyond the class of multiple polylogarithms, but depending
on a single dimensionless variable.
This is the case discussed in this paper.
Of course, there is also the case of Feynman integrals beyond the class of multiple polylogarithms and depending on several dimensionless variables.
An example for the latter would be given by the two-loop integrals associated to the process $g g \rightarrow t \bar{t}$ \cite{Czakon:2013goa,Baernreuther:2013caa,Adams:2017tga,Adams:2018bsn,Adams:2018kez}.
Let us emphasize that for the case of interest of this article we are nevertheless able to transform the system of differential equations
to an $\eps$-form \cite{Adams:2018yfj}, albeit through a non-algebraic change of the basis of the master integrals.

For the numerical evaluation of the master integrals we switch from the variable $x=p^2/m^2$ to a variable $q$, 
related to the nome squared of an elliptic curve.
We may choose $q$ such that $q$ vanishes at one of the cusps $x \in \{0,1,9,\infty\}$.
Let us call this point $j$ and the set of the remaining points $S_j=\{0,1,9,\infty\}\backslash \{j\}$.
The master integrals have a series expansion in $q$, which converges for all values $x\in {\mathbb R}\backslash S_j$, i.e. everywhere on the real line except for three points.
The $q$-expansion is a highly efficient method to evaluate numerically the master integrals for $q$ close to zero.
We give $q$-expansion for the master integrals for all possible choices $j \in \{0,1,9,\infty\}$, thus covering
the full kinematic range $x \in {\mathbb R}$ with efficient numerical evaluation routines
and thereby generalising the results of  \cite{Bogner:2017vim}.

Efficient numerical evaluations are often based on fast convergent series expansions. The difference between our result and the original work
of Sabry lies in the variables used to expand in.
With our choice the series show a significant faster convergence (and are well motivated from the underlying mathematics).
Quite recently, a purely numerical approach of solving the system of differential equations has been advocated in \cite{Lee:2017qql,Lee:2018ojn},
based on expansions around the singular points of the differential equations.
The differences and the similarities with our approach are as follows:  
Within our approach we first obtain an analytic result in terms of iterated integrals of modular forms (with a notion of weight for these iterated
integrals, such that the $\eps^j$-term of each master integrals has uniform weight $j$), and only in a second step we use efficient numerical methods
for the numerical evaluation of these iterated integrals.
The approach of \cite{Lee:2017qql,Lee:2018ojn} is numerical from the beginning.
We use a variable $q$ as expansion parameter, whereas the method of \cite{Lee:2017qql,Lee:2018ojn} uses $p^2/m^2$ or a rational function of this variable.
Common to both methods is the expansion around all singular points of the differential equations, and -- on a technical level --
the determination of the boundary constants for the expansion around the second, third and any further singular point from the boundary constants 
for the expansion around the first (or any other already known) singular point with the help of high-precision numerics and the PSLQ-algorithm \cite{Ferguson:1992}.

This paper is organised as follows:
Section~\ref{sec:lagrange} briefly reviews the Lagrange density of quantum electrodynamics and the (known)
renormalisation constants to two-loop order.
In section~\ref{sec:notation} we introduce the master integrals for the two-loop electron self-energy.
Section~\ref{sec:iterated_integrals} is a brief introduction to iterated integrals.
In section~\ref{sec:calculation} we express the bare two-loop electron self-energy in terms of master integrals.
In addition we give the counterterms from renormalisation.
In section~\ref{sec:master_integrals} we evaluate the master integrals in terms of iterated integrals of modular forms.
The iterated integrals start at a boundary point $j$ and we do this for the four choices $j \in \{0,1,9,\infty\}$.
The iterated integrals have a $q$-series expansion, which may be used for the numerical evaluation.
In section~\ref{sect:numerical} we present numerical results.
Our conclusions are given in section~\ref{sect:conclusions}.
The appendix contains useful information on the QED Feynman rules (appendix~\ref{sect:Feynman_rules}),
the numerical computation of the complete elliptic integral with the help of the arithmetic-geometric mean (appendix~\ref{sec:agm}),
the Eisenstein series of modular weight $1$ appearing in the calculation (appendix~\ref{sec:eisenstein}),
the one-loop electron self-energy (appendix~\ref{sect:oneloop}),
the coefficients appearing in the expression of the two-loop electron self-energy in terms of the master integrals
(appendix~\ref{sect:coefficients_masters}),
the boundary constants for the four cases $j \in \{0,1,9,\infty\}$ (appendix~\ref{sec:boundary_constants})
and a description of the content of the supplementary electronic file attached to this article (appendix~\ref{sect:supplement}).

% -----------------------------------------------------------------------------

\section{Lagrange density and renormalisation}
\label{sec:lagrange}

% electron has charge $-e$
The bare gauge-fixed Lagrangian for QED reads in a covariant gauge 
\bq
{\mathcal L}_{\mathrm{QED}} & = & 
 \bar{\psi}_0 ( i {\slashed \partial} - m_0 ) \psi_0
 - \frac{1}{4} \left( \partial_{\mu} A_{\nu,0} - \partial_{\nu} A_{\mu,0} \right)^2
 - \frac{1}{2 \xi_0} ( \partial^{\mu} A_{\mu,0} )^{2}
 - e_0 \bar{\psi}_0 \gamma^{\mu} A_{\mu,0} \psi_0.
\eq
We use dimensional regularisation and set $D=4-2\eps$.
Under renormalisation one redefines the fields
\bq
 A_{\mu,0} = Z_3^{\frac{1}{2}} A_\mu, 
 \;\;\;
 \psi_0 = Z_2^{\frac{1}{2}} \psi,
\eq
and the parameters
\bq
e_0 = Z_e \; \mu^\eps S_\eps^{-\frac{1}{2}} e
 = Z_3^{-\frac{1}{2}} \; \mu^\eps S_\eps^{-\frac{1}{2}} e,
 \;\;\;\;\;\;
m_0 = Z_m m,
 \;\;\;\;\;\;
\xi_0 = Z_\xi \xi
 = Z_3 \xi.
\eq
The arbitrary scale $\mu$ is introduced to keep the renormalised coupling $e$ dimensionless.
The factor $S_\eps = (4\pi)^\eps \exp(-\eps \gamma_E)$ absorbs artefacts of dimensional regularisation 
(logarithms of $4\pi$ and Euler's constant $\gamma_E$).
For convenience we set
\bq
\label{def_coupling}
 \alpha \; = \; \frac{e^2}{4 \pi},
 & &
 e^{(D)} \; = \; \mu^\eps S_\eps^{-\frac{1}{2}} e.
\eq
Substituting the above relations into the Lagrange density we obtain
\bq
 {\mathcal L}_{\mathrm{QED}} & = & {\mathcal L}_{\mathrm{renorm}} + {\mathcal L}_{\mathrm{counterterms}},
\eq
where ${\mathcal L}_{\mathrm{renorm}}$ is given by ${\mathcal L}_{\mathrm{QED}}$ 
where all bare quantities are replaced by renormalised ones (the bare coupling $e_0$ is replaced by $e^{(D)}$).
The counterterms are given by
\bq
\lefteqn{
 {\mathcal L}_{\mathrm{counterterms}} 
 = } & & \\
 & &
    \left( Z_2 - 1 \right) \bar{\psi} i {\slashed \partial} \psi
    - \left( Z_2 Z_m -1 \right) m \bar{\psi} \psi
    + \frac{1}{2} \left( Z_3 - 1 \right) A_\mu \left( g^{\mu\nu} \Box - \partial^\mu \partial^\nu \right) A_\nu
 - \left( Z_2 -1 \right) e^{(D)} \bar{\psi} \gamma^\mu A_\mu \psi.
 \nonumber
\eq
The renormalisation constants are known to very high loop order \cite{Egorian:1978zx,Tarrach:1980up,Gray:1990yh,Broadhurst:1991fy,Chetyrkin:1999qi,Melnikov:2000qh,Melnikov:2000zc,Marquard:2016dcn,Marquard:2018rwx,Luthe:2016xec,Luthe:2017ttc,Chetyrkin:2017bjc}.
Here, we only need them to order ${\mathcal O}(\alpha^2)$.
We write
\bq
 Z_3
 & = & 
 1 
 + \frac{\alpha}{4\pi} Z_3^{(1)}
 + \left(\frac{\alpha}{4\pi}\right)^2 Z_3^{(2)}
 + {\mathcal O}\left(\alpha^3\right),
 \nonumber \\
 Z_2
 & = & 
 1 
 + \frac{\alpha}{4\pi} Z_2^{(1)}
 + \left(\frac{\alpha}{4\pi}\right)^2 Z_2^{(2)}
 + {\mathcal O}\left(\alpha^3\right),
 \nonumber \\
 Z_m
 & = & 
 1 
 + \frac{\alpha}{4\pi} Z_m^{(1)}
 + \left(\frac{\alpha}{4\pi}\right)^2 Z_m^{(2)}
 + {\mathcal O}\left(\alpha^3\right).
\eq
In the on-shell scheme we have \cite{Broadhurst:1991fy}
\bq
\label{renorm_constants}
 Z_3^{(1)} & = &
 - \frac{4}{3\eps} - \frac{2}{3} \zeta_2 \eps + {\mathcal O}\left(\eps^2\right),
 \nonumber \\
 Z_2^{(1)} & = &
 - \frac{3}{\eps} - 4 - \left(\frac{3}{2} \zeta_2 + 8 \right) \eps + {\mathcal O}\left(\eps^2\right),
 \nonumber \\
 Z_m^{(1)} & = &
 - \frac{3}{\eps} - 4 - \left(\frac{3}{2} \zeta_2 + 8 \right) \eps + {\mathcal O}\left(\eps^2\right),
 \nonumber \\
 Z_3^{(2)} & = &
 - \frac{2}{\eps} - 15 + {\mathcal O}\left(\eps\right),
 \nonumber \\
 Z_2^{(2)} & = &
 \frac{9}{2 \eps^2} + \frac{55}{4\eps} + 96 \zeta_2 \ln 2
 - 24 \zeta_3 - \frac{211}{2} \zeta_2 + \frac{7685}{72} + {\mathcal O}\left(\eps\right),
 \nonumber \\
 Z_m^{(2)} & = &
 \frac{5}{2\eps^2} + \frac{155}{12\eps} + 48 \zeta_2 \ln 2
 - 12 \zeta_3 - \frac{87}{2} \zeta_2 + \frac{1169}{24} + {\mathcal O}\left(\eps\right).
\eq
For the two-loop contribution to the electron self-energy we need $Z_3$ to order $\alpha$ and
$Z_2$ and $Z_m$ to order $\alpha^2$.

% -----------------------------------------------------------------------------

\section{Definitions and notation for the master integrals}
\label{sec:notation}

There are three Feynman diagrams contributing to the two-loop electron self-energy in QED. 
\begin{figure}
\begin{center}
\includegraphics[scale=1.0]{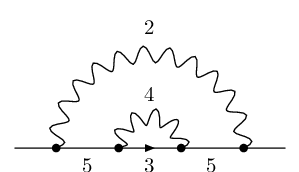}
\includegraphics[scale=1.0]{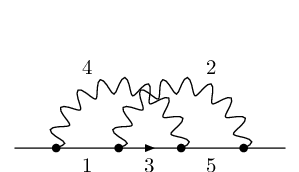}
\includegraphics[scale=1.0]{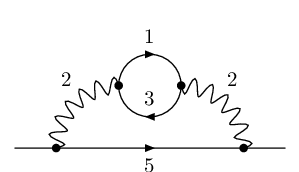}
\end{center}
\caption{
\it The Feynman graphs contributing to the two-loop electron self-energy.
}
\label{fig_diagrams}
\end{figure}
These diagrams are shown in fig.~\ref{fig_diagrams}.
We label these diagrams ``rainbow diagram'', ``kite diagram'' and ``fermion insertion diagram'', respectively.
\begin{figure}
\begin{center}
\includegraphics[scale=1.0]{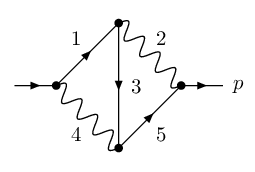}
\end{center}
\caption{
\it The kite graph. This graph is equivalent to the second graph in fig.~\ref{fig_diagrams}.
}
\label{fig_kite_graph}
\end{figure}
The second diagram (the kite diagram) can be drawn equivalently as shown in fig.~\ref{fig_kite_graph}, 
motivating the name ``kite diagram''.
The master integrals for the rainbow diagram and the fermion insertion diagram will be a subset of the master integrals
for the kite diagram.
To see this, let us first introduce the kite integral.
We set
\bq
\label{def_kite}
 I_{\nu_1 \nu_2 \nu_3 \nu_4 \nu_5}\left( D, p^2, m^2, \mu^2 \right)
 =
 \left(-1\right)^{\nu_{12345}}
 e^{2\gamma_E \eps}
 \left(\mu^2\right)^{\nu_{12345}-D}
 \int \frac{d^Dk_1}{i \pi^{\frac{D}{2}}} \frac{d^Dk_2}{i \pi^{\frac{D}{2}}}
 \frac{1}{D_1^{\nu_1} D_2^{\nu_2} D_3^{\nu_3} D_4^{\nu_4} D_5^{\nu_5}},
\eq
with the propagators
\bq
\label{def_propagators}
 D_1=k_1^2-m^2, \hspace{0.3cm}  
 D_2=k_2^2, \hspace{0.3cm}  
 D_3 = (k_1-k_2)^2-m^2, \hspace{0.3cm} 
 D_4=(k_1-p)^2, \hspace{0.3cm}  
 D_5 = (p-k_2)^2-m^2
\eq
and $\nu_{12345}=\nu_1+\nu_2+\nu_3+\nu_4+\nu_5$.
The internal momenta are denoted by $k_1$ and $k_2$, 
the internal mass by $m$, the external momentum by $p$
and the dimension of space-time by $D=4-2\eps$.
The arbitrary scale $\mu$ renders the integral dimensionless.
In the following we set $\mu=m$.
We further define
\bq
 x & = & \frac{p^2}{m^2}.
\eq
The five propagators $D_1$-$D_5$ are indicated by the numbers $1$-$5$ in fig.~\ref{fig_kite_graph}.
We note that all propagators of the rainbow diagram and the fermion loop insertion diagram are a subset of these.
In order to show this, we labelled all propagators in fig.~\ref{fig_diagrams} with the appropriate numbers.
Therefore it is sufficient to consider only the master integrals of the kite integral.
In order to present these master integrals let us first denote by $\psi_1$ and $\psi_2$ two independent solutions
of the second-order differential equation \cite{Broadhurst:1993mw,Laporta:2004rb}
\bq
\label{Picard_Fuchs_equation}
 \left[
 x \left(x-1\right) \left(x-9\right) \frac{d^2}{dx^2} 
 + \left(3x^2-20x+9\right) \frac{d}{dx}
 + x-3
 \right] \psi
 & = & 0.
\eq
Of course, this does not fully specify $\psi_1$ nor $\psi_2$, but for the moment this is all what we would like to assume
about $\psi_1$ and $\psi_2$.
The exact definitions of $\psi_1$ and $\psi_2$ will be given in section~\ref{sec:master_integrals}.
We denote the Wronskian by
\bq
\label{def_Wronskian}
 W
 & = & 
 \psi_{1} \frac{d}{dx} \psi_{2} - \psi_{2} \frac{d}{dx} \psi_{1}.
\eq
We will normalise $\psi_1$ and $\psi_2$ such that
\bq
\label{Wronskian_relation}
 W
 & = & 
 - \frac{6 \pi i}{x\left(x-1\right)\left(x-9\right)}.
\eq
We will see later that eq.~(\ref{Picard_Fuchs_equation}) is the Picard-Fuchs equation for the periods of an elliptic curve
and $\psi_1$ and $\psi_2$ will be taken as periods of an elliptic curve.
We denote the ratio of the two periods and the nome squared by
\bq
\label{def_tau}
 \tau
 \;\; = \;\;
 \frac{\psi_{2}}{\psi_{1}},
 & &
 q \;\; = \;\; e^{2 i \pi \tau}.
\eq
We further set
\bq
 \tau_n 
 \;\; = \;\;
 \frac{\tau}{n},
 & &
 q_n \;\; = \;\; e^{2 i \pi \tau_n}.
\eq
Let us now return to the master integrals.
There are eight master integrals, which we take as \cite{Remiddi:2016gno,Adams:2018yfj}
\bq
\label{def_basis}
 I_1\left(\eps,x\right)
 & = &
 4 \eps^2 \; I_{20200}\left(4-2\eps,x\right),
 \nonumber \\
 I_2\left(\eps,x\right)
 & = &
 4 \eps^2 x \; I_{20210}\left(4-2\eps,x\right),
 \nonumber \\
 I_3\left(\eps,x\right)
 & = &
 4 \eps^2 x \; I_{02210}\left(4-2\eps,x\right),
 \nonumber \\
 I_4\left(\eps,x\right)
 & = &
 4 \eps^2 \left[ 2 I_{02210}\left(4-2\eps,x\right) + \left(1-x\right) I_{02120}\left(4-2\eps,x\right) \right],
 \nonumber \\
 I_5\left(\eps,x\right)
 & = &
 4 \eps^2 x^2 \; I_{21012}\left(4-2\eps,x\right),
 \nonumber \\
 I_6\left(\eps,x\right)
 & = &
 \eps^2 \frac{\pi}{\psi_1} \; I_{10101}\left(2-2\eps,x\right),
 \nonumber \\
 I_7\left(\eps,x\right)
 & = &
 \frac{1}{\eps} \frac{\psi_1^2}{2 \pi i W} \frac{d}{dx} I_6 
 + \frac{\psi_1^2}{2 \pi i W} \frac{\left(3x^2-10x-9\right)}{2x\left(x-1\right)\left(x-9\right)} I_6,
 \nonumber \\
 I_8\left(\eps,x\right)
 & = & - 8 \eps^3 \left(1-2\eps\right) x \; I_{11111}\left(4-2\eps,x\right).
\eq
In the master integral $I_6$ the sunrise integral in $D=2-2\eps$ space-time dimensions appears.
Using dimensional-shift relations we may express this integral in terms of integrals in $D=4-2\eps$ space-time
dimensions.
We have
\bq
\lefteqn{
 I_{10101}\left(2-2\eps,x\right)
 = 
 \frac{3}{\left(x-1\right)\left(x-9\right)}
 \left[
  \left(3-x\right) I_{20200}\left(4-2\eps,x\right)
 \right.
} & & \nonumber \\
 & &
 \left.
  +
  2 \left(1-2\eps\right) \left(2-3\eps\right) I_{10101}\left(4-2\eps,x\right)
  +
 2 \left(1-2\eps\right) \left(x+3\right) I_{20101}\left(4-2\eps,x\right)
 \right].
\eq
The master integral $I_7$ involves the derivative of $I_6$. We have
\bq
 \frac{d}{dx} I_6
 & = &
 \frac{3 \eps^2}{x} \frac{\pi}{\psi_1} I_{20101}\left(2-2\eps,x\right) 
 - \left[
          \frac{\left(1+2\eps\right)}{x}  
          + \frac{1}{\psi_1} \left(\frac{d\psi_1}{dx}\right) 
 \right] \eps^2 \frac{\pi}{\psi_1} I_{10101}\left(2-2\eps,x\right),
\eq
and
\bq
\lefteqn{
 I_{20101}\left(2-2\eps,x\right)
 = 
} & & \nonumber \\
 & & 
 \frac{1}{\left(x-1\right)^2\left(x-9\right)^2}
 \left\{
  \left[ 3 \left(x-1\right) \left(x-9\right) - \eps \left(2x^3-34x^2+54x-54\right) \right] I_{20200}\left(4-2\eps,x\right)
 \right. \nonumber \\
 & & \left.
  + 2 \left(1-2\eps\right) \left(2-3\eps\right) \left[ \left(x-1\right) \left(x-9\right) - 2 \eps \left(x-3\right)\left(x+3\right) \right] I_{10101}\left(4-2\eps,x\right)
 \right. \nonumber \\
 & & \left.
 + 2 \left(1-2\eps\right) \left[ 3 \left(x-1\right)\left(x-9\right) + \eps \left(x^3-36x^2+45x+54\right) \right]  I_{20101}\left(4-2\eps,x\right)
 \right\},
\eq
which allows us to express all integrals in $D=4-2\eps$ dimensions.

Let us set $\vec{I}=(I_1,I_2,I_3,I_4,I_5,I_6,I_7,I_8)^T$.
The differential equation for $\vec{I}$ with respect to $\tau_n$ reads
\bq
\label{eps_form}
 \frac{1}{2\pi i} \frac{d}{d\tau_n} \vec{I}
 & = &
 \eps \; n \; A \; \vec{I},
\eq
where the $8 \times 8$-matrix $A$ is independent of $\eps$ and given by
\bq
\label{res_A}
 A & = & 
 \left( \begin{array}{rrrrrrrr}
 0 & 0 & 0 & 0 & 0 & 0 & 0 & 0 \\
 -g_{2,1} & g_2 & 0 & 0 & 0 & 0 & 0 & 0 \\
 0 & 0 & g_2 & g_{2,1} & 0 & 0 & 0 & 0 \\
 0 & 0 & -4 g_{2,0} + 4 g_{2,1}& -2 g_{2,1} & 0 & 0 & 0 & 0 \\
 0 & -2 g_{2,1} & 0 & 0 & 2 g_2 & 0 & 0 & 0 \\
 0 & 0 & 0 & 0 & 0 & -f_2 & 1 & 0 \\
 \frac{1}{4} f_3 & 0 & 0 & 0 & 0 & f_4 & -f_2 & 0 \\
 g_{2,1} & 0 & -2 g_{2,1} & -g_{2,1} & -2 g_{2,0} & -12 g_{3,0} + \frac{32}{3} g_{3,1}  & 0 & g_2 \\
 \end{array} \right).
 \nonumber \\
\eq
The entries of the matrix $A$ are as follows: We first define
\begin{align}
\label{integration_kernels}
 g_{2,0} 
 & = 
 \frac{1}{2 i \pi} \frac{\psi_{1}^2}{W} \frac{1}{x},
 &
 g_{3,0} 
 & = 
 \frac{1}{2 i \pi} \frac{\psi_{1}^2}{W} \frac{\psi_{1}}{\pi},
 &
 f_{4}
 & = 
 \frac{1}{2 i \pi} \frac{\psi_{1}^2}{W} 
 \left( \frac{\psi_{1}}{\pi} \right)^2 \frac{\left(x+3\right)^4}{48 x\left(x-1\right)\left(x-9\right)},
 \nonumber \\
 g_{2,1} 
 & = 
 \frac{1}{2 i \pi} \frac{\psi_{1}^2}{W} \frac{1}{x-1},
 &
 g_{3,1} 
 & = 
 \frac{1}{2 i \pi} \frac{\psi_{1}^2}{W} \frac{\psi_{1}}{\pi} \frac{x}{x-1},
 & &
 \nonumber \\
 g_{2,9} 
 & = 
 \frac{1}{2 i \pi} \frac{\psi_{1}^2}{W} \frac{1}{x-9},
 & & & &
\end{align}
and set then
\begin{align}
 f_2 & = - \frac{1}{2} g_{2,0} + g_{2,1} + g_{2,9},
 &
 g_2 & = g_{2,0} - 2 g_{2,1},
 &
 f_3 & = - \frac{1}{2} g_{3,0}.
\end{align}
All entries may be expressed as polynomials in
\bq
\label{def_b_basis}
 b_1 \; = \; \frac{\psi_1}{\pi},
 & &
 b_2 \; = \; \frac{\psi_1}{\pi} \left(x+3\right).
\eq
We have
\begin{align}
\label{kernels_b_polynomials}
 g_{2,0} 
 & = 
 4 b_1^2 - \frac{4}{3} b_1 b_2 + \frac{1}{12} b_2^2,
 &
 g_{3,0} 
 & = 
 - 12 b_1^3 + 8 b_1^2 b_2 - \frac{19}{12} b_1 b_2^2 + \frac{1}{12} b_2^3,
 &
 f_{4} 
 & = &
 \frac{1}{576} b_2^4,
 \nonumber \\
 g_{2,1} 
 & = 
 3 b_1^2 - \frac{5}{4} b_1 b_2 + \frac{1}{12} b_2^2,
 &
 g_{3,1} 
 & = 
 - 9 b_1^3 + \frac{27}{4} b_1^2 b_2 - \frac{3}{2} b_1 b_2^2 + \frac{1}{12} b_2^3,
 &
 & \nonumber \\
 g_{2,9} 
 & = 
 b_1^2 - \frac{7}{12} b_1 b_2 + \frac{1}{12} b_2^2.
 & & & &
\end{align}
Let us stress that all formulae in this section are valid for any choice of $\psi_1$ and $\psi_2$,
as long as these are two independent solutions of eq.~(\ref{Picard_Fuchs_equation})
and normalised such that eq.~(\ref{Wronskian_relation}) holds.
We will use this freedom to define four sets of master integrals, which we denote by $I_{i,j}$ 
(with $1 \le i \le 8$ and $j\in\{0,1,9,\infty\}$), corresponding to the 
different $q$-expansions around the four cusps $x \in \{0,1,9,\infty\}$.
Where it is not relevant, we will drop the additional index $j$.
The first five master integrals and the last one are identical in all four sets and differ only
in the variables they depend on.
However, in the definition of the sixth and the seventh master integral the period $\psi_1$ 
appears explicitly and these integrals are not identical in the four sets.
Of course, the dependence on our choice of $\psi_1$ is also reflected in the coefficients
expressing the two-loop self-energy as a linear combination of the master integrals, such that the final
result is independent of the choice of $\psi_1$ and $\psi_2$.
To cut the story short: This setup allows us to use in any region a choice of variables with
the best numerical convergence.

% -----------------------------------------------------------------------------

\section{Iterated integrals}
\label{sec:iterated_integrals}

We may easily solve the differential equation in eq.~(\ref{eps_form})
order-by-order in $\eps$.
The solution is expressed in terms of iterated integrals.
Let us briefly review iterated integrals \cite{Chen}.
For differential 1-forms $\omega_1$, ..., $\omega_k$ on a manifold $M$
and a path $\gamma : [a,b] \rightarrow M$ let us write 
for the pull-back of $\omega_j$ to the interval $[a,b]$
\bq
 f_j\left(\lambda\right) d\lambda & = & \gamma^\ast \omega_j.
\eq
The iterated integral is defined by
\bq
 I_{\gamma}\left(\omega_1,...,\omega_k;b\right)
 & = &
 \int\limits_a^{b} d\lambda_1 f_1\left(\lambda_1\right)
 \int\limits_a^{\lambda_1} d\lambda_2 f_2\left(\lambda_2\right)
 ...
 \int\limits_a^{\lambda_{k-1}} d\lambda_k f_k\left(\lambda_k\right).
\eq
Harmonic polylogarithms are a special case of iterated integrals \cite{Remiddi:1999ew}.
We consider two integration kernels
\bq
 f_0\left(\lambda\right) \; = \; \frac{1}{\lambda},
 & &
 f_1\left(\lambda\right) \; = \; \frac{1}{1-\lambda},
\eq
and define the harmonic polylogarithms by
\bq
 H_{m_1 m_2 ... m_k}\left(x\right) & = & \int\limits_0^x dx' f_{m_1}\left(x'\right) H_{m_2 ... m_k}\left(x'\right)
\eq
and
\bq
 H\left(x\right) \; = \; 1,
 & &
 H_{\underbrace{0 ... 0}_{k}}\left(x\right) \; = \; \frac{1}{k!} \ln^k\left(x\right).
\eq
The last equation defines harmonic polylogarithms for trailing zeros.

A second special case are iterated integrals of modular forms.
Let $f_1(\tau)$, $f_2(\tau)$, ..., $f_k(\tau)$ be modular forms of a congruence subgroup.
and assume that $f_k(\tau)$ vanishes at the cusp $\tau=i\infty$.
We define the $k$-fold iterated integral by
\bq
\label{iter_int_modular_forms}
 \iterintmodular{f_1,f_2,...,f_k}{q}
 & = &
 \left(2 \pi i \right)^k
 \int\limits_{i \infty}^{\tau} d\tau_1
 \;
 f_1\left(\tau_1\right)
 \int\limits_{i \infty}^{\tau_1} d\tau_2
 \;
 f_2\left(\tau_2\right)
 ...
 \int\limits_{i \infty}^{\tau_{k-1}} d\tau_k
 \;
 f_k\left(\tau_k\right),
 \;\;\;\;\;\;
 q \; = \; e^{2\pi i \tau}.
 \;\;\;\;\;\;
\eq
The case where $f_k(\tau)$ does not vanishes at the cusp $\tau=i\infty$ is discussed in \cite{Adams:2017ejb,Brown:2014aa}
and is similar to trailing zeros in the case of harmonic polylogarithms.
If we change the integration variables from $\tau$ to $q$ we obtain
\bq
 \iterintmodular{f_1,f_2,...,f_k}{q}
 & = &
 \int\limits_{0}^{q} \frac{dq_1}{q_1}
 \;
 \tilde{f}_1\left(q_1\right)
 \int\limits_{0}^{q_1} \frac{dq_2}{q_2}
 \;
 \tilde{f}_2\left(q_2\right)
 ...
 \int\limits_{0}^{q_{k-1}} \frac{dq_k}{q_k}
 \;
 \tilde{f}_k\left(q_k\right),
\eq
with
\bq
 \tilde{f}_j\left(q\right) & = & f_j\left(\tau\right).
\eq
Given the $q$-expansion of the modular forms $f_1$, ..., $f_k$, we may easily obtain the $q$-expansion
of the iterated integral $\iterintmodular{f_1,f_2,...,f_k}{q}$ by integrating term-by-term and multiplication
of power series.

% -----------------------------------------------------------------------------

\section{The two-loop self-energy}
\label{sec:calculation}

\subsection{The bare two-loop self-energy}

We first compute the bare two-loop electron self-energy in QED. 
We write with $\alpha=e^2/(4\pi)$ for the bare self-energy
\bq
 - i \Sigma^{(2)}_{\mathrm{bare}}
 & = &
 - i 
  \left( \frac{\alpha}{4\pi} \right)^2
  \left(
         \Sigma^{(2)}_{\mathrm{bare},V} \; {\slashed p} + \Sigma^{(2)}_{\mathrm{bare},S} \; m
  \right),
\eq
separating the part proportional to ${\slashed p}$ and the part proportional to $m$.
The quantities $\Sigma^{(2)}_{\mathrm{bare},V}$ and $\Sigma^{(2)}_{\mathrm{bare},S}$ are expressed as linear combinations of the master integrals $I_1$-$I_8$:
\bq
\label{def_linear_combination_master_integrals}
 \Sigma^{(2)}_{\mathrm{bare},V}
 \; = \;
 \sum\limits_{j=1}^8 \; c_j^V \; I_j,
 & &
 \Sigma^{(2)}_{\mathrm{bare},S}
 \; = \;
 \sum\limits_{j=1}^8 \; c_j^S \; I_j.
\eq
We work in a general covariant gauge with gauge parameter $\xi$.
The coefficients $c_j^V$ and $c_j^S$ are rational functions in $x$, $\eps$, $\xi$, $\psi_1/\pi$ and
$1/\pi \cdot d\psi_1/dx$.
For $\xi=1$ (Feynman gauge) they are listed in appendix~\ref{sect:coefficients_masters}.
For a general covariant gauge ($\xi \neq 1$) they are given  
in a supplementary electronic file attached to this article.
The master integrals $I_1$-$I_8$ satisfy a differential equation in $\eps$-form, therefore
the $\eps$-expansion of $\Sigma^{(2)}_{\mathrm{bare},V}$ and $\Sigma^{(2)}_{\mathrm{bare},S}$ is easily obtained from
eq.~(\ref{def_linear_combination_master_integrals}) by expanding in $\eps$ to the desired order.
The pole terms are rather simple.
Let us write
\bq
 \Sigma^{(2)}_{\mathrm{bare},V}
 \;\; = \;\;
 \sum\limits_{k=-2}^\infty
 \eps^k \; \Sigma^{(2,k)}_{\mathrm{bare},V},
 & &
 \Sigma^{(2)}_{\mathrm{bare},S}
 \;\; = \;\;
 \sum\limits_{k=-2}^\infty
 \eps^k \; \Sigma^{(2,k)}_{\mathrm{bare},S} .
\eq
We have
\bq
 \Sigma^{(2,-2)}_{\mathrm{bare},V}
 & = &
 - \frac{1}{2} \xi^2,
 \nonumber \\
 \Sigma^{(2,-2)}_{\mathrm{bare},S}
 & = &
 \frac{1}{2} \left(1+\xi\right) \left(5+\xi\right),
 \nonumber \\
 \Sigma^{(2,-1)}_{\mathrm{bare},V}
 & = &
    \frac{7}{4} - \frac{12\xi}{x} - \xi^2 - \frac{\xi^2}{x}
    + \left( \xi - \frac{12}{x^2} - \frac{\xi}{x^2} \right) \xi \ln\left(1-x\right),
 \nonumber \\
 \Sigma^{(2,-1)}_{\mathrm{bare},S}
 & = &
     4 + 10 \xi + 2 \xi^2
     - \left( 5 + 6 \xi + \xi^2 - \frac{23}{x} - \frac{12\xi}{x} - \frac{\xi^2}{x} \right) \ln\left(1-x\right).
\eq
The finite parts $\Sigma^{(2,0)}_{\mathrm{bare},V}$ and $\Sigma^{(2,0)}_{\mathrm{bare},S}$ will be given in section~\ref{sec:master_integrals}.

\subsection{The counterterms from renormalisation}

\begin{figure}
\begin{center}
\includegraphics[scale=1.0]{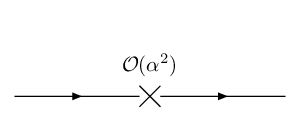}
\includegraphics[scale=1.0]{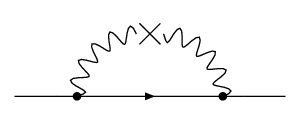}
\includegraphics[scale=1.0]{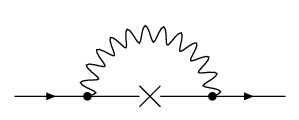}
\includegraphics[scale=1.0]{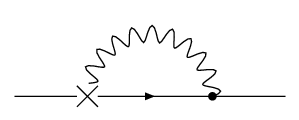}
\includegraphics[scale=1.0]{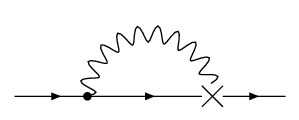}
\end{center}
\caption{
\it The Feynman graphs corresponding to the counterterms.
In the first graph we take the ${\mathcal O}(\alpha^2)$-term.
}
\label{fig_diagrams_CT}
\end{figure}

We write for the counterterms from renormalisation at order ${\mathcal O}(\alpha^2)$
\bq
 - i \Sigma^{(2)}_{\mathrm{CT}}
 & = &
 - i 
  \left( \frac{\alpha}{4\pi} \right)^2
  \left(
         \Sigma^{(2)}_{\mathrm{CT},V} \; {\slashed p} + \Sigma^{(2)}_{\mathrm{CT},S} \; m
  \right).
\eq
The relevant diagrams are shown in fig.~\ref{fig_diagrams_CT}.
$\Sigma^{(2)}_{\mathrm{CT},V}$ and $\Sigma^{(2)}_{\mathrm{CT},S}$ are given by
\bq
\label{res_counterterms_1}
 \Sigma^{(2)}_{\mathrm{CT},V}
 & = &
 - Z_2^{(2)} 
 - \left[
  Z_2^{(1)} \; \frac{1}{2 \left(1-2\eps\right)} \left( \frac{1}{x} + \frac{1-\eps}{\eps} \right) 
  + Z_m^{(1)} \; \frac{2\left(1-\eps\right)}{\left(1-2\eps\right)x}
 \right] \xi J_1
 \nonumber \\
 & &
 - \frac{\left(1-\eps\right)}{\left(1-2\eps\right)}
 \left[ 
  Z_2^{(1)} \; \frac{1}{2\eps} \left(1 - \frac{1}{x^2} \right)
  + 2 Z_m^{(1)} \left( \frac{1}{x^2} + \frac{1}{x} - \frac{1}{\eps x^2} \right)
 \right] \xi J_2,
 \nonumber \\
 \Sigma^{(2)}_{\mathrm{CT},S}
 & = &
 Z_2^{(2)} + Z_m^{(2)} + Z_2^{(1)} Z_m^{(1)}
 - \frac{1}{2 \eps \left(1-2\eps\right)} 
     \left[
              \left(3 - 2 \eps \right) Z_3^{(1)}
            - \left(3+\xi-2\eps\right) Z_2^{(1)}
 \right. \nonumber \\
 & & \left.
            - \left(3+\xi-2\eps\right) Z_m^{(1)}   
     \right] J_1
 - \frac{1}{2 \eps \left(1-2\eps\right) x} 
     \left[
            \left(3+\xi-2\eps\right) \left(1-x\right) Z_2^{(1)}
 \right. \nonumber \\
 & & \left.
            - \left(3-2\eps\right) \left(1-x\right) Z_3^{(1)}
            + \left(3+\xi-2\eps\right) \left(3-x-4\eps\right) Z_m^{(1)}
     \right] J_2.
\eq
$J_1$ and $J_2$ are the two master integrals of the one-loop calculation. They are given in appendix~\ref{sect:oneloop}.
The renormalisation constants in the on-shell scheme are given in eq.~(\ref{renorm_constants}).
In the on-shell scheme we have
\bq
\label{res_counterterms_2}
 \Sigma^{(2)}_{\mathrm{CT},V}
 & = &
 - \frac{1}{2} \left( 9 - 6 \xi \right) \frac{1}{\eps^2}
 + \left[ - \frac{55}{4} + 7 \xi + \frac{15 \xi}{x} - 3 \left( 1 - \frac{5}{x^2} \right) \xi \ln\left(1-x\right) \right] \frac{1}{\eps}
 \nonumber \\
 & &
 - \frac{7685}{72} + 18 \xi + \frac{38\xi}{x} 
 + \frac{211}{2} \zeta_2 + 3 \xi \zeta_2
 + 24 \zeta_3
 - 96 \zeta_2 \ln\left(2\right) 
 \nonumber \\
 & &
 - \left( 7 + \frac{12}{x} - \frac{23}{x^2} \right) \xi \ln\left(1-x\right)
 + 3 \left( 1 - \frac{5}{x^2} \right) \xi \left[ \mathrm{Li}_2\left(x\right) + \ln^2\left(1-x\right) \right]
 + {\mathcal O}\left(\eps\right),
 \nonumber \\
 \Sigma^{(2)}_{\mathrm{CT},S}
 & = &
 \left(2 - 6 \xi \right) \frac{1}{\eps^2}
 + \left[ 8 - 20 \xi + 2 \left( 7 + 3 \xi - \frac{16}{x} - \frac{6\xi}{x} \right) \ln\left(1-x\right) \right] \frac{1}{\eps}
 \nonumber \\
 & &
 + \frac{919}{9} - 56 \xi
 - 154 \zeta_2 - 6 \xi \zeta_2 
 - 36 \zeta_3 
 + 144 \zeta_2 \ln\left(2\right)
 + 2 \left( \frac{64}{3} + 10 \xi - \frac{82}{3x} - \frac{14\xi}{x} \right) 
 \nonumber \\
 & &
   \times \ln\left(1-x\right) 
 - 2 \left( 7 + 3 \xi - \frac{16}{x} - \frac{6\xi}{x} \right) \left[ \mathrm{Li}_2\left(x\right) + \ln^2\left(1-x\right) \right]
 + {\mathcal O}\left(\eps\right).
\eq

% -----------------------------------------------------------------------------

\section{Evaluation of the master integrals}
\label{sec:master_integrals}

\subsection{Transcendental constants}
\label{subsec:transcendental_constants}

There are a few transcendental constants appearing in the boundary constants.
The transcendental constants relevant to our calculation are
\bq
 \mbox{weight} \; 1 & : & 
 \pi, \;\; \ln\left(2\right), \;\; \ln\left(3\right),
 \nonumber \\
 \mbox{weight} \; 2 & : & 
 \mathrm{Li}_2\left(\frac{1}{3}\right), \;\;
 \mathrm{Cl}_2\left(\frac{2\pi}{3}\right),
 \nonumber \\
 \mbox{weight} \; 3 & : & 
 \zeta_3, \;\;
 \mathrm{Li}_3\left(\frac{1}{3}\right), \;\;
 \mathrm{Li}_{2,1}\left(\frac{1}{3},1\right), \;\;
\eq
and products of those.
Our convention for the notation of the multiple polylogarithms
is
\bq
 \mathrm{Li}_{n_1,n_2,...,n_k}\left(x_1,x_2,...,x_k\right)
 & = &
 \sum\limits_{j_1=1}^\infty \sum\limits_{j_2=1}^{j_1-1} ... \sum\limits_{j_k=1}^{j_{k-1}-1}
 \frac{x_1^{j_1}}{j_1^{n_1}} \frac{x_2^{j_2}}{j_2^{n_2}} ... \frac{x_k^{j_k}}{j_k^{n_k}}.
\eq
The Clausen function is defined by
\bq
 \mathrm{Cl}_2\left(\phi\right) & = &
 \frac{1}{2i} \left[ \mathrm{Li}_2\left(e^{i\phi}\right) - \mathrm{Li}_2\left(e^{-i\phi}\right) \right],
\eq
thus
\bq
 \mathrm{Cl}_2\left(\frac{2\pi}{3}\right)
 & = &
 \frac{1}{2i} \left[ \mathrm{Li}_2\left(r_3\right) - \mathrm{Li}_2\left(r_3^{-1}\right) \right],
\eq
where $r_3=\exp(2\pi i/3)$ denotes the third root of unity.
We list the relevant boundary constants for the boundary points $x \in \{0,1,9,\infty\}$ in appendix~\ref{sec:boundary_constants}.

\subsection{Integrals expressible in terms of harmonic polylogarithms}

The master integrals $I_1$-$I_5$ may be expressed (to all orders in $\eps$) in terms of harmonic polylogarithms \cite{Remiddi:2016gno,Adams:2016xah}.
The first few orders of the $\eps$-expansion of the integrals $I_1$-$I_5$ may be expressed in terms of classical polylogarithms.
We have
\bq
 I_1
 & = &
 4 + 4 \zeta_2 \eps^2 - \frac{8}{3} \zeta_3 \eps^3
 + {\mathcal O}\left(\eps^4\right),
 \nonumber \\
 I_2
 & = &
 4 H_1\left(x\right) \eps
 + \left[ 4 H_{01}\left(x\right) + 8 H_{11}\left(x\right) \right] \eps^2
 + \left[ 4 H_{001}\left(x\right) + 8 H_{011}\left(x\right) + 8 H_{101}\left(x\right) + 16 H_{111}\left(x\right) 
 \right. \nonumber \\
 & & \left.
 + 4 \zeta_2 H_1\left(x\right) \right] \eps^3
 + {\mathcal O}\left(\eps^4\right)
 \nonumber \\
 & = &
 - 4 \ln\left(1-x\right) \eps
 + \left[ 4 \;\mathrm{Li}_2\left(x\right) + 4 \ln^2\left(1-x\right) \right] \eps^2
 + \left[ 4 \; \mathrm{Li}_3\left(x\right) + 8 \; \mathrm{Li}_3\left(1-x\right)
 \right. \nonumber \\
 & & \left.
          + 4 \ln\left(x\right) \ln^2\left(1-x\right) - \frac{8}{3} \ln^3\left(1-x\right) - 12 \zeta_2 \ln\left(1-x\right) \right] \eps^3
 + {\mathcal O}\left(\eps^4\right),
 \nonumber \\
 I_3
 & = &
 - 4 H_1\left(x\right) \eps
 - \left[ 4 H_{01}\left(x\right) + 16 H_{11}\left(x\right) \right] \eps^2
 - \left[ 4 H_{001}\left(x\right) + 16 H_{011}\left(x\right) + 24 H_{101}\left(x\right) 
 \right. \nonumber \\
 & & \left. 
 + 64 H_{111}\left(x\right) 
 + 12 \zeta_2 H_1\left(x\right) \right] \eps^3
 + {\mathcal O}\left(\eps^4\right)
 \nonumber \\
 & = &
 4 \ln\left(1-x\right) \eps
 - \left[ 4 \;\mathrm{Li}_2\left(x\right) + 8 \ln^2\left(1-x\right) \right] \eps^2
 - \left[ 4 \; \mathrm{Li}_3\left(x\right) + 32 \; \mathrm{Li}_3\left(1-x\right)
 \right. \nonumber \\
 & & \left.
          + 8 \ln\left(1-x\right) \; \mathrm{Li}_2\left(x\right)
          + 16 \ln\left(x\right) \ln^2\left(1-x\right) - \frac{32}{3} \ln^3\left(1-x\right) - 44 \zeta_2 \ln\left(1-x\right) \right] \eps^3
 \nonumber \\
 & &
 + {\mathcal O}\left(\eps^4\right),
 \nonumber \\
 I_4
 & = &
 4
 + 8 H_1\left(x\right) \eps
 + \left[ 16 H_{01}\left(x\right) + 32 H_{11}\left(x\right) + 12 \zeta_2 \right] \eps^2
 + \left[ 16 H_{001}\left(x\right) + 64 H_{011}\left(x\right) 
 \right. \nonumber \\
 & & \left.
 + 48 H_{101}\left(x\right) + 128 H_{111}\left(x\right) + 24 \zeta_2 H_1\left(x\right)
 - \frac{32}{3} \zeta_3 \right] \eps^3
 + {\mathcal O}\left(\eps^4\right)
 \nonumber \\
 & = & 
 4
 - 8 \ln\left(1-x\right) \eps
 + \left[ 16 \;\mathrm{Li}_2\left(x\right) + 16 \ln^2\left(1-x\right) + 12 \zeta_2 \right] \eps^2
 + \left[ 
          16 \; \mathrm{Li}_3\left(x\right) + 32 \; \mathrm{Li}_3\left(1-x\right)
 \right. \nonumber \\
 & & \left.
          - 16  \ln\left(1-x\right) \; \mathrm{Li}_2\left(x\right)
          + 16 \ln\left(x\right) \ln^2\left(1-x\right) - \frac{64}{3} \ln^3\left(1-x\right) - 56 \zeta_2 \ln\left(1-x\right) - \frac{32}{3} \zeta_3 
   \right] \eps^3
 \nonumber \\
 & &
 + {\mathcal O}\left(\eps^4\right)
 \nonumber \\
 I_5
 & = &
 8 H_{11}\left(x\right) \eps^2
 + \left[ 16 H_{011}\left(x\right) + 8 H_{101}\left(x\right) + 48 H_{111}\left(x\right) \right] \eps^3
 + {\mathcal O}\left(\eps^4\right)
 \nonumber \\
 & = &
 4 \ln^2\left(1-x\right) \eps^2
 - \left[ 8 \ln\left(1-x\right) \; \mathrm{Li}_2\left(x\right) + 8 \ln^3\left(1-x\right) \right] \eps^3
 + {\mathcal O}\left(\eps^4\right).
\eq
The analytic continuation and the numerical evaluation of these master integrals are well understood.
The analytic continuation is dictated by Feynman's $i\delta$-prescription: $x\rightarrow x+i\delta$.
There are packages, which allow the numerical evaluation of harmonic polylogarithms $H_{m_1 ... m_k}(x)$ in double precision
and arbitrary precision \cite{Gehrmann:2001pz,Vollinga:2004sn,Maitre:2005uu,Maitre:2007kp}.

Let us also note the following alternative:
The harmonic polylogarithms with the letters $f_0$ and $f_1$ can be written as iterated integrals
of modular forms.
Since the latter are required anyhow for the problem at hand, 
we may as well treat the analytic continuation and
the numerical evaluation within the context of iterated integrals of modular forms.
We will follow this approach in this paper.

\subsection{The elliptic master integrals}

The master integrals $I_6$-$I_8$ depend on elliptic topologies and may be expressed as iterated integrals of modular
forms.
We remark that also the master integrals $I_1$-$I_5$ may be written as iterated integrals of modular forms.
This follows from the relations
\bq
 \frac{dx}{x} \;\; = \;\; 2 \pi i \; g_{2,0} \; d\tau,
 & &
 \frac{dx}{x-1} \;\; = \;\; 2 \pi i \; g_{2,1} \; d\tau.
\eq
From the maximal cut of the sunrise integral we obtain the elliptic curve
\bq
\label{def_elliptic_curve}
 E
 & : &
 w^2 - z
       \left(z + 4 \right) 
       \left[z^2 + 2 \left(1+x\right) z + \left(1-x\right)^2 \right]
 \; = \; 0.
\eq
We denote the roots of the quartic polynomial in eq.~(\ref{def_elliptic_curve}) by
\bq
 z_1 \; = \; -4,
 \;\;\;
 z_2 \; = \; -\left(1+\sqrt{x}\right)^2,
 \;\;\;
 z_3 \; = \; -\left(1-\sqrt{x}\right)^2,
 \;\;\;
 z_4 \; = \; 0.
\eq
There is an isomorphism between the elliptic curve and ${\mathbb C} / \Lambda$,
where $\Lambda$ is a lattice generated by the periods of the elliptic curve:
\bq
 \Lambda & = & \left\{ n_1 \psi_{1,0} + n_2 \psi_{2,0} \left.| n_1, n_2 \in {\mathbb Z} \right. \right\},
\eq
where $\psi_{1,0}$ and $\psi_{2,0}$ are two periods of the elliptic curve generating the lattice $\Lambda$.
It can be shown that $\psi_{1,0}$ and $\psi_{2,0}$ are two independent solutions of the 
homogeneous second-order differential equation given in eq.~(\ref{Picard_Fuchs_equation}).
Any other pair $\psi_{1,j}$ and $\psi_{2,j}$ of periods related to the first one by
\bq
 \left( \begin{array}{c}
  \psi_{2,j} \\
  \psi_{1,j} \\
 \end{array} \right)
 \;\; = \;\;
 \left( \begin{array}{cc}
  a & b \\
  c & d \\
 \end{array} \right)
 \left( \begin{array}{c}
  \psi_{2,0} \\
  \psi_{1,0} \\
 \end{array} \right),
 & &
 \left( \begin{array}{cc}
  a & b \\
  c & d \\
 \end{array} \right)
 \;\; \in \;\;
 \mathrm{SL}\left(2,{\mathbb Z}\right)
\eq
generates the same lattice.
$\psi_{1,j}$ and $\psi_{2,j}$ are again two independent solutions 
of the differential equation given in eq.~(\ref{Picard_Fuchs_equation}).
If $\psi_{1,0}$ and $\psi_{2,0}$ are normalised according to eq.~(\ref{Wronskian_relation}),
then so are $\psi_{1,j}$ and $\psi_{2,j}$.
We see that we have some freedom in choosing a pair of periods $\psi_{1,j}$ and $\psi_{2,j}$ as independent solutions
of eq.~(\ref{Picard_Fuchs_equation}). We will label different choices by the subscript $j$.
The definition of the master integrals $I_6$ and $I_7$ involves the period $\psi_{1,j}$ and hence depends on our choice
of $\psi_{1,j}$ and $\psi_{2,j}$.
Let us now discuss the dependence on our choice of $\psi_{1,j}$ and $\psi_{2,j}$
in more detail:
As already mentioned, the master integrals $I_1$-$I_5$ and $I_8$ do not depend at all on our choice of $\psi_{1,j}$ and $\psi_{2,j}$:
\bq
\label{matching_equation_123458}
 I_{i,j'} & = & I_{i,j},
 \;\;\;\;\;\;\;\;\;
 i \in \left\{ 1,2,3,4,5,8 \right\}.
\eq
The dependence of $I_{6}$ on our choice is rather simple:
\bq
\label{matching_equation_6}
 I_{6,j'}
 & = & 
 \frac{\psi_{1,j}}{\psi_{1,j'}} I_{6,j}.
\eq
The dependence of $I_{7}$ on our choice is more tricky, due to the appearance of the derivative of $d\psi_1/dx$.
One finds
\bq
\label{matching_equation_7}
 I_{7,j'}
 & = & 
 - \frac{x \left(x-1\right) \left(x-9\right)}{12 \eps \pi^2}  
 \left( \psi_{1,j} \frac{d}{dx} \psi_{1,j'} - \psi_{1,j'} \frac{d}{dx} \psi_{1,j} \right)
 I_{6,j}
 +
 \frac{\psi_{1,j'}}{\psi_{1,j}} I_{7,j}.
\eq
This relation is most easily derived by relating both the basis $I_{i,j'}$ and the basis $I_{i,j}$ to 
a basis independent of our choice of periods.
This intermediate basis does not need to be an $\eps$-basis.
A possible intermediate basis is
\bq
 \left\{ I_1, I_2, I_3, I_4, I_5, I_{10101}(2-2\eps), I_{20101}(2-2\eps), I_8 \right\}.
\eq
Eqs.~(\ref{matching_equation_123458})-(\ref{matching_equation_7})
allow us to match the master integrals with different choices of periods in regions where both choices lead to convergent
series expansions.
This can be used to determine the boundary conditions for one choice from the known boundary conditions of the other choice.
In practice we proceed as follows:
Suppose we already know the boundary constants for the choice $j$ and we would like to obtain the boundary constants for the choice
$j'$. Suppose further that there is a region where both choices lead to convergent
series expansions. Evaluating both expressions to high precision gives us numerical values to high precision for the 
boundary constants of the choice $j'$.
We may then use the PSLQ-algorithm \cite{Ferguson:1992} to match these values to a ${\mathbb Q}$-linear combination
of the transcendental constants from section~\ref{subsec:transcendental_constants}.

In the following we will discuss four choices for the pair of periods. We label them
\bq
 \left( \psi_{1,j}, \psi_{2,j} \right),
 & & 
 j \in \left\{0,1,9,\infty\right\}.
\eq
For each choice we set
\bq
 \tau_{n_j,j} \; = \; \frac{1}{n_j} \frac{\psi_{2,j}}{\psi_{1,j}},
 & &
 q_{n_j,j} \; = \; e^{2 i \pi \tau_{n_j,j}},
 \;\;\;\;\;\;\;\;\;
 n_j \in {\mathbb N}.
\eq
The values of $n_j$ will be
\bq
 n_0 \; = \; 1,
 \;\;\;\;\;\;
 n_1 \; = \; 6,
 \;\;\;\;\;\;
 n_9 \; = \; 2,
 \;\;\;\;\;\;
 n_\infty \; = \; 3.
\eq
Each of the four choices has the property that
\bq
 q_{n_j,j} \; = \; 0
 & \mbox{for} &
 x \; = \; j, 
 \;\; j \in \{0,1,9,\infty\},
\eq
i.e. $q_{n_j,j}$ vanishes at the cusp $x=j$.
We write $b_{i,j}$ if $\psi_{1,j}$ is substituted for $\psi_1$ in the generic definition of $b_i$ in eq.~(\ref{def_b_basis}).
For the $\eps$-expansion of the master integrals we write
\bq 
 I_{i,j} & = & 
 \sum\limits_{k=0}^\infty \eps^k \; I_{i,j}^{(k)},
 \;\;\;\;\;\;\;\;\;
 1 \le i \le 8,
 \;\;\;\;\;\;
 j \in \{0,1,9,\infty\}.
\eq
We will need for the two-loop self-energy up to the finite part the integral $I_{6,j}$ to order $\eps^2$, 
the integral $I_{7,j}$ to order $\eps$ and the integral $I_{8,j}$ to order $\eps^3$.
The integral $I_{6,j}$ starts at order $\eps^2$,
the integral $I_{7,j}$ starts at order $\eps$,
while the integral $I_{8,j}$ starts at order $\eps^3$.
Therefore we need in all three cases the first non-vanishing order.

\subsubsection{The cusp $p^2=0$}

We start with the cusp $x=0$.
We introduce the modulus $k$ and the complementary modulus $k'$ through
\bq
 k^2 
 \; = \;
 \frac{\left(z_3-z_2\right)\left(z_4-z_1\right)}{\left(z_3-z_1\right)\left(z_4-z_2\right)},
 & &
 k'{}^2
 \; = \;
 \frac{\left(z_2-z_1\right)\left(z_4-z_3\right)}{\left(z_3-z_1\right)\left(z_4-z_2\right)}.
\eq
Explicitly we have
\bq
 k^2 
 \; = \;
 \frac{16 \sqrt{x}}{\left(1+\sqrt{x}\right)^3 \left(3-\sqrt{x}\right)},
 & &
 k'{}^2
 \; = \;
 \frac{\left(1-\sqrt{x}\right)^3 \left(3+\sqrt{x}\right)}{\left(1+\sqrt{x}\right)^3 \left(3-\sqrt{x}\right)},
\eq
where Feynman's $i\delta$-prescription ($x\rightarrow x+i\delta$) is understood.
Our choice of periods for this case 
(which agrees with the choice made in ref.~\cite{Adams:2018yfj})
is given by
\bq
\label{def_periods_choice_0}
 \left( \begin{array}{c}
  \psi_{2,0} \\
  \psi_{1,0} \\
 \end{array} \right)
 & = &
 \frac{4}{\left(1+\sqrt{x}\right)^{\frac{3}{2}} \left(3-\sqrt{x}\right)^{\frac{1}{2}}}
 \; \gamma \;
 \left( \begin{array}{c}
  i K\left( k' \right) \\
  K\left( k \right) \\
 \end{array} \right),
\eq
where $K(k_0)$ denotes the complete elliptic integral of the first kind.
The complete elliptic integral is efficiently computed with the help of the arithmetic-geometric mean,
reviewed in appendix~\ref{sec:agm}.
The $2 \times 2$-matrix $\gamma$ is given by
\bq
 \gamma
 & = &
 \left\{ \begin{array}{ccrcccl}
  \left( \begin{array}{rr}
   1 & 0 \\
   2 & 1 \\
  \end{array} \right),
  & & -\infty & < & x & < & 3 - 2 \sqrt{3}, 
  \\
  \left( \begin{array}{rr}
   1 & 0 \\
   0 & 1 \\
  \end{array} \right),
  & & 3 - 2 \sqrt{3} & < & x & < & 1, 
  \\
  \left( \begin{array}{rr}
   1 & 0 \\
   2 & 1 \\
  \end{array} \right),
  & & 1 & < & x & < & \infty.
  \\
 \end{array} \right.
\eq
The matrix $\gamma$ ensures that the periods $\psi_{1,0}$ and $\psi_{2,0}$ vary smoothly as $x$ varies smoothly
in $x \in {\mathbb R} + i \delta$ \cite{Bogner:2017vim}.
The complete elliptic integral $K(k)$ can be viewed as a function of $k^2$: 
We set $\tilde{K}(k^2)=K(k)$.
The function $\tilde{K}(k^2)$ has a branch cut at $[1,\infty[$ in the complex $k^2$-plane.
The matrix $\gamma$ compensates for the discontinuity when we cross this branch cut.
It is relatively easy to see that $k^2$ as a function of $x$ crosses this branch cut at the point
$x=3-2\sqrt{3} \approx -0.46$, the corresponding value in the $k^2$-plane is 
$k^2=2$.
The point $x=1$ is a little bit more subtle. Let us parametrise a small path around $x=1$ by
\bq
 x\left(\phi\right) & = & 1 + \delta e^{i \left(\pi-\phi\right)},
 \;\;\;\;\;\;
 \phi \in [0,\pi],
\eq
then
\bq
 k^2
 & = &
 1 + \frac{1}{32} \delta^3 e^{3 i \left(\pi-\phi\right)}
 + {\mathcal O}\left(\delta^4\right),
\eq
and the path in $k^2$-space winds around the point $k^2=1$ by an angle $3\pi$ as 
the path in $x$-space winds around the point $x=1$ by the angle $\pi$.

Note that eq.~(\ref{def_periods_choice_0}) defines the periods $\psi_{1,0}$ and $\psi_{2,0}$
for all values $x \in {\mathbb R} + i \delta$.
The periods take values in ${\mathbb C} \cup \left\{ \infty \right\}$.

\begin{figure}
\begin{center}
\includegraphics[scale=1.0]{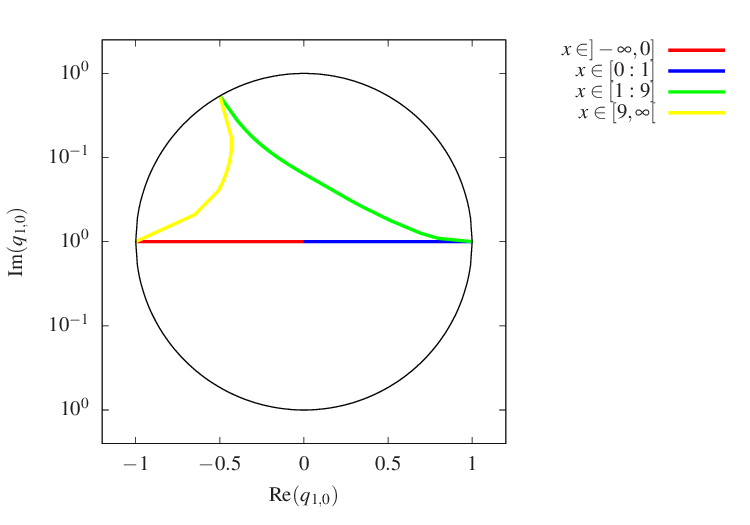}
\end{center}
\caption{\label{fig_q_0_path}
The path in $q_{1,0}$-space as $x$ ranges over ${\mathbb R}$.
We always have $|q_{1,0}| \le 1$ and $|q_{1,0}| = 1$ only at $x \in \{1,9,\infty\}$. 
}
\end{figure}

One easily verifies that $\psi_{1,0}$ and $\psi_{2,0}$ are normalised according to eq.~(\ref{Wronskian_relation}).
We set
\bq
\label{def_tau_0}
 \tau_{1,0}
 \;\; = \;\;
 \frac{\psi_{2,0}}{\psi_{1,0}},
 & &
 q_{1,0} \;\; = \;\; e^{2 i \pi \tau_{1,0}}.
\eq
The values of $\tau_{1,0}$ at the points $x \in \{0,1,9,\infty\}$ are tabulated in table~\ref{table_tau}.
\begin{table}
\begin{center}
\begin{tabular}{c|rrrr}
 $x$ & $0$ & $1$ & $9$ & $\infty$ \\
\hline
 & & & & \\
 $\tau_{1,0}$ & $i \infty$ & $0$ & $\frac{1}{3}$ & $\frac{1}{2}$ \\
 & & & & \\
 $\tau_{6,1}$ & $0$ & $i \infty$ & $-\frac{1}{2}$ & $-\frac{1}{3}$ \\
 & & & & \\
 $\tau_{2,9}$ & $\frac{1}{3}$ & $\frac{1}{2}$ & $i \infty$ & $0$ \\
 & & & & \\
 $\tau_{3,\infty}$ & $-\frac{1}{2}$ & $-\frac{1}{3}$ & $0$ & $i \infty$ \\
\end{tabular}
\caption{\label{table_tau}
The values of the variables $\tau_{1,0}$, $\tau_{6,1}$, $\tau_{2,9}$ and $\tau_{3,\infty}$ at the cusps
$x \in \{0,1,9,\infty\}$.
}
\end{center}
\end{table}
In fig.~\ref{fig_q_0_path} we plot the values of the variable $q_{1,0}$ as $x$ ranges over
${\mathbb R}$.
We see that all values of $q_{1,0}$ are inside the unit disc 
with the exception of the three points $x \in \{1,9,\infty\}$, where the corresponding $q_{1,0}$-values
are on the boundary of the unit disc.

In order to simplify the notation we write in the remaining part of this section
\bq
 \tau \;\; = \;\; \tau_{1,0},
 & &
 q \;\; = \;\; q_{1,0}.
\eq
We also use the notation $b_1$ for $b_{1,0}$ and $b_2$ for $b_{2,0}$.
Eq.~(\ref{def_tau_0}) defines $\tau$ as a function of $x$.
In a neighbourhood of $x=0$ we may invert eq.~(\ref{def_tau_0}). This gives
\bq
\label{hauptmodul_0}
 x
 & = &
 9
 \frac{\eta\left(\tau\right)^4 \eta\left(6\tau\right)^8}
      {\eta\left(3\tau\right)^4 \eta\left(2\tau\right)^8},
\eq
where $\eta$ denotes Dedekind's eta-function.
The integration kernels appearing in eq.~(\ref{integration_kernels})
are modular forms of the congruence subgroup $\Gamma_1(6)$.
In order to present the $q$-expansion of the integration kernels 
we introduce a basis $\{e_1,e_2\}$ for the modular forms of modular weight $1$ 
for the Eisenstein subspace ${\mathcal E}_1(\Gamma_1(6))$:
\bq
 e_1 \; = \; E_1\left(\tau;\chi_0,\chi_1\right),
 & &
 e_2 \; = \; E_1\left(2\tau;\chi_0,\chi_1\right),
\eq
where $\chi_0$ and $\chi_1$ denote primitive Dirichlet characters with conductors $1$ and $3$, respectively.
The Eisenstein series $E_1(\tau,\chi_0,\chi_1)$ and $E_1(2\tau,\chi_0,\chi_1)$ 
are defined in appendix~\ref{sec:eisenstein}.
All occurring integration kernels may be expressed as polynomials in $e_1$ and $e_2$. 
We first express $b_1$ and $b_2$ (defined in eq.~(\ref{def_b_basis})) 
in terms of $e_1$ and $e_2$:
\bq
 b_1
 \; = \;
 2 \sqrt{3}
 \left( e_1 + e_2 \right),
 & &
 b_2
 \; = \; 
 12 \sqrt{3} e_1.
\eq
This shows that $\{b_1,b_2\}$ is also a basis of ${\mathcal E}_1(\Gamma_1(6))$.
With the help of eq.~(\ref{kernels_b_polynomials}) we may now express the integration kernels as polynomials
in $e_1$ and $e_2$:
\bq
\label{integration_kernels_q_expansion}
 g_{2,0} 
 & = &
 - 12 \left( e_1^2 - 4 e_2^2 \right),
 \nonumber \\
 g_{2,1} 
 & = &
 - 18 \left( e_1^2 + e_1 e_2 - 2 e_2^2 \right),
 \nonumber \\
 g_{2,9} 
 & = &
 6 \left( e_1^2 - 3 e_1 e_2 + 2 e_2^2 \right),
 \nonumber \\
 g_{3,0} 
 & = &
 -72 \sqrt{3} \left( e_1^3 - e_1^2 e_2 - 4 e_1 e_2^2 + 4e_2^3 \right),
 \nonumber \\
 g_{3,1} 
 & = &
 - 108 \sqrt{3} \left( e_1^3 - 3 e_1 e_2^2 + 2 e_2^3 \right),
 \nonumber \\
 f_{4}
 & = &
 324 e_1^4.
\eq
We obtain for the first non-vanishing order of the integrals $I_{6,0}$, $I_{7,0}$ and $I_{8,0}$
\bq
 I_{6,0}^{(2)}
 & = &
 3 \, \mathrm{Cl}_2\left(\frac{2\pi}{3}\right) - \frac{1}{2}  \iterintmodular{1,g_{3,0}}{q}
 \nonumber \\
 & = &
 3 \, \mathrm{Cl}_2\left(\frac{2\pi}{3}\right)
 - 3 \sqrt{3}
   \left[
         q-\frac{5}{4}\,q^{2}+q^{3}-{\frac {11}{16}}\,q^{4}+{\frac {24}{25}}\,q^{5}
         -\frac{5}{4}\,q^{6}+{\frac {50}{49}}\,q^{7}-{\frac {53}{64}}\,q^{8}+q^{9}
   \right]
 \nonumber \\
 & &
 + {\mathcal O}\left(q^{10}\right),
 \nonumber \\
 I_{7,0}^{(1)}
 & = &
 - \frac{1}{2}  \iterintmodular{g_{3,0}}{q}
 \nonumber \\
 & = &
 - 3 \sqrt{3}
   \left[
         q-\frac{5}{2}\,q^{2}+3\,q^{3}-\frac{11}{4}\,q^{4}+{\frac {24}{5}}\,q^{5}
         -\frac{15}{2}\,q^{6}+{\frac {50}{7}}\,q^{7}-{\frac {53}{8}}\,q^{8}+9\,q^{9}
   \right]
 + {\mathcal O}\left(q^{10}\right),
 \nonumber \\
 I_{8,0}^{(3)}
 & = &
 8 \, \iterintmodular{g_{2,1},g_{2,0},g_{2,1}}{q}
 -16 \, \iterintmodular{g_{2,0},g_{2,1},g_{2,1}}{q}
 + 6 \, \iterintmodular{g_{3,0},1,g_{3,0}}{q}
 \nonumber \\
 & &
 - \frac{16}{3} \iterintmodular{g_{3,1},1,g_{3,0}}{q}
 - 8 \zeta_2 \, \iterintmodular{g_{2,1}}{q}
 + 3 \, \mathrm{Cl}_2\left(\frac{2\pi}{3}\right) \left( 
                 \frac{32}{3} \iterintmodular{g_{3,1}}{q}
                 - 12 \, \iterintmodular{g_{3,0}}{q}
          \right)
 \nonumber \\
 & = &
  324\,q^{2}+864\,q^{3}+{\frac {2025}{2}}\,q^{4}+891\,q^{5}+351\,q^{6}
 +{\frac {33372}{25}}\,q^{7}+{\frac {45074961}{19600}}\,q^{8}
 \nonumber \\
 & &
 +{\frac {8208243}{2450}}\,q^{9}
 + \zeta_2 \left[ 
 72\,q+36\,q^{2}+72\,q^{3}+18\,q^{4}+{\frac {432}{5}}\,q^{5}+36\,q^{6}
 +{\frac {576}{7}}\,q^{7}+9\,q^{8}
 \right. \nonumber \\
 & & \left.
 +72\,q^{9}
 \right]
 - 9 \sqrt{3} \mathrm{Cl}_2\left(\frac{2\pi}{3}\right) \left[
  24\,q+36\,q^{2}+72\,q^{3}+78\,q^{4}+{\frac {576}{5}}\,q^{5}+108\,q^{6}
 +{\frac {1200}{7}}\,q^{7}
 \right. \nonumber \\
 & & \left.
 +153\,q^{8}+216\,q^{9}
 \right]
 + {\mathcal O}\left(q^{10}\right).
\eq
In the supplementary electronic file attached to this article we give the $q$-expansion
for the master integrals $I_{i,0}$ up to $q^{\qorder}$ for the first four orders in $\eps$.

Let us note that all master integrals $I_1$ - $I_8$ in any order in $\eps$ may be expressed as
iterated integrals of modular forms. This includes in particular the master integrals
$I_1$ - $I_5$, which may be expressed in the simpler class of harmonic polylogarithms.
Let us further note that
\bq
 \left| q \right| & < & 1
 \;\;\;\;\;\; \mbox{for} \;\; x \in \left( {\mathbb R} \cup \left\{\infty\right\} \right) \backslash \{ 1,9,\infty \},
\eq
therefore the $q$-series converge for all values $x \in \left( {\mathbb R} \cup \left\{\infty\right\} \right) \backslash \{ 1,9,\infty \}$, i.e. for all values in ${\mathbb R} \cup \left\{\infty\right\}$
except for three points $\{1,9,\infty\}$.
With
\bq
 \psi_{1,0}
 \;\; = \;\;
 2 \sqrt{3} \pi
 \left( e_1 + e_2 \right),
 & &
 \frac{d}{dx} \psi_{1,0}
 \;\; = \;\;
 \frac{2 \pi i \, W}{\psi_{1,0}^2} \; q \frac{d}{dq} \psi_{1,0},
\eq
and eq.~(\ref{hauptmodul_0}) we may express
$\Sigma^{(2,0)}_{\mathrm{bare},V}$ and $\Sigma^{(2,0)}_{\mathrm{bare},S}$
as $q$-series.
We obtain in Feynman gauge
\bq
\label{result_q_series_case_0}
\lefteqn{
 \Sigma^{(2,0)}_{\mathrm{bare},V}
 = 
 {\frac {293}{8}}
 +511\,q
 +{\frac {61859}{16}}\,q^{2}
 +{\frac {1139579}{50}}\,q^{3}
 +{\frac {22506803}{200}}\,q^{4}
 +{\frac {2418064473}{4900}}\,q^{5}
 } & &
 \nonumber \\
 & &
 +{\frac {153385103807}{78400}}\,q^{6}
 +{\frac {13938069377}{1960}}\,q^{7}
 +{\frac {146831758723}{6125}}\,q^{8}
 +{\frac {112258408704193}{1482250}}\,q^{9}
 \nonumber \\
 & &
 + 12 \zeta_2 \left[
  \frac{1}{24}+q+\frac{11}{4}\,{q}^{2}+{\frac {23}{5}}\,{q}^{3}+{\frac {43}{10}}\,{q}^{4}-{\frac {12}{35}}\,{q}^{5}
  -{\frac {1867}{280}}\,{q}^{6}-{\frac {29}{5}}\,{q}^{7}+{\frac {49}{5}}\,{q}^{8}+{\frac {12382}{385}}\,{q}^{9}
 \right]
 \nonumber \\
 & &
 - 36 \sqrt{3} \mathrm{Cl}_2\left(\frac{2\pi}{3}\right) \left[
   1+11\,q+{\frac {335}{4}}\,{q}^{2}+{\frac {2547}{5}}\,{q}^{3}+{\frac {26057}{10}}\,{q}^{4}+{\frac {406422}{35}}\,{q}^{5}
   +{\frac {12968727}{280}}\,{q}^{6}
 \right. \nonumber \\
 & & \left.
   +{\frac {842799}{5}}\,{q}^{7}+{\frac {2841168}{5}}\,{q}^{8}
   +{\frac {138212638}{77}}\,{q}^{9}
 \right]
 + {\mathcal O}\left(q^{10}\right),
 \nonumber \\
\lefteqn{
 \Sigma^{(2,0)}_{\mathrm{bare},S}
 = 
 -72-864\,q-{\frac {12069}{2}}\,q^{2}-{\frac {290385}{8}}\,q^{3}
 -{\frac {35860023}{200}}\,q^{4}-{\frac {15981543}{20}}\,q^{5}
 } & &
 \nonumber \\
 & &
 -{\frac {31369233123}{9800}}\,q^{6}-{\frac {918608937507}{78400}}\,q^{7}-{\frac {63428578551}{1600}}\,q^{8}
 -{\frac {3075431500611}{24500}}\,q^{9}
 \nonumber \\
 & &
 + 27 \zeta_2 \left[
   \frac{4}{9}+q+{q}^{2}+\frac{3}{2}\,{q}^{3}+{\frac {7}{10}}\,{q}^{4}+\frac{3}{5}\,{q}^{5}-{\frac {3}{35}}\,{q}^{6}+{\frac {187}{140}}\,{q}^{7}+{\frac {341}{140}}\,{q}^{8}
   +{\frac {249}{70}}\,{q}^{9}
 \right]
 \nonumber \\
 & &
 + 54 \sqrt{3} \mathrm{Cl}_2\left(\frac{2\pi}{3}\right) \left[
    1+\frac{19}{2}\,q+{\frac {159}{2}}\,{q}^{2}+{\frac {2075}{4}}\,{q}^{3}+{\frac {55111}{20}}\,{q}^{4}+{\frac {125349}{10}}\,{q}^{5}
    +{\frac {3538201}{70}}\,{q}^{6}
 \right. \nonumber \\
 & & \left.
    +{\frac {51827761}{280}}\,{q}^{7}+{\frac {175368003}{280}}\,{q}^{8}+{\frac {277651999}{140}}\,{q}^{9}
 \right]
 + {\mathcal O}\left(q^{10}\right).
\eq
In the supplementary electronic file attached to this article we give the $q$-expansion
for $\Sigma^{(2,0)}_{\mathrm{bare},V}$ and $\Sigma^{(2,0)}_{\mathrm{bare},S}$ 
in an arbitrary covariant gauge up to $q^{\qorder}$.

\subsubsection{The cusp $p^2=m^2$}

\begin{figure}
\begin{center}
\includegraphics[scale=1.0]{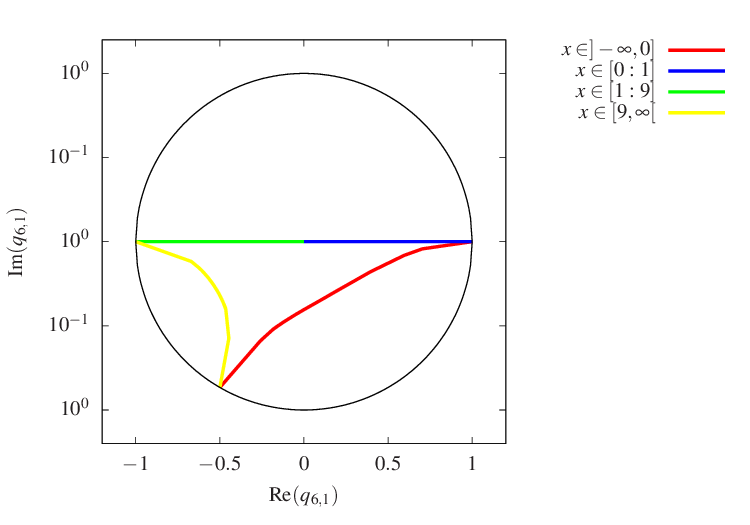}
\end{center}
\caption{\label{fig_q_1_path}
The path in $q_{6,1}$-space as $x$ ranges over ${\mathbb R}$.
We always have $|q_{6,1}| \le 1$ and $|q_{6,1}| = 1$ only at $x \in \{0,9,\infty\}$. 
}
\end{figure}

We now turn to the expansion around the cusp $x=1$.
We set
\bq
\label{def_periods_choice_1}
 \left( \begin{array}{c}
  \psi_{2,1} \\
  \psi_{1,1} \\
 \end{array} \right)
 & = &
 \left( \begin{array}{rr}
 0 & -1 \\
 1 & 0 \\
 \end{array} \right)
 \left( \begin{array}{c}
  \psi_{2,0} \\
  \psi_{1,0} \\
 \end{array} \right),
\eq
where $\psi_{1,0}$ and $\psi_{2,0}$ have been defined for all values of $x \in {\mathbb R} \cup \{\infty\}$ in eq.~(\ref{def_periods_choice_0}).
This choice agrees up to a constant phase $e^{-i \pi/2}$ with the choice made in ref.~\cite{Adams:2018kez} in the sunrise sector.
We set
\bq
\label{def_tau_1}
 \tau_{6,1}
 \;\; = \;\;
 \frac{1}{6}
 \frac{\psi_{2,1}}{\psi_{1,1}},
 & &
 q_{6,1} \;\; = \;\; e^{2 i \pi \tau_{6,1}}.
\eq
The values of $\tau_{6,1}$ at the points $x \in \{0,1,9,\infty\}$ are tabulated in table~\ref{table_tau}.
In fig.~\ref{fig_q_1_path} we plot the values of the variable $q_{6,1}$ as $x$ ranges over
${\mathbb R}$.
In order to simplify the notation we write in the remaining part of this section
\bq
 \tau \;\; = \;\; \tau_{6,1},
 & &
 q \;\; = \;\; q_{6,1},
\eq
and $e_1 = E_1\left(\tau_{6,1};\chi_0,\chi_1\right)$,
$e_2 = E_1\left(2\tau_{6,1};\chi_0,\chi_1\right)$.
We also use the notation $b_1$ for $b_{1,1}$ and $b_2$ for $b_{2,1}$.
The Hauptmodul is given by
\bq
\label{hauptmodul_1}
 x-1 & = &
 -8 \frac{\eta\left(\tau\right)^3 \eta\left(6\tau\right)^9}{\eta\left(2\tau\right)^3 \eta\left(3\tau\right)^9}.
\eq
The $q$-expansions of $b_1$ and $b_2$ are given by
\bq
 b_1 
 \; = \;
 i \left( e_1 + 2 e_2 \right),
 & &
 b_2
 \; = \;
 12 i e_2.
\eq
For the integration kernels we have
\bq
 g_{2,0} 
 & = &
 - 4 \left( e_1^2 - e_2^2 \right),
 \nonumber \\
 g_{2,1} 
 & = &
 - 3 \left( e_1^2 - e_1 e_2 - 2 e_2^2 \right), 
 \nonumber \\
 g_{2,9} 
 & = &
 - e_1^2 + 3 e_1 e_2 - 2 e_2^2,
 \nonumber \\
 g_{3,0} 
 & = &
 12 i \left( e_1^3 -2 e_1^2 e_2 - e_1 e_2^2 + 2 e_2^3 \right),
 \nonumber \\
 g_{3,1} 
 & = &
 9 i \left( e_1^3 - 3 e_1^2 e_2 + 4 e_2^3 \right),
 \nonumber \\
 f_{4}
 & = &
 36 e_2^4.
\eq
We obtain for the first non-vanishing order of the integrals $I_{6,1}$, $I_{7,1}$ and $I_{8,1}$
\bq
 I_{6,1}^{(2)}
 & = &
 -3 i \zeta_2 - 18 \, \iterintmodular{1,g_{3,0}}{q}
 \nonumber \\
 & = &
 -3 i \zeta_2
 + 12 i \left[
  q-{q}^{2}
  +\frac{1}{9} \,{q}^{3}+{q}^{4}-{\frac {24}{25}}\,{q}^{5}
  -\frac{1}{9}\,{q}^{6}+{\frac {50}{49}}\,{q}^{7}-{q}^{8}+{\frac {1}{81}}\,{q}^{9}
 \right]
 + {\mathcal O}\left(q^{10}\right),
 \nonumber \\
 I_{7,1}^{(1)}
 & = &
 - 3 \, \iterintmodular{g_{3,0}}{q}
 \nonumber \\
 & = &
 2 i \left[
   q-2\,{q}^{2}+\frac{1}{3}\,{q}^{3}+4\,{q}^{4}-{\frac {24}{5}}\,{q}^{5}
   -\frac{2}{3}\,{q}^{6}+{\frac {50}{7}}\,{q}^{7}-8\,{q}^{8}+\frac{1}{9}\,{q}^{9}
 \right]
 + {\mathcal O}\left(q^{10}\right),
 \nonumber \\
 I_{8,1}^{(3)}
 & = &
 1728\,\iterintmodular{ g_{2,1}, g_{2,0}, g_{2,1} }{q}
-3456\,\iterintmodular{ g_{2,0}, g_{2,1}, g_{2,1} }{q}
+1296\,\iterintmodular{ g_{3,0}, 1, g_{3,0} }{q}
 \nonumber \\
 & &
-1152\,\iterintmodular{ g_{3,1}, 1, g_{3,0} }{q}
+864\,\ln  \left( 2 \right) \iterintmodular{ g_{2,1}, g_{2,0} }{q}
-1728\,\ln  \left( 2 \right) \iterintmodular{ g_{2,0}, g_{2,1} }{q}
 \nonumber \\
 & &
-96\,\zeta_2\,\iterintmodular{ g_{2,1} }{q}
-432\, \ln^2\left( 2 \right)  \iterintmodular{ g_{2,0} }{q}
+216\,i\zeta_2\,\iterintmodular{ g_{3,0} }{q}
-192\,i\zeta_2\,\iterintmodular{ g_{3,1} }{q}
 \nonumber \\
 & &
+12\,\zeta_3
-48\,\zeta_2\,\ln  \left( 2 \right) 
 \nonumber \\
 & = &
 12\,\zeta_3
-48\,\zeta_2\,\ln  \left( 2 \right) 
 +192\,q+96\,{q}^{2}+{\frac {2848}{9}}\,{q}^{3}+{\frac {1088}{3}}\,{q}^{4}
 -{\frac {14464}{375}}\,{q}^{5}+{\frac {81632}{225}}\,{q}^{6}
 \nonumber \\
 & &
 +{\frac {4028176}{8575}}\,{q}^{7}
 -{\frac {222512}{735}}\,{q}^{8}+{\frac {35743076}{59535}}\,{q}^{9}
 + 64 \left[ \ln\left(q\right) + 3 \ln\left(2\right) \right]^2
   \left[
         q+{q}^{2}+\frac{1}{3}\,{q}^{3}+{q}^{4}
 \right. \nonumber \\
 & & \left.
         +\frac{6}{5}\,{q}^{5}+\frac{1}{3}\,{q}^{6}
         +{\frac {8}{7}}\,{q}^{7}+{q}^{8}+\frac{1}{9}\,{q}^{9}
   \right]
 - 192 \left[ \ln\left(q\right) + 3 \ln\left(2\right) \right]
   \left[
          q+{q}^{2}
          +{\frac {10}{9}}\,{q}^{3}+\frac{4}{3}\,{q}^{4}
 \right. \nonumber \\
 & & \left.
          +{\frac {67}{50}}\,{q}^{5}
          +{\frac {133}{90}}\,{q}^{6}+{\frac {751}{1470}}\,{q}^{7}+{\frac {23}{210}}\,{q}^{8}
          +{\frac {22919}{11340}}\,{q}^{9}
   \right]
 + 48 \zeta_2 \left[
          q-{q}^{2}-\frac{5}{3}\,{q}^{3}+\frac{5}{2}\,{q}^{4}
 \right. \nonumber \\
 & & \left.
          +\frac{6}{5}\,{q}^{5}
          -\frac{4}{3}\,{q}^{6}+{\frac {8}{7}}\,{q}^{7}
          -{\frac {19}{4}}\,{q}^{8}-{\frac {23}{9}}\,{q}^{9}
  \right]
 + {\mathcal O}\left(q^{10}\right).
\eq
In the supplementary electronic file attached to this article we give the $q$-expansion
for the master integrals $I_{i,1}$ up to $q^{\qorder}$ for the first four orders in $\eps$.
The $q$-series for the master integrals $I_{i,1}$ converge for all values 
$x \in \left( {\mathbb R} \cup \left\{\infty\right\} \right) \backslash \{ 0,9,\infty \}$.
With
\bq
 \psi_{1,1}
 \;\; = \;\;
 i \pi \left( e_1 + 2 e_2 \right),
 & &
 \frac{d}{dx} \psi_{1,1}
 \;\; = \;\;
 \frac{1}{6} \cdot
 \frac{2 \pi i \, W}{\psi_{1,1}^2} \; q \frac{d}{dq} \psi_{1,1},
\eq
and eq.~(\ref{hauptmodul_1}) we may express
$\Sigma^{(2,0)}_{\mathrm{bare},V}$ and $\Sigma^{(2,0)}_{\mathrm{bare},S}$
as $q$-series.
We obtain in Feynman gauge
\bq
\label{result_q_series_case_1}
\lefteqn{
 \Sigma^{(2,0)}_{\mathrm{bare},V}
 = 
-{\frac {363}{8}}-{\frac {1096}{3}}\,q-2386\,{q}^{2}-{\frac {2542012}{225}}\,{q}^{3}-{\frac {10420294}{225}}\,{q}^{4}
-{\frac {170189618}{875}}\,{q}^{5}
} & & \nonumber \\
 & &
-{\frac {7359947117}{7875}}\,{q}^{6}
-{\frac {1894448095382}{385875}}\,{q}^{7}
-{\frac {13120710706391}{514500}}\,{q}^{8}-{\frac {14203813439078278}{114604875}}\,{q}^{9}
 \nonumber \\
 & &
 + 12 \left[ \ln\left(q\right) + 3 \ln\left(2\right) \right]^2
 \left[
       1+{\frac {68}{3}}\,q+204\,{q}^{2}+{\frac {11012}{9}}\,{q}^{3}+{\frac {47468}{9}}\,{q}^{4}+{\frac {229528}{15}}\,{q}^{5}
       +{\frac {179068}{15}}\,{q}^{6}
 \right. \nonumber \\
 & & \left.
       -{\frac {63977248}{315}}\,{q}^{7}-{\frac {187312276}{105}}\,{q}^{8}
       -{\frac {9553315508}{945}}\,{q}^{9}
 \right]
 - 4 \left[ \ln\left(q\right) + 3 \ln\left(2\right) \right]
 \left[
       1+86\,q+981\,{q}^{2}
 \right. \nonumber \\
 & & \left.
       +{\frac {104654}{15}}\,{q}^{3}+{\frac {1077191}{30}}\,{q}^{4}+{\frac {10338842}{75}}\,{q}^{5}
       +{\frac {8907519}{25}}\,{q}^{6}+{\frac {177237254}{1225}}\,{q}^{7}
 \right. \nonumber \\
 & & \left.
       -{\frac {76489857877}{14700}}\,{q}^{8}
       -{\frac {1366150534909}{33075}}\,{q}^{9}
 \right]
 + 96 \left[ \zeta_3 -4 \zeta_2 \ln\left(2\right) \right] \left[
      \frac{1}{16}+q+9\,{q}^{2}+59\,{q}^{3}
 \right. \nonumber \\
 & & \left.
      +313\,{q}^{4}+1422\,{q}^{5}+5731\,{q}^{6}+20984\,{q}^{7}+71001\,{q}^{8}+224825\,{q}^{9}
 \right]
 + 360 \zeta_2 \left[
      {\frac {19}{720}}
      +q
 \right. \nonumber \\
 & & \left.
      +10\,{q}^{2}
      +{\frac {3049}{45}}\,{q}^{3}+{\frac {32717}{90}}\,{q}^{4}+{\frac {124466}{75}}\,{q}^{5}
      +{\frac {1508603}{225}}\,{q}^{6}+{\frac {4302456}{175}}\,{q}^{7}
      +{\frac {174855707}{2100}}\,{q}^{8}
 \right. \nonumber \\
 & & \left.
      +{\frac {1246555889}{4725}}\,{q}^{9}
 \right]
 + {\mathcal O}\left(q^{10}\right),
 \nonumber \\
\lefteqn{
 \Sigma^{(2,0)}_{\mathrm{bare},S}
 = 
50-{\frac {1612}{9}}\,q-{\frac {7334}{9}}\,{q}^{2}-{\frac {53956}{45}}\,{q}^{3}+{\frac {145429}{15}}\,{q}^{4}
+{\frac {686425828}{7875}}\,{q}^{5}+{\frac {1196280364}{2625}}\,{q}^{6}
} & & \nonumber \\
 & &
+{\frac {445266609908}{231525}}\,{q}^{7}
+{\frac {8277353161391}{1157625}}\,{q}^{8}+{\frac {2785865671205906}{114604875}}\,{q}^{9}
 - 24 \left[ \ln\left(q\right) + 3 \ln\left(2\right) \right]^2
 \nonumber \\
 & &
 \times
 \left[
       1+{\frac {40}{3}}\,q+{\frac {152}{3}}\,{q}^{2}+{\frac {776}{9}}\,{q}^{3}-{\frac {2200}{9}}\,{q}^{4}-{\frac {129712}{45}}\,{q}^{5}
       -{\frac {708808}{45}}\,{q}^{6}-{\frac {4207040}{63}}\,{q}^{7}
 \right. \nonumber \\
 & & \left.
       -{\frac {77207944}{315}}\,{q}^{8}-{\frac {770133128}{945}}\,{q}^{9}
 \right]
 + 20 \left[ \ln\left(q\right) + 3 \ln\left(2\right) \right]
 \left[
       1+{\frac {436}{15}}\,q+{\frac {406}{3}}\,{q}^{2}+{\frac {4468}{15}}\,{q}^{3}
 \right. \nonumber \\
 & & \left.
       -{\frac {4477}{15}}\,{q}^{4}-{\frac {2331472}{375}}\,{q}^{5}
       -{\frac {4624926}{125}}\,{q}^{6}-{\frac {602336248}{3675}}\,{q}^{7}-{\frac {22819804141}{36750}}\,{q}^{8}
 \right. \nonumber \\
 & & \left.
       -{\frac {350717144888}{165375}}\,{q}^{9}
 \right]
 + 24 \left[ \zeta_3 -4 \zeta_2 \ln\left(2\right) \right] \left[
       \frac{1}{4}+q+5\,{q}^{2}+19\,{q}^{3}+61\,{q}^{4}+174\,{q}^{5}+455\,{q}^{6}
 \right. \nonumber \\
 & & \left.
       +1112\,{q}^{7}+2573\,{q}^{8}+5689\,{q}^{9}
 \right]
 - 40 \zeta_2 \left[
       -\frac{3}{4}+q+{\frac {52}{5}}\,{q}^{2}+{\frac {241}{5}}\,{q}^{3}+{\frac {1631}{10}}\,{q}^{4}+{\frac {11852}{25}}\,{q}^{5}
 \right. \nonumber \\
 & & \left.
       +{\frac {31133}{25}}\,{q}^{6}+{\frac {106306}{35}}\,{q}^{7}+{\frac {4911471}{700}}\,{q}^{8}+{\frac {8143138}{525}}\,{q}^{9}
 \right]
 + {\mathcal O}\left(q^{10}\right).
\eq
In the supplementary electronic file attached to this article we give the $q$-expansion
for $\Sigma^{(2,0)}_{\mathrm{bare},V}$ and $\Sigma^{(2,0)}_{\mathrm{bare},S}$ 
in an arbitrary covariant gauge up to $q^{\qorder}$.

\subsubsection{The cusp $p^2=9m^2$}

For the expansion around the cusp $x=9$ we set
\bq
\label{def_periods_choice_9}
 \left( \begin{array}{c}
  \psi_{2,9} \\
  \psi_{1,9} \\
 \end{array} \right)
 & = &
 \left( \begin{array}{rr}
 2 & -1 \\
 3 & -1 \\
 \end{array} \right)
 \left( \begin{array}{c}
  \psi_{2,0} \\
  \psi_{1,0} \\
 \end{array} \right).
\eq
We further set
\bq
\label{def_tau_9}
 \tau_{2,9}
 \;\; = \;\;
 \frac{1}{2}
 \frac{\psi_{2,9}}{\psi_{1,9}},
 & &
 q_{2,9} \;\; = \;\; e^{2 i \pi \tau_{2,9}}.
\eq
The values of $\tau_{2,9}$ at the points $x \in \{0,1,9,\infty\}$ are tabulated in table~\ref{table_tau}.
In order to simplify the notation we write in the remaining part of this section
\bq
 \tau \;\; = \;\; \tau_{2,9},
 & &
 q \;\; = \;\; q_{2,9},
\eq
and $e_1 = E_1\left(\tau_{2,9};\chi_0,\chi_1\right)$,
$e_2 = E_1\left(2\tau_{2,9};\chi_0,\chi_1\right)$.
We also use the notation $b_1$ for $b_{1,9}$ and $b_2$ for $b_{2,9}$.
The Hauptmodul is given by
\bq
\label{hauptmodul_9}
 x-9 & = &
 72 \frac{\eta\left(2\tau\right) \eta\left(6\tau\right)^5}{\eta\left(3\tau\right) \eta\left(\tau\right)^5}.
\eq
The $q$-expansions of $b_1$ and $b_2$ are given by
\bq
 b_1 
 \; = \;
 \sqrt{3} \left( e_1 - 2 e_2 \right),
 & &
 b_2
 \; = \;
 -12 \sqrt{3} e_2.
\eq
For the integration kernels we have
\bq
 g_{2,0} 
 & = &
 12 \left( e_1^2 - e_2^2 \right),
 \nonumber \\
 g_{2,1} 
 & = &
 9 \left( e_1^2 + e_1 e_2 - 2 e_2^2 \right),
 \nonumber \\
 g_{2,9} 
 & = &
 3 \left( e_1^2 + 3 e_1 e_2 + 2 e_2^2 \right),
 \nonumber \\
 g_{3,0} 
 & = &
 - 36 \sqrt{3} \left( e_1^3 + 2 e_1^2 e_2 - e_1 e_2^2 - 2 e_2^3 \right),
 \nonumber \\
 g_{3,1} 
 & = &
 - 27 \sqrt{3} \left( e_1^3 + 3 e_1^2 e_2 - 4 e_2^3 \right),
 \nonumber \\
 f_{4}
 & = &
 324 e_2^4.
\eq
The integral $I_{8,9}^{(3)}$ is finite at $x=9$. We denote its value at $x=9$ by
\bq
\label{defC_8_9_3}
 C_{8,9}^{(3)}
 & = &
 516\,\zeta_3
-576\,\mathrm{Li}_3\left(\frac{1}{3}\right)
+576\,\mathrm{Li}_{2 1}\left(\frac{1}{3},1\right)
-120\,\ln  \left( 2 \right) \zeta_2
+96\,\ln  \left( 3 \right) \zeta_2
-96\, \ln^3\left( 3 \right)
 \nonumber \\
 & &
+72\,\ln  \left( 2 \right)  \ln^2\left( 3 \right)
+144\,\ln  \left( 2 \right) \mathrm{Li}_2\left(\frac{1}{3}\right)
-576\,\ln  \left( 3 \right) \mathrm{Li}_2\left(\frac{1}{3}\right)
-72\,\pi \,\mathrm{Cl}_2\left(\frac{2\pi}{3}\right)
 \nonumber \\
 & &
-72\,i\pi \,\zeta_2
+72\,i\pi \, \ln^2\left( 3 \right)
-48\,i\pi \,\ln  \left( 2 \right) \ln  \left( 3 \right) 
+144\,i\pi \,\mathrm{Li}_2\left(\frac{1}{3}\right).
\eq
We obtain for the first non-vanishing order of the integrals $I_{6,9}$, $I_{7,9}$ and $I_{8,9}$
\bq
 I_{6,9}^{(2)}
 & = &
 15 \, \mathrm{Cl}_2\left(\frac{2\pi}{3}\right) -12 i \zeta_2 
 + 2 \pi \, \iterintmodular{1}{q}
 - 2  \, \iterintmodular{1,g_{3,0}}{q}
 \nonumber \\
 & = &
 15 \, \mathrm{Cl}_2\left(\frac{2\pi}{3}\right) -12 i \zeta_2 
 + 2 \pi \ln\left(q\right)
 + 12 \sqrt{3} \left[
                     q+{q}^{2}+{q}^{3}+{q}^{4}+{\frac {24}{25}}\,{q}^{5}+{q}^{6}+{\frac {50}{49}}\,{q}^{7}
 \right. \nonumber \\
 & & \left.
                     +{q}^{8}+{q}^{9}
 \right]
 + {\mathcal O}\left(q^{10}\right), 
 \nonumber \\
 I_{7,9}^{(1)}
 & = &
 \pi - \iterintmodular{g_{3,0}}{q}
 \nonumber \\
 & = &
 \pi
 + 6 \sqrt{3} \left[
  q+2\,{q}^{2}+3\,{q}^{3}+4\,{q}^{4}+{\frac {24}{5}}\,{q}^{5}+6\,{q}^{6}+{\frac {50}{7}}\,{q}^{7}+8\,{q}^{8}+9\,{q}^{9}
 \right]
 + {\mathcal O}\left(q^{10}\right), 
 \nonumber \\
 I_{8,9}^{(3)}
 & = &
% depth 3
64\,\iterintmodular{ g_{2,1}, g_{2,0}, g_{2,1} }{q}
-128\,\iterintmodular{ g_{2,0}, g_{2,1}, g_{2,1} }{q}
+48\,\iterintmodular{ g_{3,0}, 1, g_{3,0} }{q}
 \nonumber \\
 & &
-{\frac {128}{3}}\,\iterintmodular{ g_{3,1}, 1, g_{3,0} }{q}
% depth 2
+96\,\ln  \left( 2 \right) \iterintmodular{ g_{2,1}, g_{2,0} }{q}
-192\,\ln  \left( 2 \right) \iterintmodular{ g_{2,0}, g_{2,1} }{q}
 \nonumber \\
 & &
-32\,i\pi \,\iterintmodular{ g_{2,1}, g_{2,0} }{q}
+64\,i\pi \,\iterintmodular{ g_{2,0}, g_{2,1} }{q}
-48\,\pi \,\iterintmodular{ g_{3,0}, 1 }{q}
+{\frac {128}{3}}\,\pi \,\iterintmodular{ g_{3,1}, 1 }{q}
 \nonumber \\
 & &
% depth 1
+96\,\zeta_2\,\iterintmodular{ g_{2,0} }{q}
-80\,\zeta_2\,\iterintmodular{ g_{2,1} }{q}
-144\, \ln^2\left( 2 \right)  \iterintmodular{ g_{2,0} }{q}
+48\, \ln^2\left( 3 \right)  \iterintmodular{ g_{2,1} }{q}
 \nonumber \\
 & &
+96\,\mathrm{Li}_2\left(\frac{1}{3}\right)\,\iterintmodular{ g_{2,1} }{q}
-360\,\mathrm{Cl}_2\left(\frac{2\pi}{3}\right)\,\iterintmodular{ g_{3,0} }{q}
+320\,\mathrm{Cl}_2\left(\frac{2\pi}{3}\right)\,\iterintmodular{ g_{3,1} }{q}
 \nonumber \\
 & &
+288\,i\zeta_2\,\iterintmodular{ g_{3,0} }{q}
-256\,i\zeta_2\, \iterintmodular{ g_{3,1} }{q}
+96\,i\pi \,\ln  \left( 2 \right) \iterintmodular{ g_{2,0} }{q}
-32\,i\pi \,\ln  \left( 3 \right) \iterintmodular{ g_{2,1} }{q} 
 \nonumber \\
 & &
% depth 0
+ C_{8,9}^{(3)}
 \nonumber \\
 & = &
 C_{8,9}^{(3)}
+864\,{q}^{3}-864\,{q}^{4}+{\frac {9504}{5}}\,{q}^{5}-{\frac {8208}{5}}\,{q}^{6}
+{\frac {424656}{175}}\,{q}^{7}-{\frac {15444}{35}}\,{q}^{8}+{\frac {2791188}{1225}}\,{q}^{9}
 \nonumber \\
 & &
-288\left[ 3\ln\left(2\right)-i\pi\right]
 \left[
       {q}^{2}+2\,{q}^{3}+\frac{5}{6}\,{q}^{4}+{\frac {49}{15}}\,{q}^{5}+{\frac {56}{15}}\,{q}^{6}+{\frac {313}{105}}\,{q}^{7}
       +{\frac {2059}{420}}\,{q}^{8}+{\frac {1217}{210}}\,{q}^{9}
 \right]
 \nonumber \\
 & &
 - 72 \sqrt{3} \pi \left[
{q}^{2}+\frac{3}{4}\,{q}^{4}+{q}^{6}+{\frac {13}{16}}\,{q}^{8}
 \right]
 + 144 \sqrt{3} \left[ \frac{15}{2} \mathrm{Cl}_2\left(\frac{2\pi}{3}\right) + \pi \ln\left(q\right) - 6 \zeta_2 \right] \left[
{q}^{2}
 \right. \nonumber \\
 & & \left.
+\frac{3}{2}\,{q}^{4}
+3\,{q}^{6}+{\frac {13}{4}}\,{q}^{8}
 \right]
 + 144 \left[ 3 \, \mathrm{Li}_2\left(\frac{1}{3}\right) + \frac{3}{2} \ln^2\left(3\right) - i \pi \ln\left(3\right) \right] \left[
q+\frac{1}{2}\,{q}^{2}+{q}^{3}+\frac{1}{4}\,{q}^{4}
 \right. \nonumber \\
 & & \left.
+\frac{6}{5}\,{q}^{5}
+\frac{1}{2}\,{q}^{6}+{\frac {8}{7}}\,{q}^{7}
+\frac{1}{8}\,{q}^{8}+{q}^{9}
 \right]
 - 192 \left[ 3 \ln^2\left(2\right) - 2 i \pi \ln\left(2\right) \right] \left[
q+{q}^{2}+\frac{1}{3}\,{q}^{3}+{q}^{4}
 \right. \nonumber \\
 & & \left.
+\frac{6}{5}\,{q}^{5}
+\frac{1}{3}\,{q}^{6}+{\frac {8}{7}}\,{q}^{7}+{q}^{8}+\frac{1}{9}\,{q}^{9}
 \right]
 + 24 \zeta_2 \left[
q+\frac{17}{2}\,{q}^{2}-{\frac {29}{3}}\,{q}^{3}+{\frac {49}{4}}\,{q}^{4}+\frac{6}{5}\,{q}^{5}-{\frac {13}{6}}\,{q}^{6}
 \right. \nonumber \\
 & & \left.
+{\frac {8}{7}}\,{q}^{7}
+{\frac {113}{8}}\,{q}^{8}-{\frac {119}{9}}\,{q}^{9}
 \right]
 + {\mathcal O}\left(q^{10}\right).
\eq
In the supplementary electronic file attached to this article we give the $q$-expansion
for the master integrals $I_{i,9}$ up to $q^{\qorder}$ for the first four orders in $\eps$.
The $q$-series for the master integrals $I_{i,9}$ converge for all values 
$x \in \left( {\mathbb R} \cup \left\{\infty\right\} \right) \backslash \{ 0,1,\infty \}$.
With
\bq
 \psi_{1,9}
 \;\; = \;\;
 \sqrt{3} \pi \left( e_1 - 2 e_2 \right),
 & &
 \frac{d}{dx} \psi_{1,9}
 \;\; = \;\;
 \frac{1}{2} \cdot
 \frac{2 \pi i \, W}{\psi_{1,9}^2} \; q \frac{d}{dq} \psi_{1,9},
\eq
and eq.~(\ref{hauptmodul_9}) we may express
$\Sigma^{(2,0)}_{\mathrm{bare},V}$ and $\Sigma^{(2,0)}_{\mathrm{bare},S}$
as $q$-series.
We obtain in Feynman gauge
\bq
\label{result_q_series_case_9}
\lefteqn{
 \Sigma^{(2,0)}_{\mathrm{bare},V}
 = 
 -{\frac {3533}{1080}}+{\frac {7984}{135}}\,q-{\frac {56998}{135}}\,{q}^{2}+{\frac {112964}{135}}\,{q}^{3}
 -{\frac {15482}{45}}\,{q}^{4}+{\frac {127198}{135}}\,{q}^{5}-{\frac {2036869}{135}}\,{q}^{6}
} & & \nonumber \\
 & &
 +{\frac {17742626}{175}}\,{q}^{7}
 -{\frac {120855851}{420}}\,{q}^{8}+{\frac {63922451764}{99225}}\,{q}^{9}
 - \frac{8}{81} C_{8,9}^{(3)} \left[
         -\frac{1}{16}+q-7\,{q}^{2}+27\,{q}^{3}
 \right. \nonumber \\
 & & \left.
         -55\,{q}^{4}+14\,{q}^{5}+243\,{q}^{6}-648\,{q}^{7}+425\,{q}^{8}+1593\,{q}^{9}
 \right]
 + 24 \sqrt{3} \left[ \frac{15}{2} \mathrm{Cl}_2\left(\frac{2\pi}{3}\right) + \pi \ln\left(q\right) 
 \right. \nonumber \\
 & & \left.
                      -6 \zeta_2 i \right] \left[
         {q}^{2}+{\frac {29}{2}}\,{q}^{4}-40\,{q}^{5}+219\,{q}^{6}-392\,{q}^{7}+{\frac {4735}{4}}\,{q}^{8}-1580\,{q}^{9}
 \right]
 - \frac{44}{27} \left\{ 2 \, \mathrm{Li}_2\left(\frac{1}{3}\right) 
 \right. \nonumber \\
 & & \left.
                         + \left[ \ln\left(3\right) - \frac{i \pi}{3} \right]^2 \right\} \left[
         -{\frac {185}{44}}+q+{\frac {575}{22}}\,{q}^{2}-{\frac {2133}{11}}\,{q}^{3}+{\frac {2453}{4}}\,{q}^{4}-{\frac {44914}{55}}\,{q}^{5}
         -{\frac {96957}{110}}\,{q}^{6}
 \right. \nonumber \\
 & & \left.
         +{\frac {2216484}{385}}\,{q}^{7}-{\frac {27707387}{3080}}\,{q}^{8}-{\frac {37395}{11}}\,{q}^{9}
 \right]
 - \frac{304}{27} \left[ \ln\left(2\right) - \frac{i \pi}{3} \right]^2 \left[
         {\frac {577}{228}}+q-{\frac {389}{19}}\,{q}^{2}
 \right. \nonumber \\
 & & \left.
         +{\frac {337}{3}}\,{q}^{3}-{\frac {5397}{19}}\,{q}^{4}+{\frac {15666}{95}}\,{q}^{5}
         +{\frac {107309}{95}}\,{q}^{6}-{\frac {2329416}{665}}\,{q}^{7}+{\frac {1762561}{665}}\,{q}^{8}+{\frac {507569}{57}}\,{q}^{9}
 \right]
 \nonumber \\
 & &
 + \frac{340}{81} \zeta_2 \left[
         -{\frac {211}{136}}+q+{\frac {997}{170}}\,{q}^{2}-{\frac {999}{17}}\,{q}^{3}+{\frac {69509}{340}}\,{q}^{4}
         -{\frac {131102}{425}}\,{q}^{5}-{\frac {164511}{850}}\,{q}^{6}
 \right. \nonumber \\
 & & \left.
         +{\frac {5470092}{2975}}\,{q}^{7}-{\frac {76934161}{23800}}\,{q}^{8}
         -{\frac {29349}{85}}\,{q}^{9}
 \right]
 - \frac{60584}{405} \left[ \ln\left(2\right) - \frac{i \pi}{3} \right] \left[
         -{\frac {2159}{15146}}+q
 \right. \nonumber \\
 & & \left.
         +{\frac {28063}{15146}}\,{q}^{2}-{\frac {22717}{7573}}\,{q}^{3}+{\frac {1344627}{30292}}\,{q}^{4}
         -{\frac {227297}{7573}}\,{q}^{5}+{\frac {1281937}{15146}}\,{q}^{6}+{\frac {29379321}{53011}}\,{q}^{7}
 \right. \nonumber \\
 & & \left.
         -{\frac {111424515}{424088}}\,{q}^{8}+{\frac {396722701}{318066}}\,{q}^{9}
 \right]
 + \frac{592}{135} \sqrt{3} \pi \left[
         -{\frac {97}{148}}+q+{\frac {416}{37}}\,{q}^{2}-{\frac {918}{37}}\,{q}^{3}+{\frac {68987}{592}}\,{q}^{4}
 \right. \nonumber \\
 & & \left.
         -{\frac {5746}{37}}\,{q}^{5}
         +{\frac {123633}{148}}\,{q}^{6}-{\frac {60021}{37}}\,{q}^{7}+{\frac {14436605}{2368}}\,{q}^{8}-{\frac {1421793}{148}}\,{q}^{9}
 \right]
 + {\mathcal O}\left(q^{10}\right),
 \nonumber \\
\lefteqn{
 \Sigma^{(2,0)}_{\mathrm{bare},S}
 = 
 {\frac {134}{3}}-{\frac {820}{3}}\,q+466\,{q}^{2}+1820\,{q}^{3}+{\frac {9085}{3}}\,{q}^{4}-{\frac {81876}{5}}\,{q}^{5}
 +{\frac {351332}{5}}\,{q}^{6}-{\frac {28705748}{175}}\,{q}^{7}
} & & \nonumber \\
 & &
 +{\frac {62966527}{175}}\,{q}^{8}-{\frac {925309578}{1225}}\,{q}^{9}
 - \frac{2}{9} C_{8,9}^{(3)} \left[
         -\frac{5}{4}+q-3\,{q}^{2}+3\,{q}^{3}+5\,{q}^{4}-18\,{q}^{5}+15\,{q}^{6}
 \right. \nonumber \\
 & & \left.
         +24\,{q}^{7}-75\,{q}^{8}+57\,{q}^{9}
 \right]
 - 72 \sqrt{3} \left[ \frac{15}{2} \mathrm{Cl}_2\left(\frac{2\pi}{3}\right) + \pi \ln\left(q\right) -6 \zeta_2 i \right] \left[
         {q}^{2}-4\,{q}^{3}+\frac{21}{2}\,{q}^{4}-30\,{q}^{5}
 \right. \nonumber \\
 & & \left.
         +85\,{q}^{6}-234\,{q}^{7}+{\frac {2311}{4}}\,{q}^{8}-1315\,{q}^{9}
 \right]
 - 36 \left\{ 2 \, \mathrm{Li}_2\left(\frac{1}{3}\right) + \left[ \ln\left(3\right) - \frac{i \pi}{3} \right]^2 \right\} \left[
        1+q-\frac{15}{2}\,{q}^{2}
 \right. \nonumber \\
 & & \left.
        +3\,{q}^{3}+{\frac {65}{4}}\,{q}^{4}-45\,{q}^{5}+{\frac {231}{10}}\,{q}^{6}+{\frac {2577}{35}}\,{q}^{7}
        -{\frac {49341}{280}}\,{q}^{8}+{\frac {1203}{14}}\,{q}^{9}
 \right]
 + \frac{832}{3} \left[ \ln\left(2\right) - \frac{i \pi}{3} \right]^2 
 \nonumber \\
 & &
 \times
 \left[
        {\frac {67}{104}}+q-{\frac {61}{13}}\,{q}^{2}+{\frac {35}{13}}\,{q}^{3}+{\frac {127}{13}}\,{q}^{4}-{\frac {346}{13}}\,{q}^{5}
        +{\frac {541}{65}}\,{q}^{6}+{\frac {28344}{455}}\,{q}^{7}-{\frac {54059}{455}}\,{q}^{8}+{\frac {905}{273}}\,{q}^{9}
 \right]
 \nonumber \\
 & &
 + 4 \zeta_2 \left[
        \frac{17}{2}+q-{\frac {87}{2}}\,{q}^{2}+3\,{q}^{3}+{\frac {425}{4}}\,{q}^{4}-261\,{q}^{5}+{\frac {879}{10}}\,{q}^{6}
        +{\frac {16473}{35}}\,{q}^{7}-{\frac {276069}{280}}\,{q}^{8}
 \right. \nonumber \\
 & & \left.
        +{\frac {4443}{14}}\,{q}^{9}
 \right]
 - \frac{1540}{9} \left[ \ln\left(2\right) - \frac{i \pi}{3} \right] \left[
        1-{\frac {1244}{385}}\,q-{\frac {822}{55}}\,{q}^{2}+{\frac {2148}{77}}\,{q}^{3}-{\frac {7489}{385}}\,{q}^{4}
        -{\frac {122904}{1925}}\,{q}^{5}
 \right. \nonumber \\
 & & \left.
        +{\frac {273198}{1925}}\,{q}^{6}-{\frac {353328}{13475}}\,{q}^{7}
        -{\frac {9629211}{26950}}\,{q}^{8}+{\frac {8197908}{13475}}\,{q}^{9}
 \right]
 - \frac{64}{3} \sqrt{3} \pi \left[
        -{\frac {11}{16}}+q+{\frac {15}{4}}\,{q}^{2}
 \right. \nonumber \\
 & & \left.
        -{\frac {69}{4}}\,{q}^{3}+{\frac {3317}{64}}\,{q}^{4}-{\frac {1989}{16}}\,{q}^{5}
        +{\frac {2361}{8}}\,{q}^{6}-{\frac {11199}{16}}\,{q}^{7}+{\frac {413043}{256}}\,{q}^{8}-{\frac {224367}{64}}\,{q}^{9}
 \right]
 + {\mathcal O}\left(q^{10}\right).
 \nonumber \\
\eq
In the supplementary electronic file attached to this article we give the $q$-expansion
for $\Sigma^{(2,0)}_{\mathrm{bare},V}$ and $\Sigma^{(2,0)}_{\mathrm{bare},S}$ 
in an arbitrary covariant gauge up to $q^{\qorder}$.

\subsubsection{The cusp $p^2=\infty$}

For the expansion around the cusp $x=\infty$ we set
\bq
\label{def_periods_choice_infty}
 \left( \begin{array}{c}
  \psi_{2,\infty} \\
  \psi_{1,\infty} \\
 \end{array} \right)
 & = &
 \left( \begin{array}{rr}
 3 & -1 \\
 -2 & 1 \\
 \end{array} \right)
 \left( \begin{array}{c}
  \psi_{2,0} \\
  \psi_{1,0} \\
 \end{array} \right).
\eq
We further set
\bq
\label{def_tau_infty}
 \tau_{3,\infty}
 \;\; = \;\;
 \frac{1}{3}
 \frac{\psi_{2,\infty}}{\psi_{1,\infty}},
 & &
 q_{3,\infty} \;\; = \;\; e^{2 i \pi \tau_{3,\infty}}.
\eq
The values of $\tau_{3,\infty}$ at the points $x \in \{0,1,9,\infty\}$ are tabulated in table~\ref{table_tau}.
In order to simplify the notation we write in the remaining part of this section
\bq
 \tau \;\; = \;\; \tau_{3,\infty},
 & &
 q \;\; = \;\; q_{3,\infty}.
\eq
and $e_1 = E_1\left(\tau_{3,\infty};\chi_0,\chi_1\right)$,
$e_2 = E_1\left(2\tau_{3,\infty};\chi_0,\chi_1\right)$.
We also use the notation $b_1$ for $b_{1,\infty}$ and $b_2$ for $b_{2,\infty}$.
The Hauptmodul is given by
\bq
\label{hauptmodul_inf}
 \frac{1}{x} & = &
 \frac{\eta\left(\tau\right)^4 \eta\left(6\tau\right)^8}{\eta\left(3\tau\right)^4 \eta\left(2\tau\right)^8}.
\eq
The $q$-expansions of $b_1$ and $b_2$ are given by
\bq
 b_1 
 \; = \;
 2 i \left( e_1 - e_2 \right),
 & &
 b_2
 \; = \;
 12 i e_1.
\eq
For the integration kernels we have
\bq
 g_{2,0} 
 & = &
 4 \left( e_1^2 - 4 e_2^2 \right),
 \nonumber \\
 g_{2,1} 
 & = &
 6 \left( e_1^2 - e_1 e_2 - 2 e_2^2 \right),
 \nonumber \\
 g_{2,9} 
 & = &
 - 2 \left( e_1^2 + 3 e_1 e_3 + 2 e_2^2 \right),
 \nonumber \\
 g_{3,0} 
 & = &
 24 i \left( e_1^3 + e_1^2 e_2 - 4 e_1 e_2^2 - 4 e_2^3 \right),
 \nonumber \\
 g_{3,1} 
 & = &
 36 i \left( e_1^3 - 3 e_1 e_2^2 - 2 e_2^3 \right),
 \nonumber \\
 f_{4}
 & = &
 36 e_1^4.
\eq
We obtain for the first non-vanishing order of the integrals $I_{6,\infty}$, $I_{7,\infty}$ and $I_{8,\infty}$
\bq
 I_{6,\infty}^{(2)}
 & = &
 - 9 i \zeta_2 
 - 3 \pi \, \iterintmodular{1}{q}
 - \frac{9}{2}  \iterintmodular{1,g_{3,0}}{q}
 \nonumber \\
 & = &
 - 9 i \zeta_2 
 - 3 \pi  \ln\left(q\right)
 + \frac{3}{2} i \ln^2\left(q\right)
 - 3 i \left[
             q+\frac{5}{4}\,{q}^{2}+\frac{1}{9}\,{q}^{3}-{\frac {11}{16}}\,{q}^{4}-{\frac {24}{25}}\,{q}^{5}+{\frac {5}{36}}\,{q}^{6}
             +{\frac {50}{49}}\,{q}^{7}
 \right. \nonumber \\
 & & \left.
             +{\frac {53}{64}}\,{q}^{8}+{\frac {1}{81}}\,{q}^{9}
 \right]
 + {\mathcal O}\left(q^{10}\right), 
 \nonumber \\
 I_{7,\infty}^{(1)}
 & = &
 - \pi - \frac{3}{2} \iterintmodular{g_{3,0}}{q}
 \nonumber \\
 & = &
 - \pi
 + i \ln\left(q\right)
 - i \left[
           q+\frac{5}{2}\,{q}^{2}+\frac{1}{3}\,{q}^{3}-\frac{11}{4}\,{q}^{4}-{\frac {24}{5}}\,{q}^{5}+\frac{5}{6}\,{q}^{6}
           +{\frac {50}{7}}\,{q}^{7}
           +{\frac {53}{8}}\,{q}^{8}+\frac{1}{9}\,{q}^{9}
 \right]
 \nonumber \\
 & &
 + {\mathcal O}\left(q^{10}\right), 
 \nonumber \\
 I_{8,\infty}^{(3)}
 & = &
% depth 3
216\,\iterintmodular{ g_{2,1}, g_{2,0}, g_{2,1} }{q}
-432\,\iterintmodular{ g_{2,0}, g_{2,1}, g_{2,1} }{q}
+162\,\iterintmodular{ g_{3,0}, 1, g_{3,0} }{q}
 \nonumber \\
 & &
-144\,\iterintmodular{ g_{3,1}, 1, g_{3,0} }{q}
% depth 2
-72\,i\pi \,\iterintmodular{ g_{2,1}, g_{2,0} }{q}
+144\,i\pi \,\iterintmodular{ g_{2,0}, g_{2,1} }{q}
% depth 1
 \nonumber \\
 & &
-72\,\zeta_2\,\iterintmodular{ g_{2,1} }{q}
+144\,\zeta_2\,\iterintmodular{ g_{2,0} }{q}
+108\,\pi \,\iterintmodular{ g_{3,0}, 1 }{q}
-96\,\pi \,\iterintmodular{ g_{3,1}, 1 }{q}
 \nonumber \\
 & &
+324\,i\zeta_2\,\iterintmodular{ g_{3,0} }{q}
-288\,i\zeta_2\,\iterintmodular{ g_{3,1} }{q}
% depth 0
+48\,\zeta_3
 \nonumber \\
 & = &
 48\,\zeta_3
+48\,q-30\,{q}^{2}+{\frac {124}{9}}\,{q}^{3}+{\frac {35}{12}}\,{q}^{4}-{\frac {47731}{375}}\,{q}^{5}
+{\frac {100861}{450}}\,{q}^{6}+{\frac {371366}{1715}}\,{q}^{7}
 \nonumber \\
 & &
-{\frac {79243781}{117600}}\,{q}^{8}
-{\frac {40995539}{1190700}}\,{q}^{9}
- 48 \left[ \ln\left(q\right) + i \pi \right] \left[
q-\frac{5}{4}\,{q}^{2}+\frac{1}{9}\,{q}^{3}+{\frac {25}{48}}\,{q}^{4}+{\frac {77}{50}}\,{q}^{5}
 \right. \nonumber \\
 & & \left.
-{\frac {157}{180}}\,{q}^{6}
-{\frac {5213}{1470}}\,{q}^{7}+{\frac {7243}{6720}}\,{q}^{8}+{\frac {37319}{11340}}\,{q}^{9}
\right]
+ 16 \left[ \ln\left(q\right) + i \pi \right]^2 \left[
q-2\,{q}^{2}-\frac{7}{6}\,{q}^{3}
 \right. \nonumber \\
 & & \left.
+10\,{q}^{4}
-{\frac {123}{10}}\,{q}^{5}-{\frac {17}{12}}\,{q}^{6}
+{\frac {205}{14}}\,{q}^{7}-11\,{q}^{8}-{\frac {17}{9}}\,{q}^{9}
\right]
 + {\mathcal O}\left(q^{10}\right).
\eq
In the supplementary electronic file attached to this article we give the $q$-expansion
for the master integrals $I_{i,\infty}$ up to $q^{\qorder}$ for the first four orders in $\eps$.
The $q$-series for the master integrals $I_{i,\infty}$ converge for all values 
$x \in \left( {\mathbb R} \cup \left\{\infty\right\} \right) \backslash \{ 0,1,9 \}$.
With
\bq
 \psi_{1,\infty}
 \;\; = \;\;
 2 i \pi \left( e_1 - e_2 \right),
 & &
 \frac{d}{dx} \psi_{1,\infty}
 \;\; = \;\;
 \frac{1}{3} \cdot
 \frac{2 \pi i \, W}{\psi_{1,\infty}^2} \; q \frac{d}{dq} \psi_{1,\infty},
\eq
and eq.~(\ref{hauptmodul_inf}) we may express
$\Sigma^{(2,0)}_{\mathrm{bare},V}$ and $\Sigma^{(2,0)}_{\mathrm{bare},S}$
as $q$-series.
We obtain in Feynman gauge
\bq
\label{result_q_series_case_inf}
\lefteqn{
 \Sigma^{(2,0)}_{\mathrm{bare},V}
 = 
 {\frac {17}{8}}-36\,q+{\frac {7375}{36}}\,{q}^{2}-{\frac {5489}{9}}\,{q}^{3}+{\frac {267081}{200}}\,{q}^{4}
 -{\frac {731467}{225}}\,{q}^{5}+{\frac {230925589}{25200}}\,{q}^{6}
} & & \nonumber \\
 & &
 -{\frac {40999299}{1750}}\,{q}^{7}
 +{\frac {9609117331}{189000}}\,{q}^{8}-{\frac {113455961056}{1157625}}\,{q}^{9}
 - \left[ \ln\left(q\right) + \pi i \right]^2 \left[
         1-{\frac {25}{2}}\,{q}^{2}+96\,{q}^{3}
 \right. \nonumber \\
 & & \left.
         -460\,{q}^{4}+{\frac {5620}{3}}\,{q}^{5}-{\frac {20675}{3}}\,{q}^{6}
         +{\frac {111072}{5}}\,{q}^{7}-{\frac {930868}{15}}\,{q}^{8}+{\frac {5408664}{35}}\,{q}^{9}
 \right]
 + \frac{3}{2} \left[ \ln\left(q\right) + \pi i \right] 
 \nonumber \\
 & &
 \times
 \left[
         1-{\frac {64}{3}}\,q+{\frac {1045}{9}}\,{q}^{2}-{\frac {1256}{3}}\,{q}^{3}+{\frac {64133}{45}}\,{q}^{4}
         -{\frac {69844}{15}}\,{q}^{5}+13816\,{q}^{6}-{\frac {58252108}{1575}}\,{q}^{7}
 \right. \nonumber \\
 & & \left.
         +{\frac {284713253}{3150}}\,{q}^{8}
         -{\frac {6779006338}{33075}}\,{q}^{9}
 \right]
 + 24 \zeta_3 \left[
        {q}^{2}-8\,{q}^{3}+36\,{q}^{4}-120\,{q}^{5}+338\,{q}^{6}-864\,{q}^{7}
 \right. \nonumber \\
 & & \left.
        +2068\,{q}^{8}-4688\,{q}^{9}
 \right]
 + \frac{1}{2} \zeta_2 \left[
       1-26\,{q}^{2}+208\,{q}^{3}-936\,{q}^{4}+3120\,{q}^{5}-8788\,{q}^{6}+22464\,{q}^{7}
 \right. \nonumber \\
 & & \left.
       -53768\,{q}^{8}+121888\,{q}^{9}
 \right]
 + {\mathcal O}\left(q^{10}\right),
 \nonumber \\
\lefteqn{
 \Sigma^{(2,0)}_{\mathrm{bare},S}
 = 
 42-136\,q+{\frac {563}{2}}\,{q}^{2}-{\frac {14051}{18}}\,{q}^{3}+{\frac {115439}{48}}\,{q}^{4}
 -{\frac {9669803}{2000}}\,{q}^{5}+{\frac {56927149}{9000}}\,{q}^{6}
} & & \nonumber \\
 & &
 -{\frac {1191806041}{171500}}\,{q}^{7}
 +{\frac {2024048811}{219520}}\,{q}^{8}-{\frac {5123197972363}{266716800}}\,{q}^{9}
 + 12 \left[ \ln\left(q\right) + \pi i \right]^2 \left[
        1-\frac{5}{2}\,q
 \right. \nonumber \\
 & & \left.
        +\frac{19}{2}\,{q}^{2}-{\frac {463}{18}}\,{q}^{3}+{\frac {1271}{18}}\,{q}^{4}-{\frac {8462}{45}}\,{q}^{5}
        +{\frac {69637}{180}}\,{q}^{6}-{\frac {735941}{1260}}\,{q}^{7}+{\frac {53987}{63}}\,{q}^{8}
        -{\frac {395291}{189}}\,{q}^{9}
 \right]
 \nonumber \\
 & &
 + 32 \left[ \ln\left(q\right) + \pi i \right] \left[
        1-{\frac {19}{8}}\,q+{\frac {181}{32}}\,{q}^{2}-{\frac {2015}{96}}\,{q}^{3}+{\frac {20233}{384}}\,{q}^{4}
        -{\frac {921139}{9600}}\,{q}^{5}+{\frac {366829}{2400}}\,{q}^{6}
 \right. \nonumber \\
 & & \left.
        -{\frac {15050761}{78400}}\,{q}^{7}
        +{\frac {64909121}{376320}}\,{q}^{8}-{\frac {1313870387}{3386880}}\,{q}^{9}
 \right]
 + 12 \zeta_3 \left[
        1+q-4\,{q}^{2}+10\,{q}^{3}-20\,{q}^{4}
 \right. \nonumber \\
 & & \left.
        +39\,{q}^{5}-76\,{q}^{6}+140\,{q}^{7}-244\,{q}^{8}+415\,{q}^{9}
 \right]
 - 6 \zeta_2 \left[
        1-6\,q+24\,{q}^{2}-60\,{q}^{3}+120\,{q}^{4}-234\,{q}^{5}
 \right. \nonumber \\
 & & \left.
        +456\,{q}^{6}-840\,{q}^{7}+1464\,{q}^{8}-2490\,{q}^{9}
 \right]
 + {\mathcal O}\left(q^{10}\right).
\eq
In the supplementary electronic file attached to this article we give the $q$-expansion
for $\Sigma^{(2,0)}_{\mathrm{bare},V}$ and $\Sigma^{(2,0)}_{\mathrm{bare},S}$ 
in an arbitrary covariant gauge up to $q^{\qorder}$.

% -----------------------------------------------------------------------------

\section{Numerical results}
\label{sect:numerical}

With the four expansions around the cusps $j \in \{0,1,9,\infty\}$ at hand, 
we now address the question, which expansion to use for a given $x$.
The four expansion parameters are
\bq
 q_{1,0}, 
 \;\;\;
 q_{6,1}, 
 \;\;\;
 q_{2,9}, 
 \;\;\;
 q_{3,\infty}.
\eq
For the absolute values of these we always have
\bq
 \left| q_{n_j,j} \right| & \le & 1,
\eq
where the value $1$ is only attained for $| q_{n_j,j} |$ at the three points $S_j=\{0,1,9,\infty\}\backslash\{j\}$.
For a fast convergence we would like to choose $j$ such that $| q_{n_j,j} |$ has a small absolute value. 
\begin{figure}
\begin{center}
\includegraphics[scale=1.0]{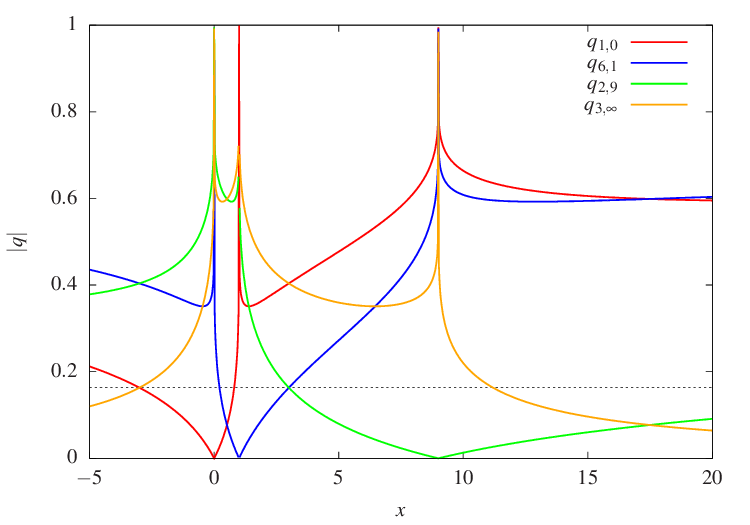}
\end{center}
\caption{\label{fig_abs_q}
The absolute values of the variables $q_{n_j,j}$ for $j \in \{0,1,9,\infty\}$ as a function of $x$.
There is always a choice such that $|q_{n_j,j}| \lessapprox 0.163$.
This value is indicated by the dashed black line.
}
\end{figure}
In fig.~\ref{fig_abs_q} we plot the absolute values of $q_{n_j,j}$ for the four choices $j \in \{0,1,9,\infty\}$.
An appropriate choice is given by
\bq
 q_{1,0} & : & -3 \; < \; x \; \lessapprox \; 0.5147 
 \nonumber \\
 q_{6,1} & : & 0.5147 \; \lessapprox \; x \; < \; 3
 \nonumber \\
 q_{2,9} & : & 3 \; < \; x \; \lessapprox \; 17.4853
 \nonumber \\
 q_{3,\infty} & : & x \; < \; -3 \;\;\; \mbox{or} \;\;\; 17.4853 \; \lessapprox \; x.
\eq
We denote the chosen variable simply by $q$.
In this way we can ensure that for all $x\in {\mathbb R}$
\bq
 \left| q \right|
  & \lessapprox &
 0.163.
\eq
The value $|q|\approx 0.163$ is indicated by a dashed line in fig.~\ref{fig_abs_q}.

Let us now discuss the precision, which can be reached by truncating the $q$-series to a certain order ${\mathcal O}(q^N)$.
For a quantity $O$ and an approximation $O_{\mathrm{approx}}$ to this quantity we define the relative precision $\delta$ of the approximation by
\bq
 \delta
 & = &
 \frac{\left| O_{\mathrm{approx}} - O \right|}{\left|O\right|}.
\eq
We consider the $\eps^0$-terms of the bare two-loop self-energy in Feynman gauge, 
i.e. the terms $\Sigma^{(2,0)}_{\mathrm{bare},V}$ and $\Sigma^{(2,0)}_{\mathrm{bare},S}$.
We may express these two quantities as linear combinations of iterated integrals of modular forms with coefficients, which
are rational functions in $x$, $\psi_1/\pi$ and $1/\pi \cdot d\psi_1/dx$.
There are now two possibilities to compute numerically the values of $\Sigma^{(2,0)}_{\mathrm{bare},V}$ and $\Sigma^{(2,0)}_{\mathrm{bare},S}$.
Within the first possibility (method A) we compute the coefficients from the known values of $x$
$\psi_1/\pi$ and $1/\pi \cdot d\psi_1/dx$, whereas we approximate the iterated integrals by their $q$-series, truncated to order 
${\mathcal O}(q^N)$.
Within the second possibility (method B) we compute $\Sigma^{(2,0)}_{\mathrm{bare},V}$ and $\Sigma^{(2,0)}_{\mathrm{bare},S}$
from their $q$-series, truncated to order ${\mathcal O}(q^N)$, i.e. we use
the analogue of eqs.~(\ref{result_q_series_case_0}), (\ref{result_q_series_case_1}), (\ref{result_q_series_case_9}) or (\ref{result_q_series_case_inf})
to the appropriate order ${\mathcal O}(q^N)$.

Let us consider a truncation of the $q$-series to order ${\mathcal O}(q^{30})$.
\begin{figure}
\begin{center}
\includegraphics[scale=0.6]{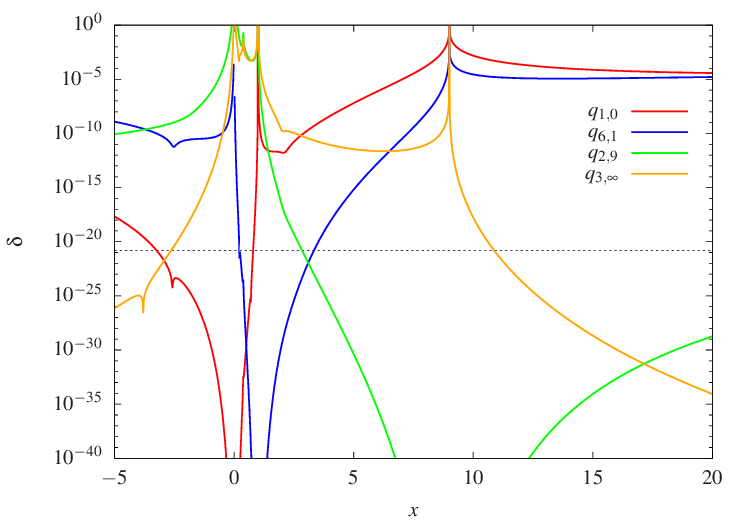}
\includegraphics[scale=0.6]{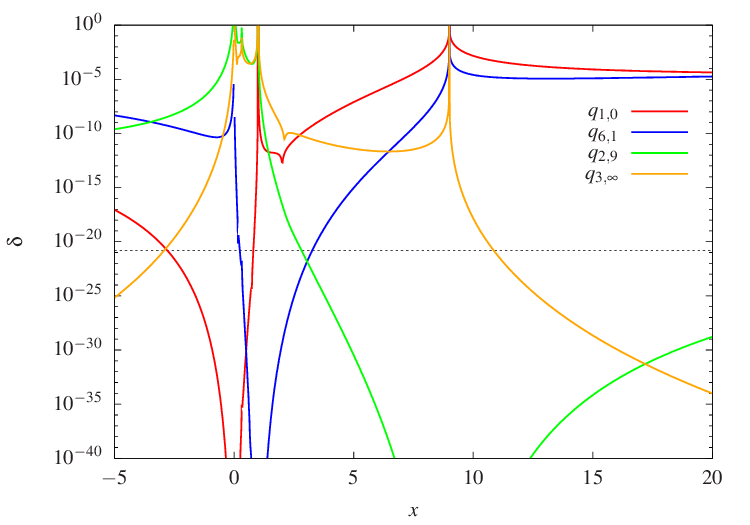}
\end{center}
\caption{\label{fig_precision}
The relative precision $\delta$ for $\Sigma^{(2,0)}_{\mathrm{bare},V}$ (left) and $\Sigma^{(2,0)}_{\mathrm{bare},S}$ (right)
obtained by truncating the $q$-series of the iterated integrals at order ${\mathcal O}(q^{30})$
for the various choices $q \in \{ q_{1,0}, q_{6,1}, q_{2,9}, q_{3,\infty} \}$.
There is always a choice such that the relative precision is below $1.5 \times 10^{-21}$. 
This value is indicated by the dashed black line.
}
\end{figure}
In fig.\ref{fig_precision} we show for method A the relative precision 
for $\Sigma^{(2,0)}_{\mathrm{bare},V}$ and $\Sigma^{(2,0)}_{\mathrm{bare},S}$
by truncating the iterated integrals to order ${\mathcal O}(q^{30})$
for the various choices $q \in \{ q_{1,0}, q_{6,1}, q_{2,9}, q_{3,\infty} \}$
as a function of $x$.
We see that for all values $x \in {\mathbb R}$ the ${\mathcal O}(q^{30})$-approximation of the iterated integrals gives us a 
relative precision on $\Sigma^{(2,0)}_{\mathrm{bare},V}$ and $\Sigma^{(2,0)}_{\mathrm{bare},S}$ better than $1.5 \times 10^{-21}$.

\begin{figure}
\begin{center}
\includegraphics[scale=0.6]{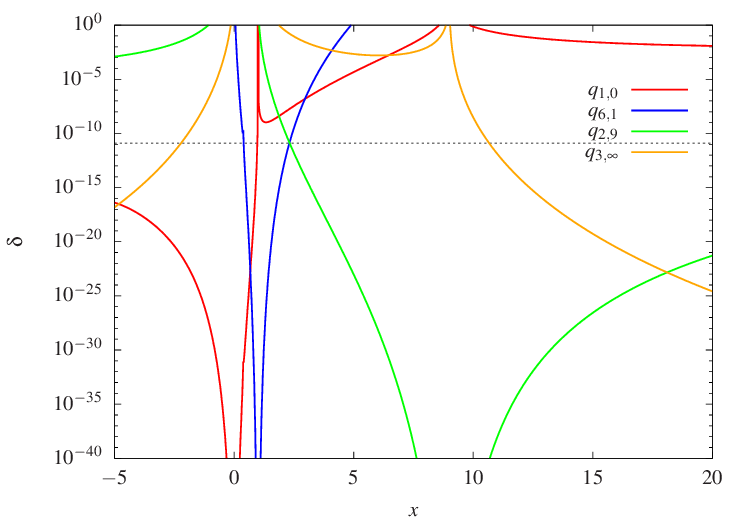}
\includegraphics[scale=0.6]{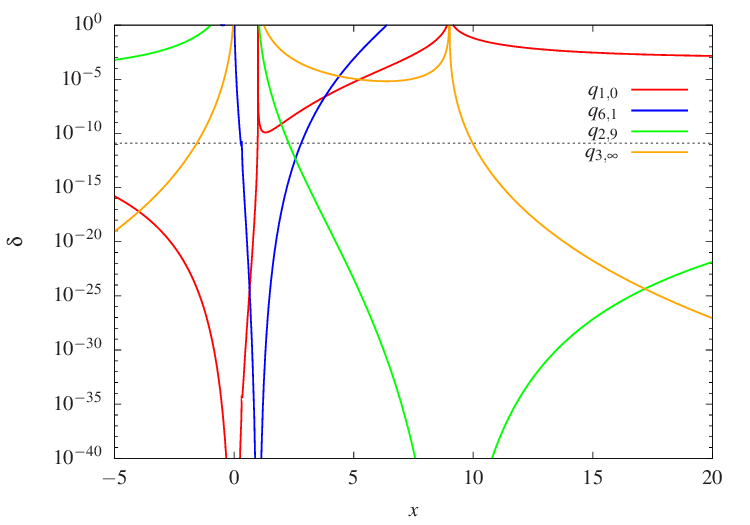}
\end{center}
\caption{\label{fig_precision_coeff_q}
The relative precision $\delta$ for $\Sigma^{(2,0)}_{\mathrm{bare},V}$ (left) and $\Sigma^{(2,0)}_{\mathrm{bare},S}$ (right)
obtained by truncating the $q$-series for the self-energies at order ${\mathcal O}(q^{30})$
for the various choices $q \in \{ q_{1,0}, q_{6,1}, q_{2,9}, q_{3,\infty} \}$.
In contrast to fig.~\ref{fig_precision}, this method gives in some regions only a relative precision of $1.3 \times 10^{-11}$.
This value is indicated by the dashed black line.
}
\end{figure}

Fig.~\ref{fig_precision_coeff_q} shows the corresponding plot for method B.
We observe, that in some regions of $x$ the relative precision is only below $1.3 \times 10^{-11}$.
The bad regions are the ones where we switch from one choice of $q$ to another, i.e. the regions where the optimal
expansion parameter $q$ is close to its maximum $|q| \approx 0.163$.
These regions are regions away from the singular points $x \in \{0,1,9,\infty\}$ of the differential equation.

The $q$-series for $\Sigma^{(2,0)}_{\mathrm{bare},V}$ and $\Sigma^{(2,0)}_{\mathrm{bare},S}$ have their virtues close
to the singular points $x \in \{0,1,9,\infty\}$.
An inspection of the formulae from appendix~\ref{sect:coefficients_masters} shows, 
that the coefficients of the iterated integrals have poles like
\bq
 \frac{1}{x^3},
 \;\;\;
 \frac{1}{x^2},
 \;\;\;
 \frac{1}{x},
 \;\;\;
 \frac{1}{\left(x-1\right)^2},
 \;\;\;
 \frac{1}{\left(x-1\right)}.
\eq
These poles don't show up in the final result for $\Sigma^{(2,0)}_{\mathrm{bare},V}$ and $\Sigma^{(2,0)}_{\mathrm{bare},S}$.
The self-energy is smooth at $x=0$ and has only logarithmic singularities at $x=1$.
Therefore these poles have to cancel.
Within method A this cancellation occurs numerically and might lead close to the singular points to numerical instabilities, 
if floating point arithmetic with a fixed mantissa is used.
On the other hand, in the $q$-series expansions of $\Sigma^{(2,0)}_{\mathrm{bare},V}$ and $\Sigma^{(2,0)}_{\mathrm{bare},S}$
these spurious poles are absent and 
the $q$-series expansions of $\Sigma^{(2,0)}_{\mathrm{bare},V}$ and $\Sigma^{(2,0)}_{\mathrm{bare},S}$
are stable as we approach the singular points.

We have investigated the reduced precision of method $B$ compared to the precision of method A in more 
detail.
The reduced precision 
is due to the slow convergence of the $q$-series of the coefficients of the master integrals.
Let us discuss the convergence of the $q$-series of the coefficients in more detail.
Empirically we found that the worst case is given by the term
\bq
 f\left(x\right) & = & \frac{1}{x^3}.
\eq
Let us assume that we want to calculate $f(x)$ for $x=-10$
from the $q$-series by expanding around the cusp $j=1$.
This is just for illustration, within method A one computes $f(x)$ for $x=-10$ directly, giving $f=-10^{-3}$.
Within method B one would use for $x=-10$ an expansion around the cusp $j=\infty$.
But $f(x)$ can be expanded around the cusp $j=1$ and we find
\bq
\label{def_example}
 f\left(x\right)
 & = &
 \frac{\eta\left(2\tau_{6,1}\right)^{12} \eta\left(3\tau_{6,1}\right)^{24}}{\eta\left(6\tau_{6,1}\right)^{12} \eta\left(\tau_{6,1}\right)^{24}}
 \;\; = \;\;
 1 + 24 q_{6,1} + 312 q_{6,1}^2 + 2888 q_{6,1}^3 + ...
\eq
For the expansion around the cusp $j=1$ we have $q_{6,1}(x=0)=1$ but
\bq
 \left| q_{6,1}\left(x=-10\right) \right| 
 & \approx & 0.485,
\eq
therefore the point $x=-10$ is in the $q_{6,1}$-space inside the unit disc (see fig.~\ref{fig_q_1_path}).
Since the expansion of $f(x)$ in the variable $q_{6,1}$ comes from an eta-quotient the series converges for all $q_{6,1}$
inside the unit disc. 
Let us now study how fast or good this series converges.
\begin{figure}
\begin{center}
\includegraphics[scale=1.0]{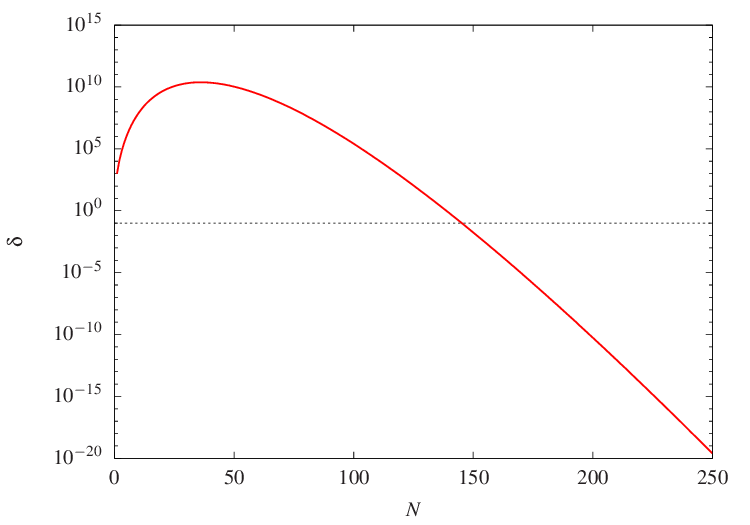}
\end{center}
\caption{\label{fig_one_over_x3}
The relative precision of the order $\left(q_{6,1}\right)^N$-truncation of the $q_{6,1}$-series of the function $f(x)=1/x^3$
at the point $x=-10$ as a function of $N$.
The dashed line indicates a relative precision of $10\%$.
}
\end{figure}
In fig.\ref{fig_one_over_x3} we plot the relative precision at the point $x=-10$ 
if we truncate in eq.~(\ref{def_example}) the $q_{6,1}$-expansion at order $N$
as a function of $N$.
We see that this series converges rather slowly. Only after including roughly $150$ terms in the $q$-expansion we reach
a precision of $10\%$.
With a truncation after $250$ terms we reach a relative precision of $10^{-20}$.
Let us note that a truncation below $150$ terms gives unreliable results.
If one only considers truncations up to $N=30$ one might be led to the false conclusion that the series is divergent.
As already mentioned, this is a constructed worst case scenario.
The actual relative precisions for the finite part of the two-loop self-energy for method A and method B 
have been given in fig.~\ref{fig_precision} and fig.~\ref{fig_precision_coeff_q}, respectively.

\begin{figure}
\begin{center}
\includegraphics[scale=0.6]{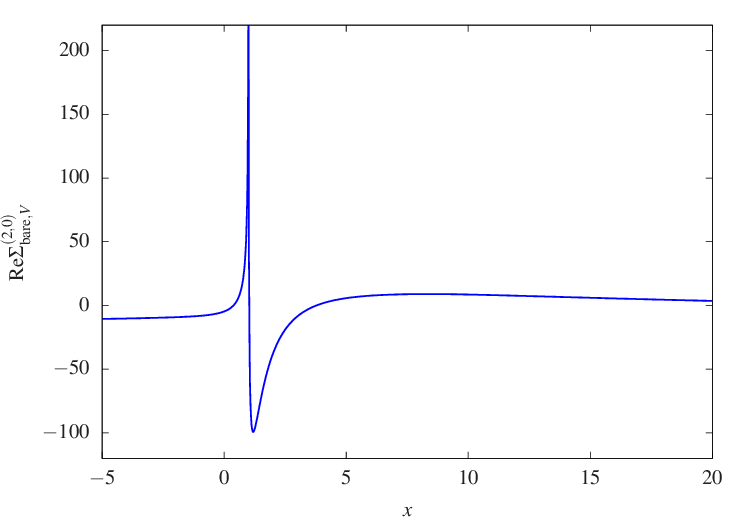}
\includegraphics[scale=0.6]{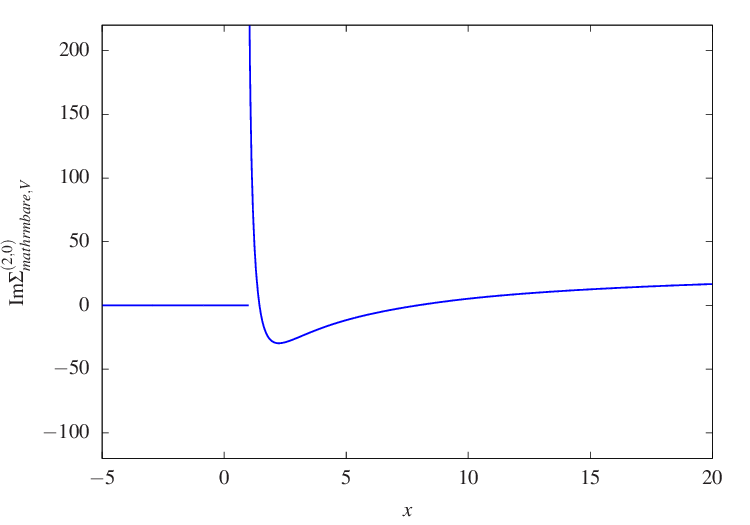}
\end{center}
\caption{\label{fig_Sigma_V}
The real part (left) and the imaginary part (right)
of the $\eps^0$-term of the bare quantity $\Sigma^{(2)}_{\mathrm{bare},V}$ in Feynman gauge.
}
\end{figure}

In fig.~\ref{fig_Sigma_V} we plot the final result for $\Sigma^{(2,0)}_{\mathrm{bare},V}$ in Feynman gauge.
We show separately the real part and the imaginary part of $\Sigma^{(2,0)}_{\mathrm{bare},V}$. 
The self-energy $\Sigma^{(2,0)}_{\mathrm{bare},V}$ is real for $x<1$.
\begin{figure}
\begin{center}
\includegraphics[scale=0.6]{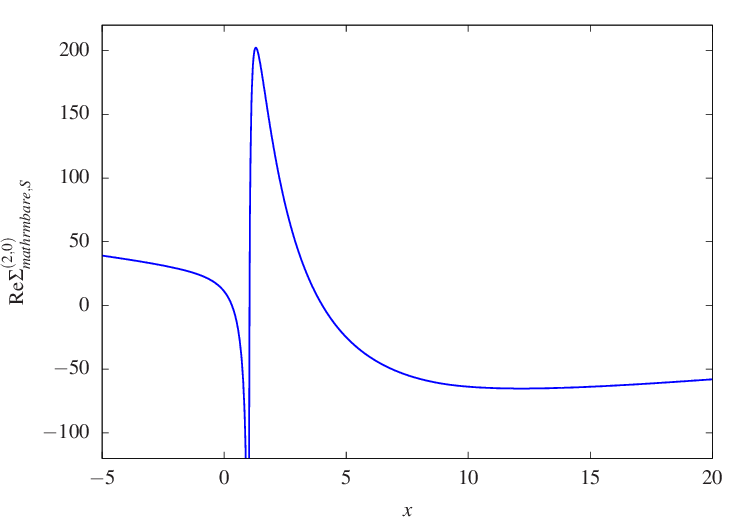}
\includegraphics[scale=0.6]{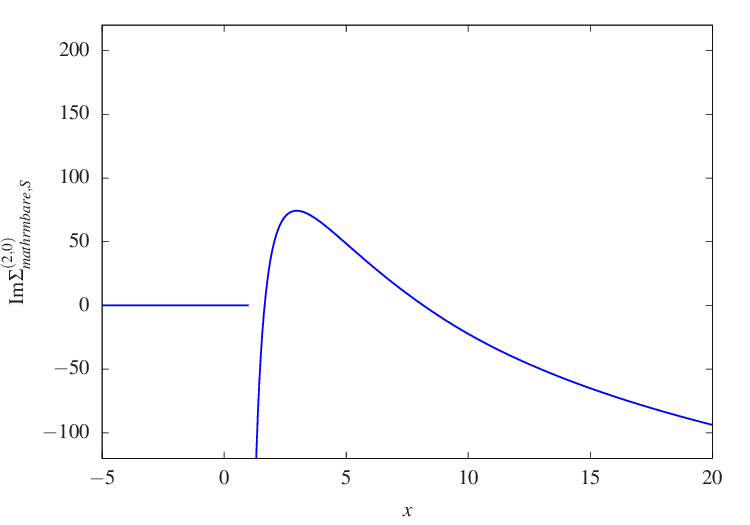}
\end{center}
\caption{\label{fig_Sigma_S}
The real part (left) and the imaginary part (right)
of the $\eps^0$-term of the bare quantity $\Sigma^{(2)}_{\mathrm{bare},S}$ in Feynman gauge.
}
\end{figure}
In fig.~\ref{fig_Sigma_S} we show the corresponding plot for $\Sigma^{(2,0)}_{\mathrm{bare},S}$.
The self-energy $\Sigma^{(2,0)}_{\mathrm{bare},S}$ is real for $x<1$.

% -----------------------------------------------------------------------------

\section{Conclusions}
\label{sect:conclusions}

In this paper we reconsidered the two-loop electron self-energy in quantum electrodynamics.
This is the simplest ``physical'' building block, where elliptic integrals make their appearance in perturbative
quantum field theory.
We expressed all relevant master integrals as iterated integrals of modular forms.
These iterated integrals have a $q$-series expansion, which converges for all real values of $x$ except for three points
out of the set $\{0,1,9,\infty\}$.
We obtained the master integrals by integrating a system of differential equations from a chosen boundary point.
By choosing different boundary points, we obtain for all values $x \in {\mathbb R}$ convergent $q$-series.
We considered explicitly the cases for the boundary points $j \in \{0,1,9,\infty\}$.
In particular we have shown that for all $x \in {\mathbb R}$ there is always a convergent $q$-series with $|q| \lessapprox 0.163$.
This allows for an efficient numerical evaluation.
In particular we find that a truncation of the $q$-series to order ${\mathcal O}(q^{30})$ gives numerically for the finite part 
of the self-energy a relative precision better than $10^{-20}$ for all real values $x$.
We expect the methods discussed here to be useful also in other precision calculations involving elliptic integrals.

\subsection*{Acknowledgements}

S.W. would like to thank David Broadhurst for his encouragement to address this problem.

\subsection*{Note added}

A typo in a computer program led to wrong expressions in eq.~(\ref{res_counterterms_1}) and eq.~(\ref{res_counterterms_2}).
These have now been corrected. Other formulae are not affected.
We thank Federico Gasparotto for bringing this to our attention.

% -----------------------------------------------------------------------------

\begin{appendix}

% -----------------------------------------------------------------------------

\section{Feynman rules}
\label{sect:Feynman_rules}

In this appendix we list the Feynman rules for QED.
The Feynman rules for the propagators are
\bq
\begin{picture}(85,20)(0,5)
 \ArrowLine(70,10)(20,10)
\end{picture} 
 & = &
 i\frac{{\slashed p}+m}{p^2-m^2},
 \nonumber \\
\begin{picture}(85,20)(0,5)
 \Photon(20,10)(70,10){-3}{5}
\end{picture} 
& = &
 \frac{i}{p^2} \left( - g^{\mu\nu} + \left(1-\xi\right) \frac{p^\mu p^\nu}{p^2} \right),
\eq
where the photon propagator is in a covariant gauge.
The vertex is given by
\bq
\begin{picture}(100,35)(0,50)
\Vertex(50,50){2}
\Photon(50,50)(80,50){3}{4}
\ArrowLine(50,50)(29,71)
\ArrowLine(29,29)(50,50)
\Text(82,50)[lc]{$\mu$}
\end{picture}
 \;\; = \;\;
 - i e^{(D)} \gamma^{\mu}.
 \\ \nonumber
\eq
The coupling $e^{(D)}$ is defined in eq.~(\ref{def_coupling}).
The Feynman rules for the counterterms are
\bq
\begin{picture}(85,20)(0,5)
 \ArrowLine(70,10)(45,10)
 \ArrowLine(45,10)(20,10)
 \Line(40,5)(50,15)
 \Line(40,15)(50,5)
\end{picture} 
 & = &
 i \left[ \left(Z_2-1\right) {\slashed p} - \left( Z_2 Z_m - 1 \right) m \right],
 \nonumber \\
\begin{picture}(85,20)(0,5)
 \Photon(20,10)(40,10){-3}{2}
 \Photon(50,10)(70,10){-3}{2}
 \Line(40,5)(50,15)
 \Line(40,15)(50,5)
\end{picture} 
& = &
 i \left( Z_3-1 \right) p^2 \left( - g^{\mu\nu} + \frac{p^\mu p^\nu}{p^2} \right),
 \nonumber \\
\begin{picture}(100,35)(0,50)
\Photon(56,50)(80,50){3}{3}
\ArrowLine(43,57)(29,71)
\ArrowLine(29,29)(43,43)
\Line(45,45)(55,55)
\Line(45,55)(55,45)
\Text(82,50)[lc]{$\mu$}
\end{picture}
& = &
 - i e^{(D)} \left(Z_2-1\right) \gamma^{\mu}.
 \\ \nonumber
\eq

% -----------------------------------------------------------------------------

\section{The arithmetic-geometric mean}
\label{sec:agm}

In this appendix we review the numerical evaluation of
the complete elliptic integral of the first kind with the help of the arithmetic-geometric mean.
Let $a_0$ and $b_0$ be two complex numbers. For $n \in {\mathbb N}_0$ one sets
\bq
 a_{n+1} \; = \; \frac{1}{2} \left( a_n+b_n \right),
 & &
 b_{n+1} \; = \; \pm \sqrt{a_n b_n}.
\eq
The sign of the square root is chosen such that \cite{Cox:1984} 
\bq
 \left| a_{n+1} - b_{n+1} \right| & \le & \left| a_{n+1} + b_{n+1} \right|,
\eq
and in case of equality one demands in addition
\bq
 \mathrm{Im}\left( \frac{b_{n+1}}{a_{n+1}} \right) & > & 0.
\eq
The sequences $(a_n)$ and $(b_n)$ converge to a common limit 
\bq
 \lim\limits_{n \rightarrow \infty} a_n
 \; = \;
 \lim\limits_{n \rightarrow \infty} b_n
 \; = \; 
 \mathrm{agm}(a_0,b_0),
\eq
known as the arithmetic-geometric mean.
The complete elliptic integral of the first kind is given by
\bq
 K\left(k\right)
 & = &
 \frac{\pi}{2 \; \mathrm{agm}\left(k',1\right)},
 \;\;\;
 k' 
 \; = \;
 \sqrt{1-k^2}.
\eq

% -----------------------------------------------------------------------------

\section{Eisenstein series}
\label{sec:eisenstein}

In this appendix we give the explicit expressions for the Eisenstein series
$E_1(\tau,\chi_0,\chi_1)$ and $E_1(2\tau,\chi_0,\chi_1)$.
$\chi_0$ and $\chi_1$ denote primitive Dirichlet characters
with conductors $1$ and $3$, respectively.
In terms of Kronecker symbols they are given by
\bq
 \chi_0 
 \; = \; 
 \left( \frac{1}{n} \right), 
 & &
 \chi_1 
 \; = \; 
 \left( \frac{-3}{n} \right).
\eq
More explicitly we have
\bq
 \chi_0\left(n\right) 
 & = & 
 1, 
 \;\;\;\;\;\; 
 \forall n \in {\mathbb Z},
 \nonumber \\
 \chi_1\left(n\right) 
 & = & 
 \left\{ \begin{array}{rl}
  0, & n = 0 \mod 3, \\
  1, & n = 1 \mod 3, \\
 -1, & n = 2 \mod 3, \\
 \end{array} \right.
\eq
$E_1(\tau,\chi_0,\chi_1)$ is given with $q=e^{2\pi i \tau}$ by
\bq
 E_1\left(\tau; \chi_0, \chi_1\right) & = & \frac{1}{6} + \sum\limits_{m=1}^{\infty} \left( \sum\limits_{d|m} \chi_1\left(d\right) \right) q^m.
\eq
In terms of the $\mathrm{ELi}$-functions, defined by
\bq
 \mathrm{ELi}_{n;m}\left(x;y;q\right) & = & \sum\limits_{j=1}^\infty \sum\limits_{k=1}^\infty \; \frac{x^j}{j^n} \frac{y^k}{k^m} q^{j k},
\eq
we have
\bq
 E_1\left(\tau; \chi_0, \chi_1\right) & = &
 \frac{1}{6} + \frac{1}{i \sqrt{3}} \left[ \mathrm{ELi}_{0,0}\left(r_3,1;q\right) - \mathrm{ELi}_{0,0}\left(r_3^{-1},1;q\right) \right],
\eq
where $r_3=\exp(2\pi i / 3)$ denotes the third root of unity.
The first few terms of $E_1(\tau,\chi_0,\chi_1)$ read
\bq
 E_1\left(\tau; \chi_0, \chi_1\right) & = &
 \frac{1}{6} + q + q^3 + q^4 + 2 q^7 + q^9 + ...
\eq
The Eisenstein series $E_1(2\tau,\chi_0,\chi_1)$ is obtained from $E_1(\tau,\chi_0,\chi_1)$ by
the substitution $\tau \rightarrow 2 \tau$ or equivalently $q \rightarrow q^2$.

% -----------------------------------------------------------------------------

\section{The one-loop self-energy}
\label{sect:oneloop}

In this appendix we consider the one-loop electron self-energy. The relevant family of (one-loop) Feynman integrals is given
by
\bq
\label{def_oneloopfamily}
 J_{\nu_1 \nu_4}\left( D, p^2, m^2, \mu^2 \right)
 & = &
 \left(-1\right)^{\nu_{14}}
 e^{\gamma_E \eps}
 \left(\mu^2\right)^{\nu_{14}-\frac{D}{2}}
 \int \frac{d^Dk_1}{i \pi^{\frac{D}{2}}}
 \frac{1}{D_1^{\nu_1} D_4^{\nu_4}},
\eq
where the propagators $D_1$ and $D_4$ have been defined in eq.~(\ref{def_propagators}).
As before we set $\mu=m$.
As a basis of master integrals we use
\bq
\label{def_basis_oneloop}
 J_1\left(\eps,x\right)
 \; = \;
 2 \eps J_{20}\left(4-2\eps,x\right),
 & &
 J_2\left(\eps,x\right)
 \; = \;
 2 \eps x J_{21}\left(4-2\eps,x\right).
\eq
The differential equation for $\vec{J}=(J_1,J_2)^T$ reads
\bq
 \frac{d}{dx} \vec{J} 
 & = &
 \eps \left( \begin{array}{cc}
 0 & 0 \\
 -\frac{1}{x-1} & \frac{1}{x} - \frac{2}{x-1} \\
 \end{array} \right)
 \vec{J},
\eq
the boundary conditions are
\bq
 J_1\left(\eps,0\right)
 \; = \;
 2 e^{\gamma_E \eps} \Gamma\left(1+\eps\right),
 & &
 J_2\left(\eps,0\right)
 \; = \;
 0.
\eq
We have
\bq
 J_1\left(\eps,x\right)
 & = &
 2 + \zeta_2 \eps^2 
 + {\mathcal O}\left(\eps^3\right),
 \nonumber \\
 J_2\left(\eps,x\right)
 & = &
 - 2 \eps \ln\left(1-x\right)
 + 2 \left[ \mathrm{Li}_2\left(x\right) + \ln^2\left(1-x\right) \right] \eps^2
 + {\mathcal O}\left(\eps^3\right).
\eq
For the bare one-loop electron self-energy we write
\bq
 - i \Sigma^{(1)}_{\mathrm{bare}}
 & = &
 - i 
  \frac{\alpha}{4\pi}
  \left(
         \Sigma^{(1)}_{\mathrm{bare},V} \; {\slashed p} + \Sigma^{(1)}_{\mathrm{bare},S} \; m
  \right),
\eq
with
\bq
 \Sigma^{(1)}_{\mathrm{bare},V}
 & = &
 - \frac{\xi}{2\left(1-2\eps\right)} \left( \frac{1}{\eps} + \frac{1}{x} - 1 \right) J_1
 + \frac{\left(1-\eps\right)\xi}{2\eps\left(1-2\eps\right)} \left( \frac{1}{x^2} - 1 \right) J_2,
 \nonumber \\
 \Sigma^{(1)}_{\mathrm{bare},S}
 & = &
 \frac{3+\xi-2\eps}{2\eps\left(1-2\eps\right)} J_1
 - \frac{3+\xi-2\eps}{2\eps\left(1-2\eps\right)} \left( \frac{1}{x} - 1 \right) J_2.
\eq
Expanded in $\eps$ we obtain
\bq
 \Sigma^{(1)}_{\mathrm{bare},V}
 & = &
 - \frac{\xi}{\eps} 
 - \xi \left(\frac{1}{x^2} -1\right) \ln\left(1-x\right)
 - \frac{\xi}{x} - \xi
 + {\mathcal O}\left(\eps\right),
 \nonumber \\
 \Sigma^{(1)}_{\mathrm{bare},S}
 & = &
 \frac{3+\xi}{\eps}
 + \left(3+\xi\right) \left(\frac{1}{x} - 1 \right) \ln\left(1-x\right) + 4 + 2 \xi
 + {\mathcal O}\left(\eps\right).
\eq
The counterterms from renormalisation are given by
\bq
 - i \Sigma^{(1)}_{\mathrm{CT}}
 & = &
 - i 
  \frac{\alpha}{4\pi}
  \left[
         - Z_2^{(1)} \; {\slashed p} + \left( Z_2^{(1)} + Z_m^{(1)} \right)  \; m
  \right].
\eq
The renormalisation constants $Z_2^{(1)}$ and $Z_m^{(1)}$ in the on-shell scheme are given in eq.~(\ref{renorm_constants}).
In the on-shell scheme we have
\bq
 - i \Sigma^{(1)}_{\mathrm{CT}}
 & = &
 - i 
  \frac{\alpha}{4\pi}
  \left[
         \left( \frac{3}{\eps} + 4 \right)  \; {\slashed p} - \left( \frac{6}{\eps} + 8  \right)  \; m
  \right]
 + {\mathcal O}\left(\eps\right).
\eq

% -----------------------------------------------------------------------------

\section{The coefficients of the master integrals}
\label{sect:coefficients_masters}

In this appendix we give the coefficients $c_j^V$ and $c_j^S$ ($j=1,...,8$), defined by eq.~(\ref{def_linear_combination_master_integrals}) in Feynman gauge ($\xi=1$).
These coefficients occur when we express the two-loop self-energy in terms of the master integrals $I_1,...,I_8$.
The coefficients in a general covariant gauge ($\xi \neq 1$) are given
in a supplementary electronic file attached to this article.
\bq
 c_1^V & = &
 {\frac {75-322\,\eps+509\,{\eps}^{2}-372\,{\eps}^{3}+114\,{\eps}^{4}}
        {60 \eps\, \left( 1-\eps \right) \left( 1-2\,\eps \right) ^{2} \left( 1-3\,\eps \right) \left( 2-3\,\eps \right) }}
 -
 {\frac {16}{5   \left( 1-2\,\eps \right)  \left( 1+2\,\eps \right) \left( x-1 \right) }},
 \nonumber \\
 & &
 +
 {\frac {30-345\,\eps+1439\,{\eps}^{2}-2965\,{\eps}^{3}+3279\,{\eps}^{4}-1980\,{\eps}^{5}+540\,{\eps}^{6}}
        {30 {\eps}^{2} \left( 1-\eps \right) \left( 1-2\,\eps \right) ^{2} \left( 1-3\,\eps \right)  \left( 2-3\,\eps \right)  x }},
 \nonumber \\
 c_2^V & = &
 {\frac {15+2\,x\eps-21\,\eps}
        {60 {\eps}^{2} \left( 1-2\,\eps \right) }}
 -
 {\frac {19-10\,\eps}
        {10 \eps\, \left( 1-2\,\eps \right) x }}
 +
 {\frac {165-167\,\eps+60\,{\eps}^{2}}
        { 60 {\eps}^{2} \left( 1-2\,\eps \right) {x}^{2} }},
 \nonumber \\
 c_3^V & = &
 {\frac { \left( 1-\eps \right) \left( 1+2\,\eps \right)  }
        { 4 {\eps}^{2} \left( 1-2\,\eps \right) ^{2}}}
 -
 {\frac {4-14\,\eps+9\,{\eps}^{2}+{\eps}^{3}+4\,{\eps}^{4}}
        { 2 {\eps}^{2} \left( 1-2\,\eps \right) ^{2} \left( 1-3\,\eps \right)  \left( 2-3\,\eps \right) x}}
 -
 {\frac {4-21\,\eps+18\,{\eps}^{2}+13\,{\eps}^{3}-10\,{\eps}^{4}}
        { 4 {\eps}^{2}  \left( 1-2\,\eps \right) ^{2} \left( 1-3\,\eps \right) \left( 2-3\,\eps \right) {x}^{2}}},
 \nonumber \\
 c_4^V & = &
 -
 {\frac { \left( 1-\eps \right)  \left( 1-4\,\eps+14\,{\eps}^{2}-16\,{\eps}^{3} \right) }
        { 4 {\eps}^{2} \left( 1-2\,\eps \right) ^{2} \left( 1-3\,\eps \right)  \left( 2-3\,\eps \right) }}
 -
 {\frac {2-8\,\eps+{\eps}^{2}+12\,{\eps}^{3}-4\,{\eps}^{4}}
        { 2 {\eps}^{2}  \left( 1-2\,\eps \right) ^{2} \left( 1-3\,\eps \right) \left( 2-3\,\eps \right) x}},
 \nonumber \\
 c_5^V & = &
 {\frac {1-\eps}
        {2 \eps\, \left( 1-2\,\eps \right) ^{2}}}
 -
 {\frac { \left( 1-\eps \right) \left( 1+\eps \right)  }
        {{\eps}^{2} \left( 1-2\,\eps \right) ^{2} x }}
 +
 {\frac { \left( 1-\eps \right) \left( 4+\eps \right)  }
        { 2 {\eps}^{2} \left( 1-2\,\eps \right) ^{2}{x}^{2}}}
 -
 {\frac {1-\eps}
        { {\eps}^{2} \left( 1-2\,\eps \right) ^{2}{x}^{3}}},
 \nonumber \\
 c_6^V & = &
 \left[
 -
 {\frac {712-1943\,\eps+1307\,{\eps}^{2}-108\,{\eps}^{3}}
        { 30 {\eps}^{2} \left( 1-2\,\eps \right)  \left( 1-3\,\eps \right)  \left( 2-3\,\eps \right) }}
 +
 {\frac { \left( 238 -525\,\eps+183\,{\eps}^{2}+150\,{\eps}^{3} \right) x}
        { 90 {\eps}^{2} \left( 1-2\,\eps \right) \left( 1-3\,\eps \right)   \left( 2-3\,\eps \right) }}
 \right. \nonumber \\
 & & \left.
 -
 {\frac { \left( 2+\eps \right) {x}^{2}}
        { 90 {\eps}^{2} \left( 1-2\,\eps \right) }}
 +
 {\frac {64 \left(1+2\,\eps\right)}
        { 15 {\eps}^{2} \left( 1-2\,\eps \right)  \left( x-1 \right) }}
 +
 {\frac {3 \left( 74-181\,\eps+151\,{\eps}^{2}-162\,{\eps}^{3} \right)}
        { 10 {\eps}^{2} \left( 1-2\,\eps \right)  \left( 1-3\,\eps \right) \left( 2-3\,\eps \right) x }}
 \right. \nonumber \\
 & & \left.
 -
 {\frac {9 \left( 17-38\,\eps+11\,{\eps}^{2} \right)}
        {10 \eps  \left( 1-2\,\eps \right)  \left( 1-3\,\eps \right) \left( 2-3\,\eps \right) {x}^{2} }}
 -
 {\frac {256}
        { 15 \eps \left( 1-2\,\eps \right)  \left( x-1 \right) ^{2}}}
 \right] \frac{\psi_1}{\pi}
 \nonumber \\
 & &
 +
 \left[
 {\frac {851-1761\,\eps+564\,{\eps}^{2}}
       { 15 {\eps}^{2} \left( 1-2\,\eps \right)  \left( 1-3\,\eps \right) \left( 2-3\,\eps \right) }}
 -
 {\frac { \left( 507-1151\,\eps+640\,{\eps}^{2} \right) x}
        { 15 {\eps}^{2}  \left( 1-2\,\eps \right)  \left( 1-3\,\eps \right) \left( 2-3\,\eps \right) }}
 \right. \nonumber \\
 & & \left.
 +
 {\frac { \left( 155-393\,\eps+264\,{\eps}^{2} \right) {x}^{2}}
        { 45 {\eps}^{2}  \left( 1-2\,\eps \right) \left( 1-3\,\eps \right)  \left( 2-3\,\eps \right) }}
 -
 {\frac {{x}^{3}}
        { 45 {\eps}^{2} \left( 1-2\,\eps \right) }}
 +
 {\frac {256}
        { 15 {\eps}^{2} \left( 1-2\,\eps \right)  \left( x-1 \right) }}
 \right. \nonumber \\
 & & \left.
 -
 {\frac {9 \left( 17-38\,\eps+11\,{\eps}^{2}\right)}
        { 5 {\eps}^{2}\left( 1-2\,\eps \right)  \left( 1-3\,\eps \right) \left( 2-3\,\eps \right) x }}
 \right] \frac{1}{\pi} \frac{d\psi_1}{dx},
 \nonumber \\
 c_7^V & = &
 \left[
 -
 {\frac {4}
        {15 \eps\, \left( 1-2\,\eps \right) }}
 -
 {\frac {4 \left( 19-187\,\eps+230\,{\eps}^{2} \right)}
        { 5 \eps \left( 1-2\,\eps \right)  \left( 1-3\,\eps \right) \left( 2-3\,\eps \right) x }}
 +
 {\frac {128}
        { 5 \eps \left( 1-2\,\eps \right)  \left( x-1 \right) }}
 \right. \nonumber \\
 & & \left.
 -
 {\frac {12 \left( 17-38\,\eps+11\,{\eps}^{2} \right)}
        {5 \eps  \left( 1-2\,\eps \right)  \left( 1-3\,\eps \right) \left( 2-3\,\eps \right) {x}^{2} }}
 -
 {\frac {128}
        { 5 \eps \left( 1-2\,\eps \right)  \left( x-1 \right) ^{2}}}
 \right] \frac{\pi}{\psi_1},
 \nonumber \\
 c_8^V & = &
 -
 {\frac {1-\eps}
        {4 {\eps}^{2} \left( 1-2\,\eps \right) x}}
 +
 {\frac { \left( 1+\eps \right)  \left( 2-\eps \right) }
        {4 {\eps}^{3} \left( 1-2\,\eps \right) {x}^{2}}},
 \nonumber \\
 c_1^S & = &
 {\frac {2-14\,\eps+20\,{\eps}^{2}-9\,{\eps}^{3}}
        { 2 {\eps}^{2} \left( 1-2\,\eps \right) ^{2} \left( 1-3\,\eps \right) }}
 +
 {\frac {16}
        { 3 \left( 1-2\,\eps \right)  \left( 1+2\,\eps \right) \left( x-1 \right)  }},
 \nonumber \\
 c_2^S & = &
 {\frac {3+8\,\eps-4\,{\eps}^{2}}
        { 3 {\eps}^{2} \left( 1-2\,\eps \right) ^{2}}}
 -
 {\frac {21-49\,\eps+44\,{\eps}^{2}-12\,{\eps}^{3}}
        {3 {\eps}^{2} \left( 1-2\,\eps \right) ^{2} x}},
 \nonumber \\
 c_3^S & = &
 -
 {\frac { \left( 1-\eps \right) ^{2}}
        { {\eps}^{2} \left( 1-2\,\eps \right) ^{2}}}
 +
 {\frac {2-4\,\eps+{\eps}^{3}}
        { {\eps}^{2} \left( 1-2\,\eps \right) ^{2} \left( 1-3\,\eps \right) x}},
 \nonumber \\
 c_4^S & = &
 {\frac {1+\eps-7\,{\eps}^{2}+4\,{\eps}^{3}}
        { 2 {\eps}^{2} \left( 1-2\,\eps \right) ^{2} \left( 1-3\,\eps \right) }},
 \nonumber \\
 c_5^S & = &
 {\frac {3-3\,\eps+{\eps}^{2}}
        { 2 {\eps}^{2} \left( 1-2\,\eps \right) ^{2}}}
 -
 {\frac {3-3\,\eps+{\eps}^{2}}
        {{\eps}^{2} \left( 1-2\,\eps \right) ^{2} x }}
 +
 {\frac {3-3\,\eps+{\eps}^{2}}
        { 2 {\eps}^{2} \left( 1-2\,\eps \right) ^{2}{x}^{2}}},
 \nonumber \\
 c_6^S & = &
 \left[
 {\frac {2 \left( 157-167\,\eps+33\,{\eps}^{2} \right)}
        { 9 {\eps}^{2} \left( 1-2\,\eps \right) \left( 1-3\,\eps \right)  }}
 -
 {\frac { \left( 34+8\,\eps-15\,{\eps}^{2} \right) x}
        { 9 {\eps}^{2}  \left( 1-2\,\eps \right) \left( 1-3\,\eps \right) }}
 -
 {\frac {9 \left( 2+\eps \right)}
        { \eps \left( 1-2\,\eps \right) \left( 1-3\,\eps \right)  x}}
 \right. \nonumber \\
 & & \left.
 -
 {\frac {64}
        { 9 {\eps}^{2} \left( 1-2\,\eps \right)  \left( x-1 \right) }}
 +
 {\frac {256}
        { 9 \eps \left( 1-2\,\eps \right)  \left( x-1 \right) ^{2}}}
 \right] \frac{\psi_1}{\pi}
 \nonumber \\
 & &
 +
 \left[
 -
 {\frac {2 \left(290-303\,\eps \right)}
        { 9 {\eps}^{2} \left( 1-2\,\eps \right) \left( 1-3\,\eps \right) }}
 +
 {\frac { 4 \left( 134-87\,\eps \right) x}
        { 9 {\eps}^{2}  \left( 1-2\,\eps \right) \left( 1-3\,\eps \right) }}
 -
 {\frac { 2 \left( 26-15\,\eps \right) {x}^{2}}
        { 9 {\eps}^{2}  \left( 1-2\,\eps \right) \left( 1-3\,\eps \right) }}
 \right. \nonumber \\
 & & \left.
 -
 {\frac {256}
        { 9 {\eps}^{2} \left( 1-2\,\eps \right)  \left( x-1 \right) }}
 \right] \frac{1}{\pi} \frac{d\psi_1}{dx},
 \nonumber \\
 c_7^S & = &
 \left[
 -
 {\frac {24 \left( 2+\eps \right) }
        { \eps \left( 1-2\,\eps \right) \left( 1-3\,\eps \right)  x}}
 -
 {\frac {64}
        { 3 \eps \left( 1-2\,\eps \right)  \left( x-1 \right) }}
 +
 {\frac {128}
        { 3 \eps\left( 1-2\,\eps \right)  \left( x-1 \right) ^{2}}}
 \right] \frac{\pi}{\psi_1},
 \nonumber \\
 c_8^S & = &
 {\frac {1+\eps-{\eps}^{2}}
        { 4 {\eps}^{3} \left( 1-2\,\eps \right) }}
 +
 {\frac {1-\eps+{\eps}^{2}}
        {4 {\eps}^{3} \left( 1-2\,\eps \right) x}}.
\eq

% -----------------------------------------------------------------------------

\section{Boundary constants}
\label{sec:boundary_constants}

In this appendix we list the relevant boundary constants when we integrate the differential equation eq.~(\ref{eps_form})
from one of the points $x \in \{0,1,9,\infty\}$.
For the finite part of the two-loop electron self-energy we need the master integrals $I_1$-$I_6$ to order $\eps^2$,
the master integral $I_7$ to order $\eps^1$ and the master integral $I_8$ to order $\eps^3$.
We denote by $C_{i,j}^{k}$ the boundary constant for the $\eps^k$-th term of the master integral $I_i$ at boundary point $j$.

For the boundary point $x=0$ we have
\begin{align}
 C_{1,0}^{(0)} = & 4,  & C_{1,0}^{(1)} = & 0, & C_{1,0}^{(2)} = & 4 \zeta_2, & & \nonumber \\ 
 C_{2,0}^{(0)} = & 0,  & C_{2,0}^{(1)} = & 0, & C_{2,0}^{(2)} = & 0, & & \nonumber \\ 
 C_{3,0}^{(0)} = & 0,  & C_{3,0}^{(1)} = & 0, & C_{3,0}^{(2)} = & 0, & & \nonumber \\ 
 C_{4,0}^{(0)} = & 4,  & C_{4,0}^{(1)} = & 0, & C_{4,0}^{(2)} = & 12 \zeta_2, & & \nonumber \\ 
 C_{5,0}^{(0)} = & 0,  & C_{5,0}^{(1)} = & 0, & C_{5,0}^{(2)} = & 0, & & \nonumber \\ 
 C_{6,0}^{(0)} = & 0,  & C_{6,0}^{(1)} = & 0, & C_{6,0}^{(2)} = & 3 \, \mathrm{Cl}_2\left(\frac{2\pi}{3}\right), & & \nonumber \\ 
 C_{7,0}^{(0)} = & 0,  & C_{7,0}^{(1)} = & 0, & & & & \nonumber \\ 
 C_{8,0}^{(0)} = & 0,  & C_{8,0}^{(1)} = & 0, & C_{8,0}^{(2)} = & 0, & C_{8,0}^{(3)} = & 0. \nonumber \\ 
\end{align}
For the boundary point $x=1$ we have
\begin{align}
 C_{1,1}^{(0)} = & 4,  & C_{1,1}^{(1)} = & 0, & C_{1,1}^{(2)} = & 4 \zeta_2, & & \nonumber \\ 
 C_{2,1}^{(0)} = & 0,  & C_{2,1}^{(1)} = & -12 \ln\left(2\right), & C_{2,1}^{(2)} = & 4 \zeta_2 + 36 \ln^2\left(2\right), & & \nonumber \\ 
 C_{3,1}^{(0)} = & 0,  & C_{3,1}^{(1)} = & 12 \ln\left(2\right), & C_{3,1}^{(2)} = & -4 \zeta_2 - 72 \ln^2\left(2\right), & & \nonumber \\ 
 C_{4,1}^{(0)} = & 4,  & C_{4,1}^{(1)} = & -24 \ln\left(2\right), & C_{4,1}^{(2)} = & 28 \zeta_2 + 144 \ln^2\left(2\right), & & \nonumber \\ 
 C_{5,1}^{(0)} = & 0,  & C_{5,1}^{(1)} = & 0, & C_{5,1}^{(2)} = & 36 \ln^2\left(2\right), & & \nonumber \\ 
 C_{6,1}^{(0)} = & 0,  & C_{6,1}^{(1)} = & 0, & C_{6,1}^{(2)} = & -3 \zeta_2 i, & & \nonumber \\ 
 C_{7,1}^{(0)} = & 0,  & C_{7,1}^{(1)} = & 0, & & & & \nonumber \\ 
 C_{8,1}^{(0)} = & 0,  & C_{8,1}^{(1)} = & 0, & C_{8,1}^{(2)} = & 0, & C_{8,1}^{(3)} = & 12 \zeta_3 -48 \zeta_2 \ln\left(2\right). \nonumber \\ 
\end{align}
For the boundary point $x=9$ we have
\begin{align}
 C_{1,9}^{(0)} = & 4,  & C_{1,9}^{(1)} = & 0,                               & C_{1,9}^{(2)} = & 4 \zeta_2, \nonumber \\ 
 C_{2,9}^{(0)} = & 0,  & C_{2,9}^{(1)} = & 4 \pi i - 12 \ln\left(2\right),  & C_{2,9}^{(2)} = & -8\zeta_2+36\ln^2\left(2\right) - 12 \ln^2\left(3\right) -24 \, \mathrm{Li}_2\left(\frac{1}{3}\right) \nonumber \\ 
 & & & & & - 24 \pi \ln\left(2\right) i + 8 \pi \ln\left(3\right) i, \nonumber \\ 
 C_{3,9}^{(0)} = & 0,  & C_{3,9}^{(1)} = & -4 \pi i + 12 \ln\left(2\right), & C_{3,9}^{(2)} = & 32 \zeta_2-72\ln^2\left(2\right) + 12 \ln^2\left(3\right) +24 \, \mathrm{Li}_2\left(\frac{1}{3}\right) \nonumber \\ 
 & & & & & + 48 \pi \ln\left(2\right) i - 8 \pi \ln\left(3\right) i, \nonumber \\ 
 C_{4,9}^{(0)} = & 4,  & C_{4,9}^{(1)} = & 8 \pi i - 24 \ln\left(2\right),  & C_{4,9}^{(2)} = & -20\zeta_2+144\ln^2\left(2\right) - 48 \ln^2\left(3\right) -96 \, \mathrm{Li}_2\left(\frac{1}{3}\right) \nonumber \\ 
 & & & & & - 96 \pi \ln\left(2\right) i + 32 \pi \ln\left(3\right) i, \nonumber \\ 
 C_{5,9}^{(0)} = & 0,  & C_{5,9}^{(1)} = & 0,                               & C_{5,9}^{(2)} = & -24\zeta_2+36\ln^2\left(2\right) - 24 \pi \ln\left(2\right) i, \nonumber \\ 
 C_{6,9}^{(0)} = & 0,  & C_{6,9}^{(1)} = & 0,                               & C_{6,9}^{(2)} = & 15 \, \mathrm{Cl}_2\left(\frac{2\pi}{3}\right) - 12 \zeta_2 i, \nonumber \\ 
 C_{7,9}^{(0)} = & 0,  & C_{7,9}^{(1)} = & \pi, & & \nonumber \\ 
 C_{8,9}^{(0)} = & 0,  & C_{8,9}^{(1)} = & 0,                               & C_{8,9}^{(2)} = & 0. & \nonumber \\ 
\end{align}
The boundary constant $C_{8,9}^{(3)}$ is rather long and has already been given in eq.~(\ref{defC_8_9_3}):
\bq
 C_{8,9}^{(3)}
 & = &
 516\,\zeta_3
-576\,\mathrm{Li}_3\left(\frac{1}{3}\right)
+576\,\mathrm{Li}_{2 1}\left(\frac{1}{3},1\right)
-120\,\ln  \left( 2 \right) \zeta_2
+96\,\ln  \left( 3 \right) \zeta_2
-96\, \ln^3\left( 3 \right)
 \nonumber \\
 & &
+72\,\ln  \left( 2 \right)  \ln^2\left( 3 \right)
+144\,\ln  \left( 2 \right) \mathrm{Li}_2\left(\frac{1}{3}\right)
-576\,\ln  \left( 3 \right) \mathrm{Li}_2\left(\frac{1}{3}\right)
-72\,\pi \,\mathrm{Cl}_2\left(\frac{2\pi}{3}\right)
 \nonumber \\
 & &
-72\,i\pi \,\zeta_2
+72\,i\pi \, \ln^2\left( 3 \right)
-48\,i\pi \,\ln  \left( 2 \right) \ln  \left( 3 \right) 
+144\,i\pi \,\mathrm{Li}_2\left(\frac{1}{3}\right).
\eq
For the boundary point $x=\infty$ we have
\begin{align}
 C_{1,\infty}^{(0)} = & 4,  & C_{1,\infty}^{(1)} = & 0, & C_{1,\infty}^{(2)} = & 4 \zeta_2, & & \nonumber \\ 
 C_{2,\infty}^{(0)} = & 0,  & C_{2,\infty}^{(1)} = & 4 \pi i, & C_{2,\infty}^{(2)} = & -16 \zeta_2, & & \nonumber \\ 
 C_{3,\infty}^{(0)} = & 0,  & C_{3,\infty}^{(1)} = & -4 \pi i, & C_{3,\infty}^{(2)} = & 40 \zeta_2, & & \nonumber \\ 
 C_{4,\infty}^{(0)} = & 4,  & C_{4,\infty}^{(1)} = & 8 \pi i, & C_{4,\infty}^{(2)} = & -52 \zeta_2, & & \nonumber \\ 
 C_{5,\infty}^{(0)} = & 0,  & C_{5,\infty}^{(1)} = & 0, & C_{5,\infty}^{(2)} = & -24 \zeta_2, & & \nonumber \\ 
 C_{6,\infty}^{(0)} = & 0,  & C_{6,\infty}^{(1)} = & 0, & C_{6,\infty}^{(2)} = & -9 \zeta_2 i, & & \nonumber \\ 
 C_{7,\infty}^{(0)} = & 0,  & C_{7,\infty}^{(1)} = & -\pi, & & & & \nonumber \\ 
 C_{8,\infty}^{(0)} = & 0,  & C_{8,\infty}^{(1)} = & 0, & C_{8,\infty}^{(2)} = & 0, & C_{8,\infty}^{(3)} = & 48 \zeta_3. \nonumber \\ 
\end{align}

% -----------------------------------------------------------------------------

\section{Supplementary material}
\label{sect:supplement}

Attached to this article is an electronic file in ASCII format with {\tt Maple} syntax, defining the quantities
\begin{flushleft}
 \verb|c_V|, \; \verb|c_S|, 
 \\
 \verb|I_case_0|, \; \verb|I_case_1|, \; \verb|I_case_9|, \; \verb|I_case_inf|, 
 \\
 \verb|Sigma_V_case_0|, \; \verb|Sigma_S_case_0|, \; \verb|Sigma_V_case_1|, \; \verb|Sigma_S_case_1|, 
 \\
 \verb|Sigma_V_case_9|, \; \verb|Sigma_S_case_9|, \; \verb|Sigma_V_case_inf|, \; \verb|Sigma_S_case_inf|.
\end{flushleft}
\verb|c_V| and \verb|c_S| are vectors, giving the coefficients $c_j^V$ and $c_j^S$ appearing in eq.~(\ref{def_linear_combination_master_integrals}) in a general covariant gauge with gauge parameter $\xi$.
\verb|I_case_0| is a vector, giving the results for the eight master integrals $I_1$-$I_8$ with boundary point $j=0$
as an expansion in $\eps$ to the relevant order ($I_1$-$I_6$ to order $\eps^2$, $I_7$ to order $\eps$
and $I_8$ to order $\eps^3$)
and an expansion in $q$ to order ${\mathcal O}(q^{\qorder})$.
\verb|I_case_1|, \verb|I_case_9| and \verb|I_case_inf| are similar, but for the boundary points
$j=1$, $j=9$ and $j=\infty$, respectively.
\verb|Sigma_V_case_0| and \verb|Sigma_S_case_0| give the $\eps^0$-parts
$\Sigma^{(2,0)}_{\mathrm{bare},V}$ and $\Sigma^{(2,0)}_{\mathrm{bare},S}$ of the bare two-loop self-energy as
an expansion in $q$ to order ${\mathcal O}(q^{\qorder})$ in a general covariant gauge.
The quantities \verb|Sigma_V_case_1|, \verb|Sigma_S_case_1|, \verb|Sigma_V_case_9|, \verb|Sigma_S_case_9|, 
\verb|Sigma_V_case_inf|, and \verb|Sigma_S_case_inf| are similar,
but for the boundary points $j=1$, $j=9$ and $j=\infty$, respectively.

\end{appendix}

%------------------------------------------------------------------------------
% references
{\footnotesize
\bibliography{/home/stefanw/notes/biblio}

\begin{thebibliography}{10}

\bibitem{Sabry:1962}
A.~Sabry,
\newblock Nucl. Phys. {\bf 33}, 401 (1962).

\bibitem{Laporta:2004rb}
S.~Laporta and E.~Remiddi,
\newblock Nucl. Phys. {\bf B704}, 349 (2005), hep-ph/0406160.
%%CITATION = HEP-PH/0406160;%%

\bibitem{MullerStach:2011ru}
S.~M{\"u}ller-Stach, S.~Weinzierl, and R.~Zayadeh,
\newblock Commun. Num. Theor. Phys. {\bf 6}, 203 (2012), arXiv:1112.4360.
%%CITATION = 1112.4360;%%

\bibitem{Adams:2013nia}
L.~Adams, C.~Bogner, and S.~Weinzierl,
\newblock J. Math. Phys. {\bf 54}, 052303 (2013), arXiv:1302.7004.
%%CITATION = ARXIV:1302.7004;%%

\bibitem{Bloch:2013tra}
S.~Bloch and P.~Vanhove,
\newblock J. Numb. Theor. {\bf 148}, 328 (2015), arXiv:1309.5865.
%%CITATION = ARXIV:1309.5865;%%

\bibitem{Remiddi:2013joa}
E.~Remiddi and L.~Tancredi,
\newblock Nucl.Phys. {\bf B880}, 343 (2014), arXiv:1311.3342.
%%CITATION = ARXIV:1311.3342;%%

\bibitem{Adams:2014vja}
L.~Adams, C.~Bogner, and S.~Weinzierl,
\newblock J. Math. Phys. {\bf 55}, 102301 (2014), arXiv:1405.5640.
%%CITATION = ARXIV:1405.5640;%%

\bibitem{Bloch:2014qca}
S.~Bloch, M.~Kerr, and P.~Vanhove,
\newblock Compos. Math. {\bf 151}, 2329 (2015), arXiv:1406.2664.
%%CITATION = ARXIV:1406.2664;%%

\bibitem{Adams:2015gva}
L.~Adams, C.~Bogner, and S.~Weinzierl,
\newblock J. Math. Phys. {\bf 56}, 072303 (2015), arXiv:1504.03255.
%%CITATION = ARXIV:1504.03255;%%

\bibitem{Adams:2015ydq}
L.~Adams, C.~Bogner, and S.~Weinzierl,
\newblock J. Math. Phys. {\bf 57}, 032304 (2016), arXiv:1512.05630.
%%CITATION = ARXIV:1512.05630;%%

\bibitem{Sogaard:2014jla}
M.~Søgaard and Y.~Zhang,
\newblock Phys. Rev. {\bf D91}, 081701 (2015), arXiv:1412.5577.
%%CITATION = ARXIV:1412.5577;%%

\bibitem{Bloch:2016izu}
S.~Bloch, M.~Kerr, and P.~Vanhove,
\newblock Adv. Theor. Math. Phys. {\bf 21}, 1373 (2017), arXiv:1601.08181.
%%CITATION = ARXIV:1601.08181;%%

\bibitem{Remiddi:2016gno}
E.~Remiddi and L.~Tancredi,
\newblock Nucl. Phys. {\bf B907}, 400 (2016), arXiv:1602.01481.
%%CITATION = ARXIV:1602.01481;%%

\bibitem{Adams:2016xah}
L.~Adams, C.~Bogner, A.~Schweitzer, and S.~Weinzierl,
\newblock J. Math. Phys. {\bf 57}, 122302 (2016), arXiv:1607.01571.
%%CITATION = ARXIV:1607.01571;%%

\bibitem{Bonciani:2016qxi}
R.~Bonciani {\em et~al.},
\newblock JHEP {\bf 12}, 096 (2016), arXiv:1609.06685.
%%CITATION = ARXIV:1609.06685;%%

\bibitem{vonManteuffel:2017hms}
A.~von Manteuffel and L.~Tancredi,
\newblock JHEP {\bf 06}, 127 (2017), arXiv:1701.05905.
%%CITATION = ARXIV:1701.05905;%%

\bibitem{Adams:2017tga}
L.~Adams, E.~Chaubey, and S.~Weinzierl,
\newblock Phys. Rev. Lett. {\bf 118}, 141602 (2017), arXiv:1702.04279.
%%CITATION = ARXIV:1702.04279;%%

\bibitem{Adams:2017ejb}
L.~Adams and S.~Weinzierl,
\newblock Commun. Num. Theor. Phys. {\bf 12}, 193 (2018), arXiv:1704.08895.
%%CITATION = ARXIV:1704.08895;%%

\bibitem{Bogner:2017vim}
C.~Bogner, A.~Schweitzer, and S.~Weinzierl,
\newblock Nucl. Phys. {\bf B922}, 528 (2017), arXiv:1705.08952.
%%CITATION = ARXIV:1705.08952;%%

\bibitem{Ablinger:2017bjx}
J.~Ablinger {\em et~al.},
\newblock J. Math. Phys. {\bf 59}, 062305 (2018), arXiv:1706.01299.
%%CITATION = ARXIV:1706.01299;%%

\bibitem{Remiddi:2017har}
E.~Remiddi and L.~Tancredi,
\newblock Nucl. Phys. {\bf B925}, 212 (2017), arXiv:1709.03622.
%%CITATION = ARXIV:1709.03622;%%

\bibitem{Bourjaily:2017bsb}
J.~L. Bourjaily, A.~J. McLeod, M.~Spradlin, M.~von Hippel, and M.~Wilhelm,
\newblock Phys. Rev. Lett. {\bf 120}, 121603 (2018), arXiv:1712.02785.
%%CITATION = ARXIV:1712.02785;%%

\bibitem{Hidding:2017jkk}
M.~Hidding and F.~Moriello,
\newblock JHEP {\bf 01}, 169 (2019), arXiv:1712.04441.
%%CITATION = ARXIV:1712.04441;%%

\bibitem{Broedel:2017kkb}
J.~Broedel, C.~Duhr, F.~Dulat, and L.~Tancredi,
\newblock JHEP {\bf 05}, 093 (2018), arXiv:1712.07089.
%%CITATION = ARXIV:1712.07089;%%

\bibitem{Broedel:2017siw}
J.~Broedel, C.~Duhr, F.~Dulat, and L.~Tancredi,
\newblock Phys. Rev. {\bf D97}, 116009 (2018), arXiv:1712.07095.
%%CITATION = ARXIV:1712.07095;%%

\bibitem{Broedel:2018iwv}
J.~Broedel, C.~Duhr, F.~Dulat, B.~Penante, and L.~Tancredi,
\newblock JHEP {\bf 08}, 014 (2018), arXiv:1803.10256.
%%CITATION = ARXIV:1803.10256;%%

\bibitem{Adams:2018yfj}
L.~Adams and S.~Weinzierl,
\newblock Phys. Lett. {\bf B781}, 270 (2018), arXiv:1802.05020.
%%CITATION = ARXIV:1802.05020;%%

\bibitem{Adams:2018bsn}
L.~Adams, E.~Chaubey, and S.~Weinzierl,
\newblock Phys. Rev. Lett. {\bf 121}, 142001 (2018), arXiv:1804.11144.
%%CITATION = ARXIV:1804.11144;%%

\bibitem{Adams:2018kez}
L.~Adams, E.~Chaubey, and S.~Weinzierl,
\newblock JHEP {\bf 10}, 206 (2018), arXiv:1806.04981.
%%CITATION = ARXIV:1806.04981;%%

\bibitem{Broedel:2018qkq}
J.~Broedel, C.~Duhr, F.~Dulat, B.~Penante, and L.~Tancredi,
\newblock JHEP {\bf 01}, 023 (2019), arXiv:1809.10698.
%%CITATION = ARXIV:1809.10698;%%

\bibitem{Bourjaily:2018yfy}
J.~L. Bourjaily, A.~J. McLeod, M.~von Hippel, and M.~Wilhelm,
\newblock Phys. Rev. Lett. {\bf 122}, 031601 (2019), arXiv:1810.07689.
%%CITATION = ARXIV:1810.07689;%%

\bibitem{Bourjaily:2018aeq}
J.~L. Bourjaily, A.~J. McLeod, M.~von Hippel, and M.~Wilhelm,
\newblock JHEP {\bf 08}, 184 (2018), arXiv:1805.10281.
%%CITATION = ARXIV:1805.10281;%%

\bibitem{Besier:2018jen}
M.~Besier, D.~Van~Straten, and S.~Weinzierl,
\newblock Commun. Num. Theor. Phys. {\bf 13}, 253 (2019), arXiv:1809.10983.
%%CITATION = ARXIV:1809.10983;%%

\bibitem{Mastrolia:2018uzb}
P.~Mastrolia and S.~Mizera,
\newblock JHEP {\bf 02}, 139 (2019), arXiv:1810.03818.
%%CITATION = ARXIV:1810.03818;%%

\bibitem{Ablinger:2018zwz}
J.~Ablinger, J.~Blümlein, P.~Marquard, N.~Rana, and C.~Schneider,
\newblock Nucl. Phys. {\bf B939}, 253 (2019), arXiv:1810.12261.
%%CITATION = ARXIV:1810.12261;%%

\bibitem{Broadhurst:1993mw}
D.~J. Broadhurst, J.~Fleischer, and O.~Tarasov,
\newblock Z.Phys. {\bf C60}, 287 (1993), arXiv:hep-ph/9304303.
%%CITATION = HEP-PH/9304303;%%

\bibitem{Fleischer:1994ef}
J.~Fleischer and O.~V. Tarasov,
\newblock Z. Phys. {\bf C64}, 413 (1994), arXiv:hep-ph/9403230.
%%CITATION = HEP-PH/9403230;%%

\bibitem{Caffo:2008aw}
M.~Caffo, H.~Czyz, M.~Gunia, and E.~Remiddi,
\newblock Comput. Phys. Commun. {\bf 180}, 427 (2009), arXiv:0807.1959.
%%CITATION = 0807.1959;%%

\bibitem{Pozzorini:2005ff}
S.~Pozzorini and E.~Remiddi,
\newblock Comput. Phys. Commun. {\bf 175}, 381 (2006), arXiv:hep-ph/0505041.
%%CITATION = HEP-PH/0505041;%%

\bibitem{Passarino:2016zcd}
G.~Passarino,
\newblock Eur. Phys. J. {\bf C77}, 77 (2017), arXiv:1610.06207.
%%CITATION = ARXIV:1610.06207;%%

\bibitem{Kotikov:1990kg}
A.~V. Kotikov,
\newblock Phys. Lett. B {\bf 254}, 158 (1991).

\bibitem{Kotikov:1991pm}
A.~V. Kotikov,
\newblock Phys. Lett. B {\bf 267}, 123 (1991),
\newblock [Erratum: Phys.Lett.B 295, 409--409 (1992)].

\bibitem{Remiddi:1997ny}
E.~Remiddi,
\newblock Nuovo Cim. {\bf A110}, 1435 (1997), hep-th/9711188.
%%CITATION = HEP-TH 9711188;%%

\bibitem{Gehrmann:1999as}
T.~Gehrmann and E.~Remiddi,
\newblock Nucl. Phys. {\bf B580}, 485 (2000), hep-ph/9912329.
%%CITATION = NUPHA,B580,485;%%

\bibitem{Argeri:2007up}
M.~Argeri and P.~Mastrolia,
\newblock Int. J. Mod. Phys. {\bf A22}, 4375 (2007), arXiv:0707.4037.
%%CITATION = 0707.4037;%%

\bibitem{MullerStach:2012mp}
S.~M{\"u}ller-Stach, S.~Weinzierl, and R.~Zayadeh,
\newblock Commun.Math.Phys. {\bf 326}, 237 (2014), arXiv:1212.4389.
%%CITATION = ARXIV:1212.4389;%%

\bibitem{Henn:2013pwa}
J.~M. Henn,
\newblock Phys. Rev. Lett. {\bf 110}, 251601 (2013), arXiv:1304.1806.
%%CITATION = ARXIV:1304.1806;%%

\bibitem{Henn:2014qga}
J.~M. Henn,
\newblock J. Phys. {\bf A48}, 153001 (2015), arXiv:1412.2296.
%%CITATION = ARXIV:1412.2296;%%

\bibitem{Tancredi:2015pta}
L.~Tancredi,
\newblock Nucl. Phys. {\bf B901}, 282 (2015), arXiv:1509.03330.
%%CITATION = ARXIV:1509.03330;%%

\bibitem{Ablinger:2015tua}
J.~Ablinger {\em et~al.},
\newblock Comput. Phys. Commun. {\bf 202}, 33 (2016), arXiv:1509.08324.
%%CITATION = ARXIV:1509.08324;%%

\bibitem{Bosma:2017hrk}
J.~Bosma, K.~J. Larsen, and Y.~Zhang,
\newblock Phys. Rev. {\bf D97}, 105014 (2018), arXiv:1712.03760.
%%CITATION = ARXIV:1712.03760;%%

\bibitem{Bern:2000ie}
Z.~Bern, L.~Dixon, and A.~Ghinculov,
\newblock Phys. Rev. {\bf D63}, 053007 (2001), hep-ph/0010075.
%%CITATION = PHRVA,D63,053007;%%

\bibitem{Bern:2000dn}
Z.~Bern, L.~Dixon, and D.~A. Kosower,
\newblock JHEP {\bf 01}, 027 (2000), hep-ph/0001001.
%%CITATION = JHEPA,0001,027;%%

\bibitem{Anastasiou:2000kg}
C.~Anastasiou, E.~W.~N. Glover, C.~Oleari, and M.~E. Tejeda-Yeomans,
\newblock Nucl. Phys. {\bf B601}, 318 (2001), hep-ph/0010212.
%%CITATION = HEP-PH 0010212;%%

\bibitem{Anastasiou:2000ue}
C.~Anastasiou, E.~W.~N. Glover, C.~Oleari, and M.~E. Tejeda-Yeomans,
\newblock Nucl. Phys. {\bf B601}, 341 (2001), hep-ph/0011094.
%%CITATION = HEP-PH 0011094;%%

\bibitem{Anastasiou:2000mv}
C.~Anastasiou, E.~W.~N. Glover, C.~Oleari, and M.~E. Tejeda-Yeomans,
\newblock Phys. Lett. {\bf B506}, 59 (2001), hep-ph/0012007.
%%CITATION = HEP-PH 0012007;%%

\bibitem{Anastasiou:2001sv}
C.~Anastasiou, E.~W.~N. Glover, C.~Oleari, and M.~E. Tejeda-Yeomans,
\newblock Nucl. Phys. {\bf B605}, 486 (2001), hep-ph/0101304.
%%CITATION = HEP-PH 0101304;%%

\bibitem{Glover:2001af}
E.~W.~N. Glover, C.~Oleari, and M.~E. Tejeda-Yeomans,
\newblock Nucl. Phys. {\bf B605}, 467 (2001), hep-ph/0102201.
%%CITATION = HEP-PH 0102201;%%

\bibitem{Bern:2002tk}
Z.~Bern, A.~De~Freitas, and L.~Dixon,
\newblock JHEP {\bf 03}, 018 (2002), hep-ph/0201161.
%%CITATION = HEP-PH 0201161;%%

\bibitem{Garland:2001tf}
L.~W. Garland, T.~Gehrmann, E.~W.~N. Glover, A.~Koukoutsakis, and E.~Remiddi,
\newblock Nucl. Phys. {\bf B627}, 107 (2002), hep-ph/0112081.
%%CITATION = HEP-PH 0112081;%%

\bibitem{Garland:2002ak}
L.~W. Garland, T.~Gehrmann, E.~W.~N. Glover, A.~Koukoutsakis, and E.~Remiddi,
\newblock Nucl. Phys. {\bf B642}, 227 (2002), hep-ph/0206067.
%%CITATION = HEP-PH 0206067;%%

\bibitem{Moch:2002hm}
S.~Moch, P.~Uwer, and S.~Weinzierl,
\newblock Phys. Rev. {\bf D66}, 114001 (2002), hep-ph/0207043.
%%CITATION = HEP-PH 0207043;%%

\bibitem{Czakon:2013goa}
M.~Czakon, P.~Fiedler, and A.~Mitov,
\newblock Phys.Rev.Lett. {\bf 110}, 252004 (2013), arXiv:1303.6254.
%%CITATION = ARXIV:1303.6254;%%

\bibitem{Baernreuther:2013caa}
P.~B{\"a}rnreuther, M.~Czakon, and P.~Fiedler,
\newblock JHEP {\bf 02}, 078 (2014), arXiv:1312.6279.
%%CITATION = ARXIV:1312.6279;%%

\bibitem{Lee:2017qql}
R.~N. Lee, A.~V. Smirnov, and V.~A. Smirnov,
\newblock JHEP {\bf 03}, 008 (2018), arXiv:1709.07525.
%%CITATION = ARXIV:1709.07525;%%

\bibitem{Lee:2018ojn}
R.~N. Lee, A.~V. Smirnov, and V.~A. Smirnov,
\newblock JHEP {\bf 07}, 102 (2018), arXiv:1805.00227.
%%CITATION = ARXIV:1805.00227;%%

\bibitem{Ferguson:1992}
H.~R.~P. Ferguson and D.~H. Bailey,
\newblock RNR Technical Report RNR-91-032  (1992).

\bibitem{Egorian:1978zx}
E.~Egorian and O.~V. Tarasov,
\newblock Teor. Mat. Fiz. {\bf 41}, 26 (1979),
\newblock [Theor. Math. Phys.41,863(1979)].
%%CITATION = TMFZA,41,26;%%

\bibitem{Tarrach:1980up}
R.~Tarrach,
\newblock Nucl. Phys. {\bf B183}, 384 (1981).
%%CITATION = NUPHA,B183,384;%%

\bibitem{Gray:1990yh}
N.~Gray, D.~J. Broadhurst, W.~Grafe, and K.~Schilcher,
\newblock Z. Phys. {\bf C48}, 673 (1990).
%%CITATION = ZEPYA,C48,673;%%

\bibitem{Broadhurst:1991fy}
D.~J. Broadhurst, N.~Gray, and K.~Schilcher,
\newblock Z. Phys. {\bf C52}, 111 (1991).
%%CITATION = ZEPYA,C52,111;%%

\bibitem{Chetyrkin:1999qi}
K.~Chetyrkin and M.~Steinhauser,
\newblock Nucl.Phys. {\bf B573}, 617 (2000), arXiv:hep-ph/9911434.
%%CITATION = HEP-PH/9911434;%%

\bibitem{Melnikov:2000qh}
K.~Melnikov and T.~v. Ritbergen,
\newblock Phys.Lett. {\bf B482}, 99 (2000), arXiv:hep-ph/9912391.
%%CITATION = HEP-PH/9912391;%%

\bibitem{Melnikov:2000zc}
K.~Melnikov and T.~van Ritbergen,
\newblock Nucl. Phys. {\bf B591}, 515 (2000), arXiv:hep-ph/0005131.
%%CITATION = HEP-PH/0005131;%%

\bibitem{Marquard:2016dcn}
P.~Marquard, A.~V. Smirnov, V.~A. Smirnov, M.~Steinhauser, and D.~Wellmann,
\newblock Phys. Rev. {\bf D94}, 074025 (2016), arXiv:1606.06754.
%%CITATION = ARXIV:1606.06754;%%

\bibitem{Marquard:2018rwx}
P.~Marquard, A.~V. Smirnov, V.~A. Smirnov, and M.~Steinhauser,
\newblock Phys. Rev. {\bf D97}, 054032 (2018), arXiv:1801.08292.
%%CITATION = ARXIV:1801.08292;%%

\bibitem{Luthe:2016xec}
T.~Luthe, A.~Maier, P.~Marquard, and Y.~Schröder,
\newblock JHEP {\bf 01}, 081 (2017), arXiv:1612.05512.
%%CITATION = ARXIV:1612.05512;%%

\bibitem{Luthe:2017ttc}
T.~Luthe, A.~Maier, P.~Marquard, and Y.~Schroder,
\newblock JHEP {\bf 03}, 020 (2017), arXiv:1701.07068.
%%CITATION = ARXIV:1701.07068;%%

\bibitem{Chetyrkin:2017bjc}
K.~G. Chetyrkin, G.~Falcioni, F.~Herzog, and J.~A.~M. Vermaseren,
\newblock JHEP {\bf 10}, 179 (2017), arXiv:1709.08541,
\newblock [Addendum: JHEP12,006(2017)].
%%CITATION = ARXIV:1709.08541;%%

\bibitem{Chen}
K.-T. Chen,
\newblock Bull. Amer. Math. Soc. {\bf 83}, 831 (1977).

\bibitem{Remiddi:1999ew}
E.~Remiddi and J.~A.~M. Vermaseren,
\newblock Int. J. Mod. Phys. {\bf A15}, 725 (2000), hep-ph/9905237.
%%CITATION = IMPAE,A15,725;%%

\bibitem{Brown:2014aa}
F.~{Brown},
\newblock (2014), arXiv:1407.5167.

\bibitem{Gehrmann:2001pz}
T.~Gehrmann and E.~Remiddi,
\newblock Comput. Phys. Commun. {\bf 141}, 296 (2001), hep-ph/0107173.
%%CITATION = HEP-PH 0107173;%%

\bibitem{Vollinga:2004sn}
J.~Vollinga and S.~Weinzierl,
\newblock Comput. Phys. Commun. {\bf 167}, 177 (2005), hep-ph/0410259.
%%CITATION = HEP-PH 0410259;%%

\bibitem{Maitre:2005uu}
D.~Maitre,
\newblock Comput. Phys. Commun. {\bf 174}, 222 (2006), hep-ph/0507152.
%%CITATION = HEP-PH/0507152;%%

\bibitem{Maitre:2007kp}
D.~Maitre,
\newblock Comput. Phys. Commun. {\bf 183}, 846 (2012), arXiv:hep-ph/0703052.
%%CITATION = HEP-PH/0703052;%%

\bibitem{Cox:1984}
D.~A. Cox,
\newblock Enseign. Math. {\bf 30}, 275 (1984).

\end{thebibliography}
\bibliographystyle{/home/stefanw/latex-style/h-physrev5}
}

\end{document}